\documentclass{aa}  
\usepackage{tabularx}
\usepackage{graphicx}
\usepackage[normalem]{ulem}
\usepackage{longtable}
\usepackage[table]{xcolor}
\usepackage{booktabs}
\usepackage{graphicx}
\usepackage{multirow}
\usepackage{hyperref}
\usepackage{float}
\usepackage{subcaption}
\usepackage{txfonts}
\usepackage{placeins}
\usepackage{lipsum}
\usepackage{soul}
\usepackage{makecell}
\usepackage{array}
\usepackage{orcidlink}
\usepackage{subcaption}  
\usepackage{lscape}      
\usepackage{lineno}
\usepackage{booktabs}                           
\usepackage{hyperref} 
\hypersetup{
    colorlinks=true,
    linkcolor=blue, 
    filecolor=magenta,
    urlcolor=cyan,
    citecolor=blue, 
    pdftitle={TDR
},
    pdfauthor={A. Chopra et al.},
}
\authorrunning{Chopra et al.}

\begin{document}
\nolinenumbers

   \title{Detectability of Gravitational-wave counterparts of EP-FXTs observed during the O4b LIGO-Virgo-KAGRA Observing Run}
   \titlerunning{Search for GW counterpart of FXTs}
   \authorrunning{Chopra et al 2026}

   \author{
   {A.} Chopra\orcidlink{0000-0001-5679-0695}\inst{1,2}\thanks{\email{ansh.chopra@gssi.it}},
   {S.} Ronchini\orcidlink{0000-0003-0020-687X}\inst{1,2}\thanks{\email{samuele.ronchini@gssi.it}},
   {B.} Banerjee\orcidlink{0000-0002-8008-2485}\inst{1, 2}\thanks{\email{biswajit.banerjee@gssi.it}},
   {M.} Branchesi\orcidlink{0000-0003-1643-0526}\inst{1,2},
   {S.} Ascenzi\orcidlink{0000-0001-5116-6789}\inst{1, 2},
   {M. E.} Ravasio\orcidlink{0000-0003-3193-4714}\inst{3,4,5,6},
   {P. G.} Jonker\orcidlink{0000-0001-5679-0695}\inst{5},
   {A. J.} Levan\orcidlink{0000-0001-7821-9369}\inst{5, 7}}
   
   \institute{Gran Sasso Science Institute (GSSI), I-67100 L'Aquila, Italy
   \and
   INFN, Laboratori Nazionali del Gran Sasso, I-67100 Assergi, Italy
   \and
   Institute of Space Sciences (ICE, CSIC), Campus UAB, Carrer de Can Magrans s/n, Barcelona, E-08193, Spain
   \and 
   Institut d’Estudis Espacials de Catalunya (IEEC), Edifici RDIT, Campus UPC, Castelldefels (Barcelona), E-08860, Spain
   \and 
   Department of Astrophysics/IMAPP, Radboud University, 6525 AJ Nijmegen, The Netherlands
   \and
   INAF -- Osservatorio Astronomico di Brera, via Emilio Bianchi 46, I-23807 Merate (LC), Italy
   \and
   Department of Physics, University of Warwick, Coventry, CV4 7AL, UK
   }

\abstract
{Fast X-ray transients (FXTs) detected by the Einstein Probe mission have emerged as a rapidly growing class of extragalactic transients, whose physical origin remains uncertain. Compact binary coalescence (CBC) systems 
have been proposed as one possible progenitor for at least a subset of these events, making FXTs promising targets for multi-messenger studies with gravitational-wave (GW) observations. This work presents the first systematic investigation of GW counterparts to FXTs observed by Einstein Probe and assesses the detectability of associated CBC. We focus on FXTs detected during the second half of the fourth observing run (O4b) of the Advanced LIGO-Virgo-KAGRA detector network by searching for temporal coincidences with GW candidates from the fifth Gravitational-Wave Transient Catalog (GWTC-5). 
We analyze a sample of 47 FXTs, including 11 with measured redshifts, and assess the significance of the association between FXTs and GW candidates using a ranking statistic.
We find no significant GW counterpart associated with any FXT in our sample. In the absence of a detection, we place 90\% exclusion-distance constraints under the assumptions of binary neutron star and neutron star–black hole progenitor scenarios. For observations with the full LIGO and Virgo detector network, the typical median exclusion distances are $\sim178$ Mpc and $\sim349$ Mpc, respectively. These constraints disfavor a nearby compact-binary merger origin.
Longer periods of joint observations by the LVK and Einstein Probe, combined with improved GW detector sensitivity, will enhance the prospects for identifying genuine GW–FXT associations and place tighter constraints on the progenitor scenarios of these events.
}


   \maketitle
   \nolinenumbers

\section{Introduction}

Fast X-ray transients (FXTs) are a class of high-energy transients characterized by durations ranging from a few seconds to several hours \citep[e.g.,][]{2017MNRAS.467.4841B, 2019Natur.568..198X, 2020ApJ...896...39A, 2023A&A...675A..44Q, 2024GCN.36405....1L}. The first well-localized FXT was serendipitously discovered in the Swift/XRT observation XRO 080109 by \citet{2008Natur.453..469S}. Subsequent systematic searches of archival observations from the Chandra X-ray Observatory  identified additional FXTs localized to arcsecond accuracies \citep{2013ApJ...779...14J, Quirola, 2023A&A...675A..44Q}. Since then, FXTs have also been detected by other X-ray facilities, including XMM-Newton, eROSITA, and NuSTAR \citep{2020ApJ...896...39A,  2025A&A...693A..62G, 2026ApJ...999...75B}. The launch of Einstein Probe \citep[EP;][]{2025SCPMA..6839501Y} has dramatically increased the discovery rate of FXTs. Its Wide-field X-ray Telescope (WXT), operating in the 0.5-4 keV band with a field of view of about 3600 deg$^2$, is uncovering a rapidly growing population of FXTs localized to a few arcminutes.
While observations of local EP events have shown many to be related to the core collapse of massive stars (e.g. \citealt{2025ApJ...982L..47V, 2025ApJ...988L..13R, 2026arXiv260606213C, 2026arXiv260610002M}), many other progenitor scenarios have also been widely discussed in the literature. These include the tidal disruption of a white dwarf by an intermediate-mass black hole \citep{2013ApJ...779...14J, 2025ApJ...990L..28L, 2025A&A...703L...2O, 2026arXiv260617230E}, the explosion of massive stars \citep{2008Natur.453..469S},  and compact binary coalescences (CBCs) such as neutron star - black hole (NSBH) and binary neutron star (BNS) systems \citep{2019Natur.568..198X, 2026A&A...705A.233B}. However, their relative contributions to the observed population is still uncertain. FXTs might also represent some unknown astrophysical phenomena.

In the case of a CBC origin, several processes are able to generate X-ray transients with the typical duration of an FXT. The merger remnant can power a relativistic jet \citep{1989Natur.340..126E, 1992ApJ...395L..83N} through the neutrino–antineutrino annihilation process \citep{Eichler:1989ve, 2011MNRAS.410.2302Z}, or by the  Blandford–Znajek mechanism \citep{1977MNRAS.179..433B}. Part of its kinetic energy can be internally dissipated, powering the so-called Gamma-ray burst (GRB) \emph{prompt emission}, detectable in the MeV band by observers aligned with the jet axis \citep{2004RvMP...76.1143P, 2006RPPh...69.2259M, 2015PhR...561....1K}. In this case, a FXT can emerge 
in the tail of the prompt emission \citep{2000ApJ...541L..51K, 2020ApJ...893...88O, 2026A&A...708A.190I}, and may also be observed off-axis \citep{2005astro.ph.11699D, 2020A&A...641A..61A, 2026ApJ..1001..177C}.
Moreover, the FXT rate is difficult to reconcile with beamed emission, suggesting that most FXTs can not be on-axis GRB-like systems, and motivates emission channels with a broader angular distribution.
For events where the jet does not emerge successfully, and instead suppressed by the dense post-merger ejecta, more isotropic X-ray emission can occur in correspondence to the breakout of a relativistic shock, generated by the interaction of the chocked jet with the ejecta \citep{2017ApJ...848L...6L, 2017Sci...358.1559K, 2018Natur.554..207M, 2018MNRAS.479..588G, 2025PhRvD.111f3031G}. FXTs could also represent the X-ray flare due to the late-time activity of the central engine \citep{2006ApJ...642..354Z, 2007RSPTA.365.1213B, 2011MNRAS.413.2031M, 2011MNRAS.417.2144M, 2016ApJ...829...72C}.

A BNS merger may leave behind a long-lived millisecond magnetar \citep{1994MNRAS.270..480T, 1998PhRvL..81.4301D, 2001ApJ...552L..35Z, 2013ApJ...771L..26G, 2022ApJ...927..211L, 2024A&A...683A.243Q}, whose spin-down energy can power a wind whose emission can peak in the X-rays, with a typical duration compatible with the one of FXTs. 
Magnetar remnants have been invoked as possible GRB central engines \citep{1992Natur.357..472U, 1994MNRAS.270..480T} and to explain several X-ray features observed in GRBs and related transients, including plateau phases in GRB afterglows and late-time X-ray flares powered by magnetar spin-down \citep{2009ApJ...702.1171C, 2010MNRAS.402..705L, 2008MNRAS.385.1455M, 2011A&A...526A.121D, 
2011MNRAS.413.2031M, 2013ApJ...763L..22Z, 2013MNRAS.430.1061R, 2013MNRAS.431.1745G, 2016ApJ...829...72C, 2019Natur.568..198X, 2022ASSL..465..245D}.
Recently, the soft X-ray emission detected from the compact-merger candidate GRB 230307A \citep{2025NSRev..12E.401S, 2024Natur.626..737L, 2026ApJ..1000...97D} by LEIA\footnote{Lobster Eye Imager for Astronomy is the pathfinder of the Einstein Probe mission, and has a FoV of 340 deg$^{2}$.} as well as the extended emission of a short Gamma-Ray Burst GRB250704B, identified as EP250704a \citep{2026arXiv260115732F} have been interpreted as the emergence of a magnetar-powered component. 
The spin-down energy dissipated in the wind can be absorbed and thermalized by the matter ejected during the merger and post-merger phase of the BNS coalescence. This matter, heated and accelerated by the energy injection, is expected to re-emit quasi-isotropically, powering a transient potentially observable in X-rays even for a large inclination of the binary system (\citealp{2013ApJ...776L..40Y, 2014MNRAS.439.3916M, 2016ApJ...819...14S, 2016ApJ...819...15S}; see also \citealp[for a review]{2021JPlPh..87a8402A}).
Consequently, even in the absence of a coincident gamma-ray detection, several models predict the existence of fast-evolving X-ray transients, which could be promising 
candidates for the electromagnetic counterparts of neutron-star mergers \citep[see][]{2026MNRAS.545f2021J}, although such events are expected to represent only a subset of the broader EP-FXT population.

CBCs are also known to produce gravitational-wave (GW) emission in the frequency band of the Laser
Interferometer Gravitational-Wave Observatory \citep[LIGO;][]{2015CQGra..32g4001L}, Virgo \citep{2015CQGra..32b4001A} and KAGRA \citep{2021PTEP.2021eA101A} obsvervatories, making FXTs promising targets for 
multi-messenger searches. The LIGO-Virgo-KAGRA collaboration \citep{2018LRR....21....3A} searches for GW transients through two complementary strategies: all-sky, all-time searches that scan the full sky continuously without prior assumptions on source location or time \citep{2017PhRvD..95d2003A, 2019PhRvD.100b4017A, 2021PhRvD.104l2004A, 2025PhRvD.112j2005A}, and targeted searches that exploit the known sky position and trigger time provided by an external electromagnetic observation (O1:  \citealt{2017ApJ...841...89A}; O2: \citealt{2019ApJ...886...75A}; O3a: \citealt{2021ApJ...915...86A}; O3b: \citealt{2022ApJ...928..186A}). The latter benefit from a reduced search parameter space, as constraining both the time and sky location of the putative source lowers the detection threshold and increases the effective sensitive volume compared to an untriggered search \citep{2013PhRvD..87f4033D}. GW candidates identified by the online all-sky pipelines are reported in real time through the GW Candidate Event Database (GraceDB\footnote{https://gracedb.ligo.org/})
and can be cross-matched against X-ray transient catalogues to search for spatial and temporal coincidences. For the off-line targeted search, two analysis approaches are adopted, depending on the assumed progenitor. For merger-driven GRBs, modelled searches are performed with \texttt{PyGRB} \citep{2011PhRvD..83h4002H, 2014PhRvD..90l2004W}, which applies matched-filter techniques using CBC waveforms to identify the expected signals. Long GRBs and transients with uncertain GW morphology are instead analyzed with coherent burst pipelines such as X-Pipeline \citep{2010NJPh...12e3034S, 2012PhRvD..86b2003W}, which combine data from multiple detectors and search for statistically significant coherent excesses above the noise across the network detectors.

Two FXTs detected by EP/WXT, EP250704a \citep{2026arXiv260115732F} and EP260527a \citep{2026GCN.44722....1S}, have been associated with short GRB GRB250704B and GRB260527A, respectively. Their soft X-ray durations are, however, much longer than the prompt gamma-ray durations usually associated with short GRBs, showing that FXT duration alone is not a reliable criterion for identifying 
CBC-related events. In contrast, a larger sample of EP-FXT shows a clear association with long GRBs \citetext{EP240315a: \citealp{2025NatAs...9..564L}; EP240801a: \citealp{2025ApJ...988L..34J}; EP250702a: \citealp{2026SciBu..71..538L}}, while a few others hint at a similar origin from the afterglow observed in the optical band, despite the lack of a clear detection in the MeV band \citep{2026MNRAS.545f2062E}. However, the majority of these events have no identified gamma-ray counterpart \citetext{\citealp{2026MNRAS.545f2021J, 2026MNRAS.545f2064Q}; Ravasio et al. 2026 in prep.}, although some show evidence of collapsar origin through associations with a broad-lined Type Ic supernova \citep{2025ApJ...982L..47V, 2025ApJ...988L..60S, 2026arXiv260606213C}. 
Thus, the EP-FXT population is heterogeneous, with GRB- and collapsar-related events contributing a substantial fraction. 
Nevertheless, FXTs without an identified gamma-ray counterpart remain compelling targets for GW follow-up, as a CBC origin cannot be excluded based on the electromagnetic observations alone. Moreover, LVK targeted searches have so far been performed for GRB triggers, while a systematic search for GW counterparts to FXTs without observed MeV emission has not yet been carried out.

In this work, we investigate the presence of GW emission associated with FXTs observed by EP during the second half of the fourth observing run (O4b) of LVK, excluding those associated with observed GRB MeV emission, in order to avoid overlap with the LVK searches targeted on GRB triggers.
We perform a search for temporally and spatially coincident GW candidates and evaluate the association significance. We also investigate the detectability of CBC signals using the Targeted Detectability Range  \citep[TDR;][]{2026arXiv260521578R} framework, which enables us to estimate exclusion distances for BNS and NSBH merger scenarios in the absence of a GW detection in all-sky, all-time GW observations.

The paper is organized as follows. In Sect.~\ref{Sec:2}, we describe the selection criteria for the analyzed sample of EP FXTs and the definition of the temporal search windows for GW emission. Sect.~\ref{Sec:4} presents the methodology used to evaluate the significance of coincident GW candidates. In Sect.~\ref{Sec:5}, we present the exclusion distance estimates using the TDR.  Finally, we summarize our results, discuss their implications and conclude in Sect.~\ref{Sec:6} and ~\ref{Sec:conclusion}, respectively.

\begin{table*}[!t]
\caption{Einstein Probe fast X-ray transients used in this work. The table reports the source name, the trigger time, duration, sky position, discovery circular or telegram, GW detector availability during the adopted search window, and redshift when available. The search window for each source is defined as $[T_0-1000~{\rm s}, T_0+T_{100}]$, where $T_0$ is the Einstein Probe trigger time and $T_{100}$ is the reported transient duration. The time intervals indicated with the (*) symbol have partial coverage. This table defines the source sample used for the temporal-coincidence search in Tab.~\ref{ep_gw_candidates} and for the targeted detectability-range calculations shown in Fig.~\ref{D90_evol}.}
\label{IFO-list}
\centering
\small
\setlength{\tabcolsep}{7pt}
\renewcommand{\arraystretch}{0.95}
\begin{tabular}{lcrrrcrrrr}
\toprule
Name & T$_{0}$ & T$_{100}$ & RA&DEC & Citation & \multicolumn{3}{c}{duration of online IFO [s]} & Redshift (z)\\
\cmidrule(lr){7-9}
EP- & UTC & [s] & [deg]& [deg] &  & H1 & L1 & V1 & [Citation] \\
\midrule
240413a & 14:39:37 & 200 & 228.795 & $-18.799$ & \href{https://gcn.nasa.gov/circulars/36086}{36086} & [-1000, 200] & [-1000, 200]  & [-1000, -848]* & - \\
 &&&&&&&&[-592, 200]*&\\
240414a & 09:50:12 & - & 191.498 & $-9.695$ & \href{https://gcn.nasa.gov/circulars/36091}{36091} & - & [-1000, 60]  & [-1000, 60] & 0.40 [\href{https://gcn.nasa.gov/circulars/36110?view=index&query=EP240414a&startDate=&endDate=&sort=circularID}{36110}] \\
240416a & 02:42:13 & 200 & 203.151 & $-13.611$ & \href{https://gcn.nasa.gov/circulars/36138}{36138} & -- & [-1000, 200]  & [-1000, -816]* & -  \\
&&&&&&&&[-560, -80]*&\\
&&&&&&&&[184, 200]*&\\
240417a & 15:12:33 & 1500 & 177.442 & $-15.437$ & \href{https://gcn.nasa.gov/circulars/36161}{36161} & [-1000, 928]* & [-1000, -144]* & [-1000, -784]* & - \\
240420a & 12:04:28 & 100 & 228.729 & 14.802 & \href{https://gcn.nasa.gov/circulars/36194}{36194} & -- & [-1000, 100]  & [-1000, 100] & 0.82 [\href{https://gcn.nasa.gov/circulars/36202?view=index&query=EP240420a&startDate=&endDate=&sort=circularID}{36202}] \\
240426b & 14:19:06 & 300 & 173.787 & $-40.741$ & \href{https://gcn.nasa.gov/circulars/36330}{36330} & [-1000, 300]  & [-1000, 300]  & [-1000, 300] & -  \\
240506a & 05:01:39 & 50 & 213.987 & $-16.688$ & \href{https://gcn.nasa.gov/circulars/36405}{36405} & [-1000, 50]  & [-1000, 50]  & [-1000, 50] & 0.12 $^{\dagger}$  \\
240527a$^{\S}$  & 08:41:28 & - & 13.186 & $-72.483$ & \href{https://www.astronomerstelegram.org/?read=16631}{16631$^{\ddagger}$} & [-1000, 60]  & [-1000, 60]  & [-1000, 60] & - \\
240618a & 05:43:43 & 100 & 281.623 & 23.828 & \href{https://gcn.nasa.gov/circulars/36690}{36690} & [-1000, -720]*  & [-1000, 100]  & [-1000, 100] & - \\
240625a & 01:48:23 & 300 & 310.76 & $-15.966$ & \href{https://gcn.nasa.gov/circulars/36757}{36757} & [-1000, 300]  & [-1000, 300]  & [-1000, 300] & -  \\
240626a & 06:28:28 & 160 & 263.023 & $-13.051$ & \href{https://gcn.nasa.gov/circulars/36766}{36766} & [-1000, 160]  & [-1000, 160]  & [-1000, 160] & -  \\
240702a & 00:50:05 & 50 & 328.203 & $-38.98$ & \href{https://gcn.nasa.gov/circulars/36801}{36801} & [-1000, 50]  & -- & [-1000, 50] & -  \\
240703b & 05:24:26 & 600 & 279.536 & $-57.401$ & \href{https://gcn.nasa.gov/circulars/36810}{36810} & -- & [-1000, -432]* & [-1000, -912]* & - \\
&&&&&&&[176, 600]*&[-624, 600]*&\\
 & & & & & & [176, 600]* & [-624, 600]* & - \\
240703c & 18:15:00 & 1000 & 289.264 & $-30.325$ & \href{https://gcn.nasa.gov/circulars/36818}{36818} & -- & -- & -- & - \\
240708a & 23:28:23 & 1300 & 345.963 & $-22.84$ & \href{https://gcn.nasa.gov/circulars/37838}{37838} & [-1000, 1300]  & [-1000, 1300]  & [-1000, 1300] & 0.37$^{\ast}$  \\
240709a & 13:35:48 & - & 7.91 & $-56.76$ &  \href{https://www.astronomerstelegram.org/?read=16704}{16704$^{\ddagger}$} & [-1000, 60]  & -- & [-1000, 60] & -  \\
240806a & 04:47:53 & 150 & 11.496 & 5.085 & \href{https://gcn.nasa.gov/circulars/37063}{37063} & -- & [-1000, 150]  & [-1000, 150] & 2.818 [\href{https://gcn.nasa.gov/circulars/37087?view=index&query=EP240806a&startDate=&endDate=&sort=circularID}{37087}]  \\
240809a & 16:53:08 & - & 268.268 & $-45.864$ & \href{https://www.astronomerstelegram.org/?read=16765}{16765$^{\ddagger}$} & -- & [-1000, 60]  & [-1000, 60] & -  \\
240816a & 03:28:42 & 200 & 292.925 & $-54.412$ &  \href{https://gcn.nasa.gov/circulars/37188}{37188} & -- & -- & [-1000, 200]  & - \\
240816b & 01:44:27 & 50 & 16.013 & 15.398 &  \href{https://gcn.nasa.gov/circulars/37185}{37185} & -- & -- & [-1000, 50]  & - \\
240820a & 00:54:47 & 250 & 16.221 & $-34.698$ & \href{https://gcn.nasa.gov/circulars/37214}{37214} & -- & [-1000, 250]  & [-1000, 250] & -  \\
240908a & 17:28:27 & 950 & 13.992 & 8.089 & \href{https://gcn.nasa.gov/circulars/37443}{37443} & [-304, 950]*  & [-1000, 688]*  & [-1000, 950]  & - \\
240918a & 11:21:52 & 170 & 289.393 & 46.128 & \href{https://gcn.nasa.gov/circulars/37554}{37554} & [-1000, 170]  & -- & [-1000, 170] & -  \\
240918b & 15:40:00 & 200 & 258.66 & 66.739 & \href{https://gcn.nasa.gov/circulars/37555}{37555} & [-1000, 200]  & [-1000, 200]  & [-1000, -560]*  & -\\
&&&&&&&&[-304, 200]*&\\
240918c & 18:06:47 & 100 & 281.338 & $-13.167$ & \href{https://gcn.nasa.gov/circulars/37555}{37555} &  [-1000, 100] & [-1000, 100] & [-1000, 16]* & - \\
241021a & 05:07:56 & 100 & 28.852 & 5.957 & \href{https://gcn.nasa.gov/circulars/37834}{37834} & [-1000, 100]  & -- & [-1000, 100] & 0.748 [\href{https://gcn.nasa.gov/circulars/37852?view=index&query=EP241021a&startDate=&endDate=&sort=circularID}{37852}] \\
241026b & 18:14:30 & 100 & 56.403 & 41.031 & \href{https://gcn.nasa.gov/circulars/37902}{37902} & -- & -- & [-1000, 100]  & - \\
241101a & 23:52:49 & 100 & 37.763 & 22.731 & \href{https://gcn.nasa.gov/circulars/38039}{38039} & -- & [-1000, 100]  & [-1000, 100] & -  \\
241103a & 01:23:37 & 60 & 27.757 & 18.959 & \href{https://gcn.nasa.gov/circulars/38058}{38051} & -- & [-1000, 60]  & [-1000, 60] & -  \\
241107a & 14:10:24 & 400 & 35.016 & 3.329 & \href{https://gcn.nasa.gov/circulars/38112}{38112} & -- & [-1000, 400]  & [-1000, 400] & 0.456 [\href{https://gcn.nasa.gov/circulars/38126?view=index&query=EP241107a&startDate=&endDate=&sort=circularID}{38126}] \\
241113a & 19:09:19 & - & 131.981 & 52.367 & \href{https://gcn.nasa.gov/circulars/38211}{38211} & -- & [-1000, 60]  & -- & 1.53 [\href{https://gcn.nasa.gov/circulars/38449?view=index&query=EP241113a&startDate=&endDate=&sort=circularID}{38449}] \\
241119a & 17:53:20 & 200 & 84.116 & 3.832 & \href{https://gcn.nasa.gov/circulars/38281}{38281} & -- & -- & -- & - \\
241125a & 00:06:06 & 150 & 48.561 & 37.677 & \href{https://gcn.nasa.gov/circulars/38318}{38318} & [-1000, 150]  & [-1000, 150]  & [-1000, 150] & -  \\
241126a & 19:39:41 & 60 & 33.744 & 11.705 & \href{https://gcn.nasa.gov/circulars/38337}{38337} & -- & -- & [-1000, 60] & - \\
241201a & 20:59:16 & 230 & 282.596 & 66.081 & \href{https://gcn.nasa.gov/circulars/38415}{38415} & [-1000, 230]  & [-1000, 230]  & [-1000, 230] & - \\
241202b & 15:12:55 & 140 & 45.302 & 2.441 & \href{https://gcn.nasa.gov/circulars/38426}{38426} & [-1000, 140]  & [-1000, -496]*  & [-1000, -304]* & - \\
241206a & 16:34:47 & 400 & 34.702 & 38.914 & \href{https://gcn.nasa.gov/circulars/38457}{38457} & [-1000, 400]  & [-1000, 400]  & [-1000, 400] & -  \\
241208a & 16:36:13 & 50 & 127.812 & 49.082 & \href{https://gcn.nasa.gov/circulars/38477}{38477} & -- & -- & [-1000, 50]  & - \\
241212a & 11:21:52 & - & 153.817 & 60.068 & \href{https://www.astronomerstelegram.org/?read=16943}{16943$^{\ddagger}$} & [-1000, 60] & [-1000, 60] & [-1000, 60] & - \\
241217a & 05:34:10 & - & 46.957 & 30.901 & \href{https://gcn.nasa.gov/circulars/38624}{38624} & -- & -- & [-1000, 60] & 4.59 [\href{https://gcn.nasa.gov/circulars/38593?view=index&query=EP241217a&startDate=&endDate=&sort=circularID}{38593}] \\
241223a & 07:21:23 & 80 & 74.804 & 7.11 & \href{https://gcn.nasa.gov/circulars/38660}{38660} & -- & -- & -- & - \\
241231b & 19:27:29 & - & 100.064 & 16.171 & \href{https://gcn.nasa.gov/circulars/38778}{38778} & [-1000, 60]  & [-1000, 60]  & [-1000, -336]* & -   \\
&&&&&&&&[-80, 60]* \\
250101a & 05:52:21 & 2500 & 85.575 & 0.352 & \href{https://gcn.nasa.gov/circulars/38778}{38778} & [-1000, 2500]  & [-1000, 2500]  & [-1000, 2500] & -   \\

250108a & 12:30:28 & 2500 & 55.623 & $-22.509$ & \href{https://gcn.nasa.gov/circulars/38861}{38861} & [-1000, 2500]  & [-1000, 1872]*  & [-1000, -816]* & 0.176 [\href{https://gcn.nasa.gov/circulars/38908?view=index&query=EP250108a&startDate=&endDate=&sort=circularID}{38908}]  \\
&&&&&&&&[-560, -432]*&\\
&&&&&&&&[-144, 2500]*&\\
250109b & 08:06:40 & 200 & 118.611 & $-14.651$ & \href{https://www.astronomerstelegram.org/?read=16974}{16974$^{\ddagger}$} & [-1000, 200]  & -- & [-1000, 200] & -  \\
250111a & 01:20:24 & 83 & 97.181 & 56.898 & \href{https://gcn.nasa.gov/circulars/38905}{38905} & [-1000, 83]  & -- & [-1000, 83] & -  \\
250125a & 02:36:19 & 74 & 175.353 & $-21.704$ & \href{https://gcn.nasa.gov/circulars/39028}{39028} & [-1000, 74]  & [-1000, 74]  & [-1000, 74] & 2.89 [\href{https://gcn.nasa.gov/circulars/39027?view=index&query=EP250125a&startDate=&endDate=&sort=circularID}{39027}] \\

\hline
\end{tabular}
\begin{minipage}{\textwidth}
\footnotesize
$^{\dagger}$ \citet{2026ApJ...999..239L}.  $^{\ast}$ \citet{2026arXiv260627048V}.\\
$^{\ddagger}$ Sources reported on ATel.\\
$^{\S}$ CXOU J005245.0-722844.
\end{minipage}
\end{table*}

\section{FXT Sample and GW Search Window} \label{Sec:2}
The O4b observing run lasted from 2024-04-10 15:00:00 UTC (1396796418 GPS) to 2025-01-28 17:00:00 UTC (1422118818 GPS). During this period the EP mission was actively discovering FXTs. We construct a sample of EP-detected FXTs observed during O4b, excluding events that are associated with GRBs (target of dedicated LVK searches) or flaring stars. 

For each source, we collect the publicly available information reported through the General Coordinates Network (GCN) and Astronomer's Telegrams (ATel), including the sky position, trigger or start time, and duration of the transient. The final sample contains 47 EP-FXTs, including 11 with measured redshifts from optical follow-ups. FXTs with uncertain redshift are particularly interesting, as an association with CBC mergers remains plausible despite the lack of a detected MeV counterpart. They may correspond to GRBs for which the prompt MeV emission was not observed because no gamma-ray instruments were covering the source location at the time of the event, intrinsically low-luminosity GRBs, off-axis events, or mergers leading to the formation of a long-lived magnetar remnant.

We search for GW candidates in the 5th Gravitational Wave Transient Catalog (GWTC-5, \citealt{2026arXiv260527225T})\footnote{\url{https://zenodo.org/records/20276130}} 
that are in temporal proximity to any of the EP-FXTs with our sample. The choice of the temporal window needs to be informed by our priors on the expected time difference between the GW trigger time and the EP trigger time. In the assumption of a CBC origin of the FXT, this is the difference between the merger time and the onset time of the X-ray emission detected by EP. We define the GW search time window in the interval $[t_{FXT} - t_a; t_{FXT} + t_b]$, being $t_{FXT}$ the publicly available EP trigger time.

In order to make a conservative choice of $t_a$, we need to consider all possible scenarios where the X-ray emission is delayed with respect to the GW trigger. Since the WXT light curves are not publicly available, we cannot determine whether the detected X-ray emission corresponds to the onset of a prompt soft X-ray component, a steep-decay phase, or an early afterglow component connected to a GRB jet. Moreover, the absence of an estimate of the X-ray time variability prevents us from favoring one scenario over the other.
In a merger-driven GRB scenario, where the GW signal and prompt MeV emission are expected to be nearly simultaneous, the lower-energy X-ray emission can last longer \citep{2025ApJ...988L..34J, 2025NSRev..12E.401S} and may remain detectable after the harder prompt emission has faded. \citet{2026A&A...708A.190I} showed that this steep-decay component of short GRBs can be detected by EP/WXT as an X-ray transient lasting from several hundred seconds up to timescales of order $10^{3}$ s, if the GRB is seen on-axis. If the jet produced by the NS merger is oriented off-axis with respect to the line of sight, the dissipation due to the GRB prompt emission appears as a soft X-ray transient, peaking at $10^2-10^3$ s after the merger \footnote{In the case of a structured jet for a viewing angle $\theta_v$=$3\times\theta_c$, where $\theta_c$=3$^\circ$ is the half-opening angle of the core of the jet. See \citet{2020A&A...641A..61A}.  }\citep{2020A&A...641A..61A, 2026ApJ..1001..177C}.

Similar delays are also expected in the case of X-ray flares, commonly observed in GRBs \citep{2006ChJAA...6..513G,2011MNRAS.417.2144M,2018ApJ...858...34M} or the spin-down emission of fast-rotating magnetars \citep{2013MNRAS.430.1061R}. The presence of a magnetar after a NS merger can be responsible for an X-ray feature known as "internal" plateau, attributed to the internal dissipation inside the wind powered by the central engine. If driven by a magnetar wind, the corresponding X-ray transient is expected to be less beamed and more isotropic with respect to the GRB jet. The time scales of "internal" plateaus observed in short GRBs (few tens up to $\sim$ 10$^3$ s, \citealt{2015ApJ...805...89L}) can be used to infer an upper limit on the typical delay between the GW signal and the onset of the X-ray emission.
Magnetar-remnant models of NS--NS mergers predict sGRB-less X-ray transients \emph{e.g.} \citep{2016ApJ...819...14S, 2016ApJ...819...15S}. In some of those, the detectability also depends on viewing geometry and ejecta opacity \citep{2017ApJ...835....7S,2019ApJ...886..129S}, while recent models of X-ray emission from BNS mergers show that the observed signal depends strongly on the progenitor, remnant, emission component, and observer line of sight \citep{2026ApJ..1001..177C}. There might exist exotic scenarios where the large optical depth of the emitting region can exceed the 10$^3$ s delay (e.g. \citealt{2017ApJ...835....7S}), however, the choice of $t_a =$ 10$^3$ s encompasses the great majority of the scenarios described so far. Moreover, we do not extend the search beyond 10$^3$ s in order to avoid an unnecessarily large amount of false-positive GW candidates in the search window, which would reduce the effectiveness of the search.

Regarding the choice of $t_b$, we need to consider that there are also scenarios where X-ray emission can occur prior to the merger itself. These scenarios involve resonant shattering of NS crusts \citep{2012PhRvL.108a1102T} or magnetospheric interactions between the NS and the companion \citep{2021ApJ...921...92B,2023MNRAS.519.3923C,2026arXiv260214300S}. These effects can power an X-ray/$\gamma-$ray emission $\mathcal{O}(1-10)$ s before the merger. Moreover, the EP trigger time reported publicly may be affected by uncertainty, since a deeper search may find a signal earlier in time. Therefore, in order to be conservative, we set $t_b=T_{100}$, where the $T_{100}$ is the EP transient duration and is typically of the order of $10-100$ s.
For 8 of the 47 FXTs, no publicly available duration is reported; for these cases, we set $T_{100}=60~\mathrm{s}$, following the typical time window adopted for the generic GW transient searches externally triggered electromagnetic transients with X-Pipeline (see \citealt{2022ApJ...928..186A}). In conclusion, taking into account the wide plethora of theoretical priors on the expected GW-FXT time delay, we adopt the GW search temporal window $[t_{FXT} - 1000\, s; t_{FXT} + T_{100}]$. Table~\ref{IFO-list} shows the trigger time, duration, and position of our sample, along with the ATel and GCN circular numbers.

\section{GW candidates associated with EP-FXTs} \label{Sec:4}
\subsection{GWTC-5 events} \label{sec:3.1}

In the LVK low-latency alert framework, searches for coincidences between GW candidates and external astrophysical triggers are performed online by RAVEN\footnote{\url{https://lscsoft.docs.ligo.org/raven/}}, the Rapid on-source VOEvent Coincidence Monitor \citep{2022APS..APRY13002P}, which combines timing and sky-localization information to estimate a joint false-alarm rate for possible multi-messenger associations \citep{2024PNAS..12116474C}. However, RAVEN was not configured to process EP-FXT alerts during O4b. We therefore carried out an offline coincidence search between the GWTC-5 candidate list and the EP-FXT sample considered in this work.

The GWTC-5 catalog reports a total of 1857 GW candidates from the O4b run. As our analysis focuses on the CBC progenitor scenario, we exclude the GW candidates found by the coherent WaveBurst (cWB) pipeline \citep{2021SoftX..1400678D} from the search.\footnote{One cWB candidate, GW240918\_175452 (superevent\_id S240918gl), is temporally coincident with EP240918c within our adopted search window. We discuss this candidate separately in Appendix~\ref{app}.} We find that 7 CBC candidates fall within the search windows defined for our EP-FXT sample. 
The GW candidates have been identified by PyCBC \citep{2021ApJ...923..254D}, GstLAL \citep{2024PhRvD.109d2008E} and MBTA \citep{2021CQGra..38i5004A} pipelines.
Assuming that the GW candidates are uniformly distributed in time over O4b, we can estimate the expected number of random temporal coincidences with the FXT sample. After excluding cWB-only events, the O4b sample contains 1619 (\(N_{\rm GW}\)) GW candidates over a duration \(T_{\rm O4b}= 22460968\,{\rm s}\) when atleast one interferometer was operating \citep{2026arXiv260527223T}. For the 47 FXTs considered here, the total temporal search window is
\[
T_{\rm win}
=
47\times1000\,{\rm s}
+
\sum_i T_{{\rm FXT},i}
=
62837\,{\rm s},
\]
is the cumulative duration of the temporal search windows. Under the null hypothesis of random temporal associations, the number of coincidences follows a Poisson distribution with mean \(\lambda\) given by 

\[
\lambda
=
N_{\rm GW}\frac{T_{\rm win}}{T_{\rm O4b}}
=
4.53.
\]
The probability of obtaining seven or more temporal coincidences is therefore
\[
p(K\geq7)
=
1-
\sum_{k=0}^{6}
\frac{\lambda^k e^{-\lambda}}{k!}
\simeq
0.172~~(17.2\%)
\]

This indicates that the number of temporal coincidences alone is not sufficient to provide statistically significant evidence for a physical association.

\begin{figure*}[t]
\centering
\includegraphics[width=0.4\linewidth]{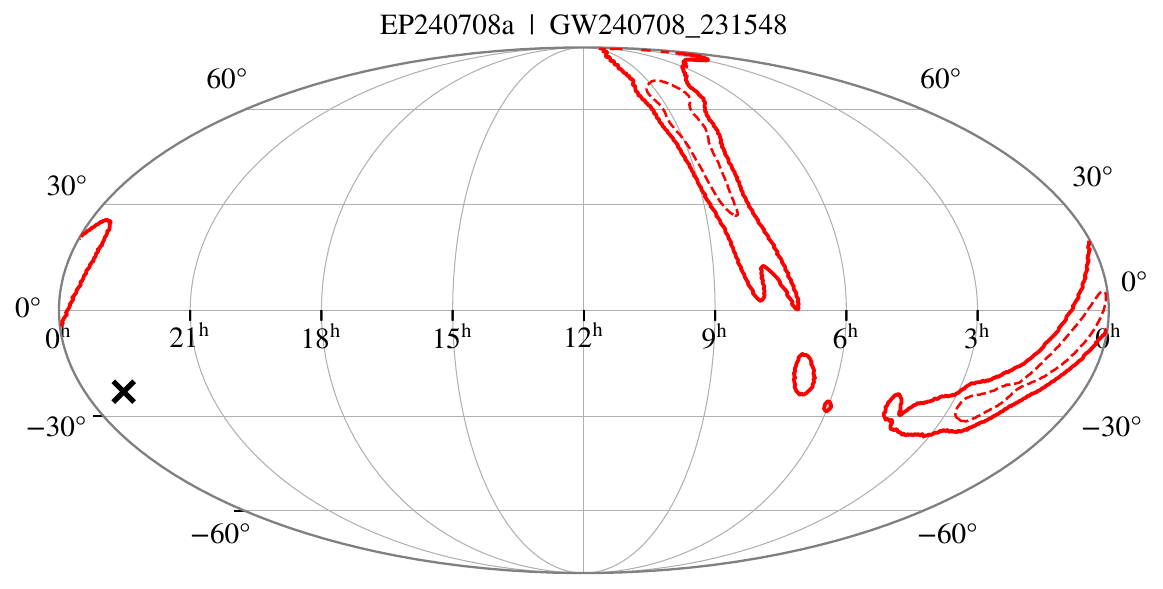}
\vspace{0.5cm}
\includegraphics[width=0.4\linewidth]
{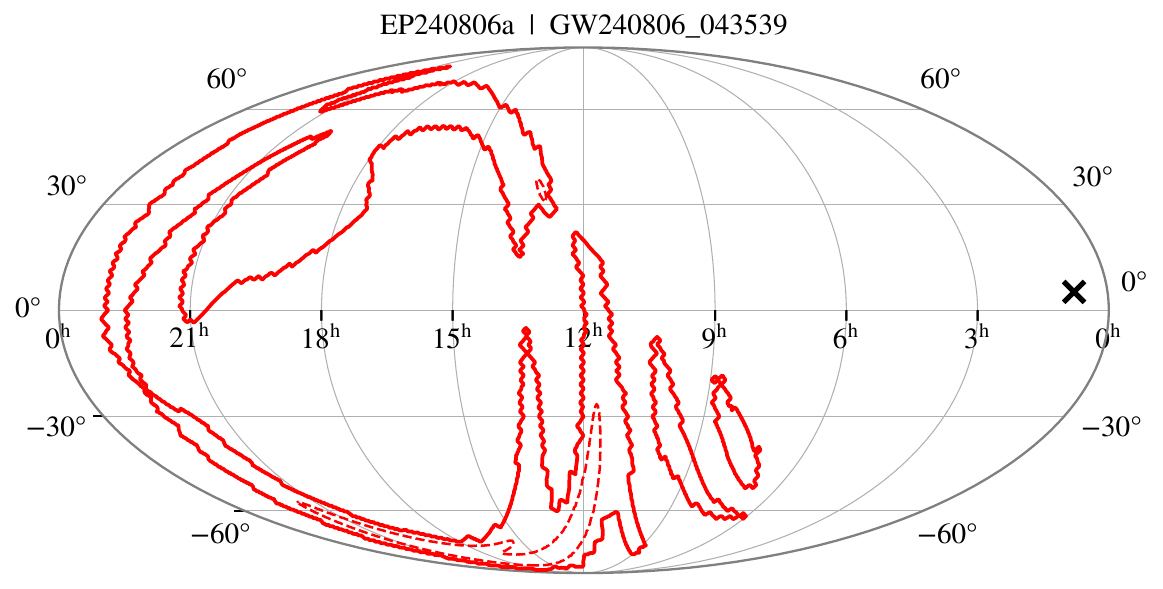}
\vspace{0.5cm}
\includegraphics[width=0.4\linewidth]{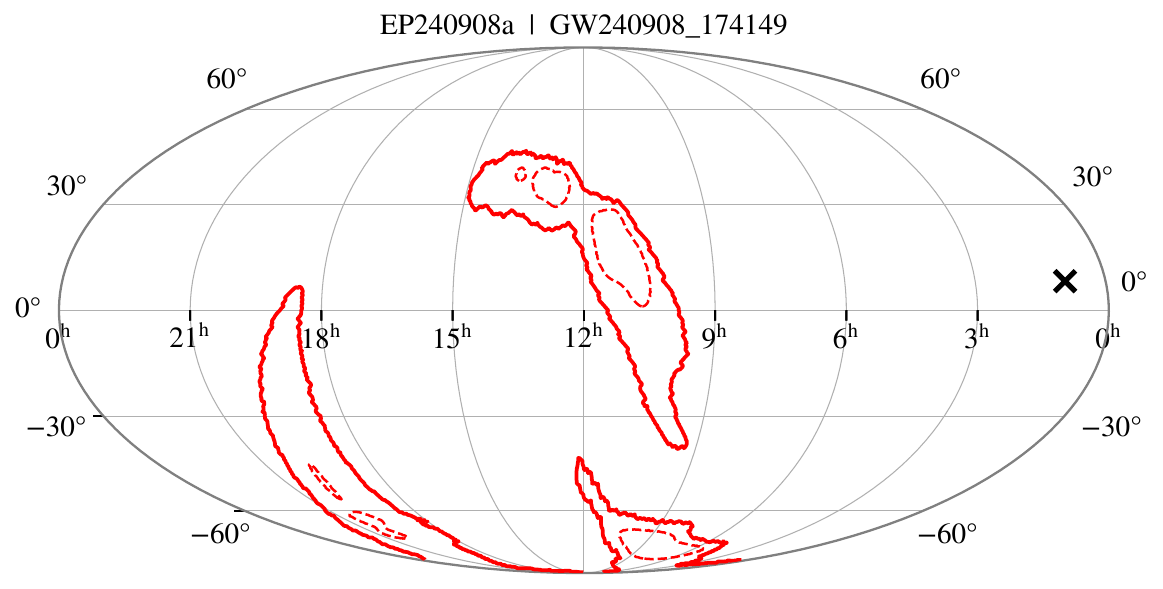}
\includegraphics[width=0.4\linewidth]{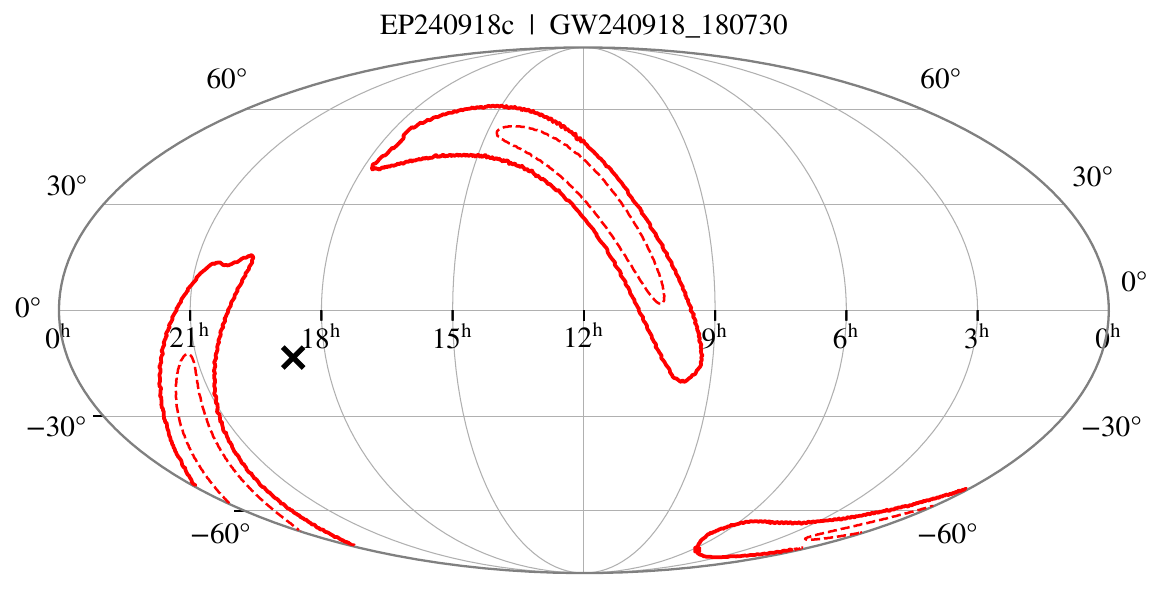}
\includegraphics[width=0.4\linewidth]{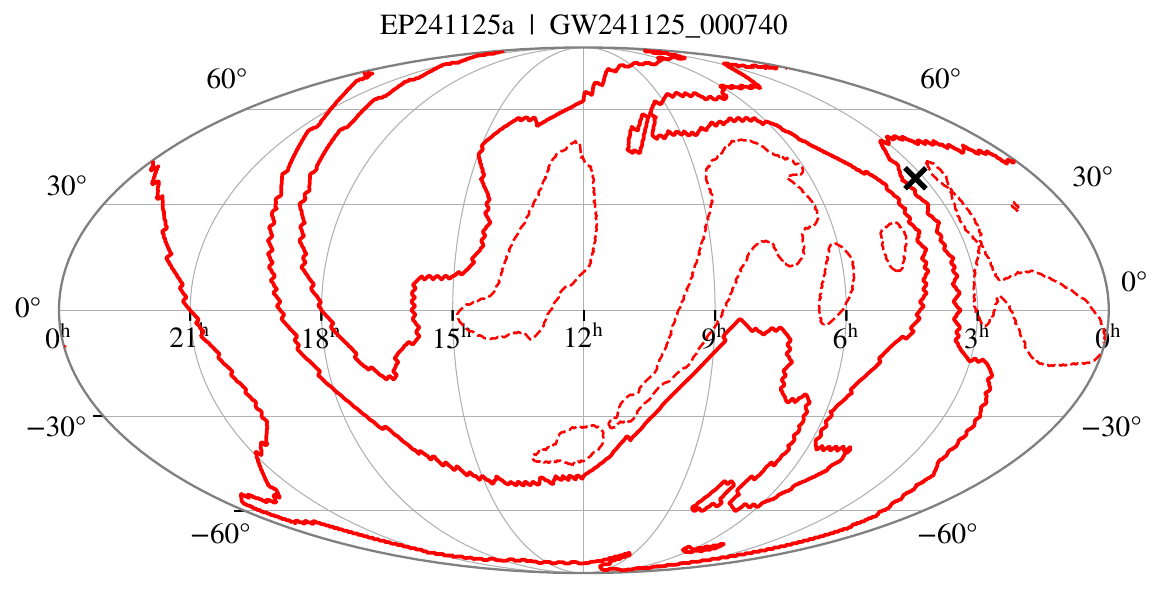}
\vspace{0.5cm}
\includegraphics[width=0.4\linewidth]{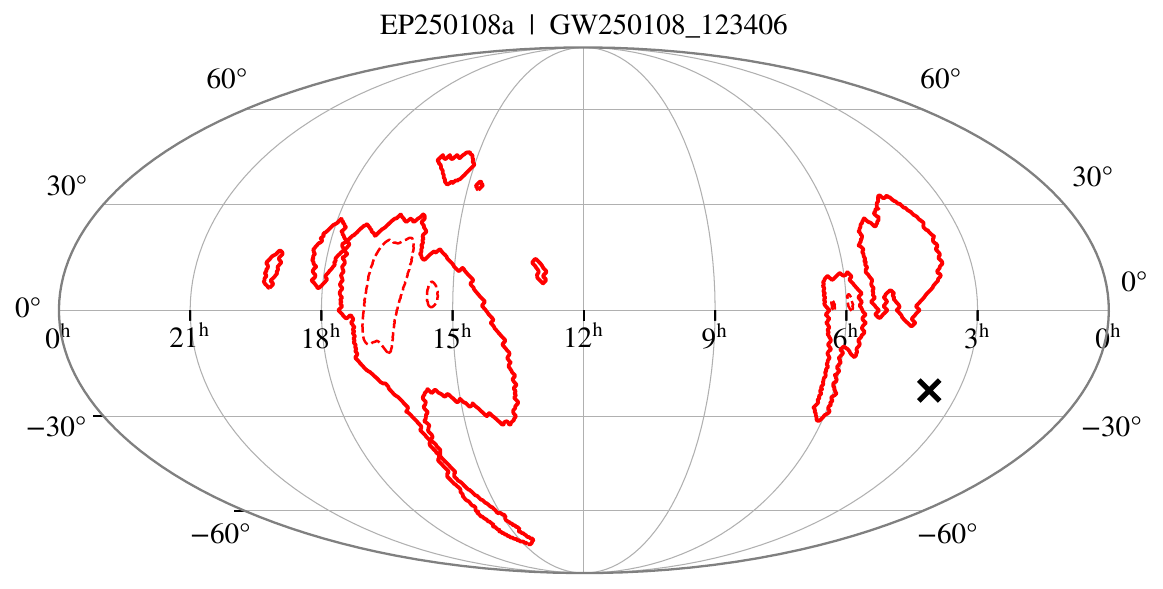}
\includegraphics[width=0.4\linewidth]{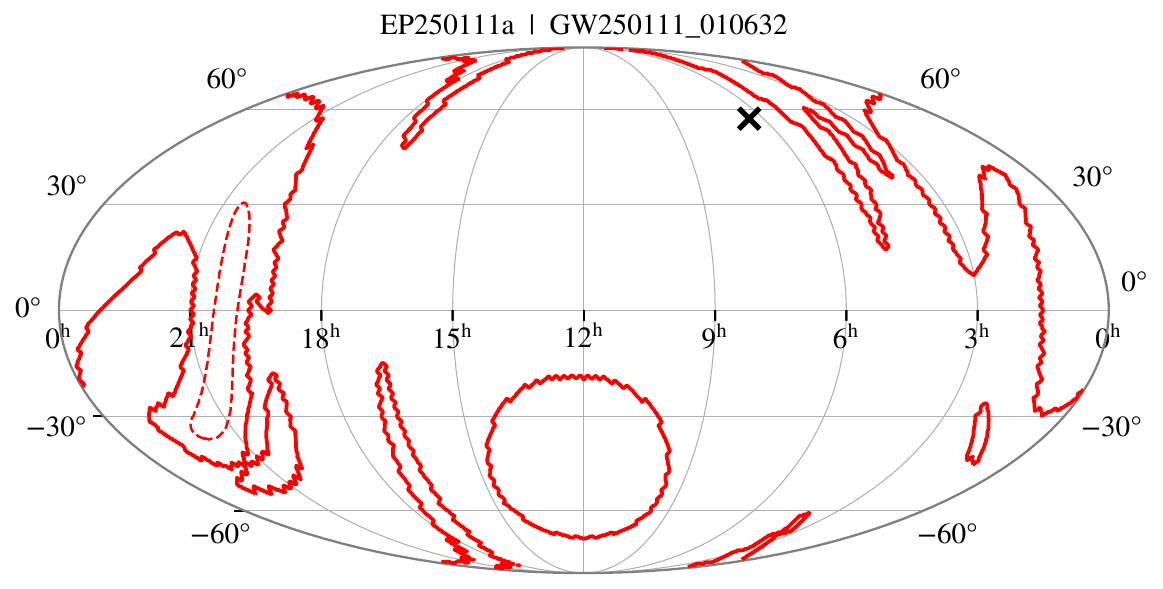}
\captionof{figure}{Sky localization for the seven temporally coincident Einstein Probe and GW candidate pairs listed in Tab.~\ref{ep_gw_candidates}. In each panel, the black cross marks the Einstein Probe FXT position, while the dashed and solid contours show the 50\% and 90\% credible regions of the corresponding GW sky localization, respectively. The figure illustrates that most candidate pairs show little or no meaningful spatial overlap. These sky localizations are further used to quantify the spatial-coincidence term entering the ranking statistic described in Sec.~\ref{ranking statistic}.}
\label{skymap}
\end{figure*}

We further examine the spatial overlap between the EP-FXT positions and the GW localization regions for these remaining candidates. The corresponding sky  maps are shown in Fig.~\ref{skymap}, along with the sky position of the EP transients. The main properties of the  temporally coincident EP-FXT/GW candidates are summarized in 
Table~\ref{ep_gw_candidates}.

\begin{table*}
\centering
\caption{Candidate GW superevents found within the temporal window of Einstein Probe fast X-ray transient transients listed in Tab.~\ref{IFO-list}. For each pair, the table reports the Einstein Probe trigger time, the associated GW superevent, the time offset between the GW trigger and the X-ray trigger, the false alarm rate, network signal-to-noise ratio, search pipeline, and detector network operating at the GW trigger time. These seven temporal coincidences form the candidate set used for the spatial-overlap inspection in Fig.~\ref{skymap} and for the ranking-statistic analysis in Tab.~\ref{rank}.}

\begin{tabular}{c c c c c c c c}
\toprule
EP\_event & T$_{0,\rm EP}$  & Superevent\_ID & Offset ($T_{\rm GW}-T_0$) & FAR  & SNR & Pipeline & IFOs \\
 & [UTC] &  & [sec] & [yr$^{-1}$] &  &  & \\
\midrule
EP240708a & 23:28:23 & S240708ex & -754.7  & 448.27 & 8.0 & gstlal & H1L1V1 \\
EP240806a & 04:47:53 & S240806f & -733.8  & 1.91 & 8.4 & pycbc & L1V1 \\
EP240908a & 17:28:27 & S240908fo & +802.2  & 3.08 & 8.9 & MBTA & H1L1V1 \\
EP240918c & 18:06:47 & S240918gi & +43.8  & 233.51 & 6.8 & pycbc & H1L1 \\
EP241125a & 00:06:06 & S241125fj & +94.9  & 316.67 & 9.0 & MBTA & H1L1V1 \\
EP250108a & 12:30:28 & S250108dc & +218.1 & 297.42 & 9.3 & MBTA & H1L1V1 \\
EP250111a & 01:20:24 & S250111cg & -831.1 & 155.18 & 8.8 & MBTA & H1V1 \\ 

\bottomrule
\end{tabular}
\label{ep_gw_candidates}
\end{table*}

\subsection{Joint ranking statistic} \label{ranking statistic}

Joint searches for electromagnetic counterparts to GW candidates have previously been conducted primarily in the context of gamma-ray transients. These include sensitive searches for GRB counterparts to GW candidates using Fermi-GBM and Swift-BAT data \citep{2020CQGra..37q5001S,2020ApJ...893..100H,2023ApJ...956...56P,2024ApJ...964..149F,2025ApJ...980..207R}. In such analyses, the key statistical problem is to determine whether a measured temporal and spatial coincidence is more likely to arise from a common astrophysical origin or from the random overlap of unrelated candidates. In our case, the external electromagnetic triggers are EP-FXTs, and we use a joint ranking statistic to quantify the relative significance of each candidate EP--GW association.

To rank the temporal and spatial overlap between EP-FXTs events and candidate GW superevents, we follow the methodology presented in \cite{2022CQGra..39h5010P}. The ranking statistic is derived starting from their eq. 14. Since the EP candidates are highly significant, we approximate the relative noise Bayes factor $B_{n/s}(EP) \rightarrow 0$. We then define a 
ranking statistic $\Lambda$ that goes like the odds ratio $\mathcal{O}_{c / \text { uncor }}\left(x_a, x_b\right)$, hence obtaining

\begin{equation}
\label{lambda}
\Lambda =
\frac{I_{\Omega}}
{1 + \mathrm{BCI}^{-1}} ,
\end{equation}
where $I_{\Omega}$ is the sky overlap between the GW and the FXT localizations. Since the localization of the FXT is always at the arcminute precision, the overlap is:
\begin{equation}
    I_{\Omega}\propto\int  \rm p_{GW}\left(\Omega \right) p_{FXT}\left(\Omega \right) d \Omega =  p_{GW} (RA_{FXT},Dec_{FXT}),
\end{equation}
namely the probability density of the GW sky-localization map evaluated at the position of the EP source. The quantity $\mathrm{BCI}$ is the Bayes factor of the GW candidate, which compares the hypothesis that the data contain a coherent signal across the detector network with the hypothesis that the observed excess power is produced by incoherent noise transients in the individual detectors. We obtain this quantity from the GW localization map fits file associated with each candidate event. Larger values of $\mathrm{BCI}$ therefore indicate stronger support for a coherent network signal. According to Eq. (20) of \cite{2022CQGra..39h5010P}, additional constant terms should appear at the denominator, connected to the expected astrophysical rates of GWs and FXTs, as well as rate of GW noise events. However, the inclusion of these constants and their numerical value does not change the relative ordering of the ranking itself, and hence have no impact on ranking our candidate GW-FXT associations. We therefore drop these constants from Eq.~\ref{lambda}.

\subsection{Simulating the background distribution}

To evaluate whether the observed EP--GW associations are more significant than expected from chance coincidence, we construct an empirical background distribution for the ranking statistic. We draw 1000 random GW candidates from the GWTC-5 catalog, excluding the superevents already associated with the EP-FXT sample, and retaining only those whose preferred search is a CBC pipeline. This ensures that the background is built from the same class of GW candidates relevant for the foreground analysis, while avoiding contamination from the EP-associated sample itself. We do not impose any selection cut on the FAR of the GW candidate drawn from the GWTC-5 catalog. In this way the background population is a mixture of associations between noise GW events and real FXTs plus astrophysical GW events and real FXTs. The second component represents the chance of having two real events in temporal proximity but coming from independent astrophysical events.  For each selected GW candidate, we generate 1000 sky positions uniformly distributed over the sphere and evaluate the ranking statistic for each GW-FXT couple. In this way, we obtain $10^6$ realizations of the ranking statistic with the null hypothesis in which EP-FXTs and GW candidates are unrelated.

\subsection{Significance of FXT-GW associations}

The significance of a given FXT-GW couple is quantified through the tail probability, namely the fraction of background realizations whose ranking statistic is greater than or equal to that of the foreground candidate.
For a foreground value $\Lambda_{\rm fg}$, the corresponding $p$-value is estimated as the fraction of the background realizations satisfying $\Lambda_{\rm bkg} \ge \Lambda_{\rm fg}$. Small $p$-values thus identify foreground associations that lie in the extreme right tail of the null distribution.

\begin{figure}[t]
    \includegraphics[width=1\linewidth]{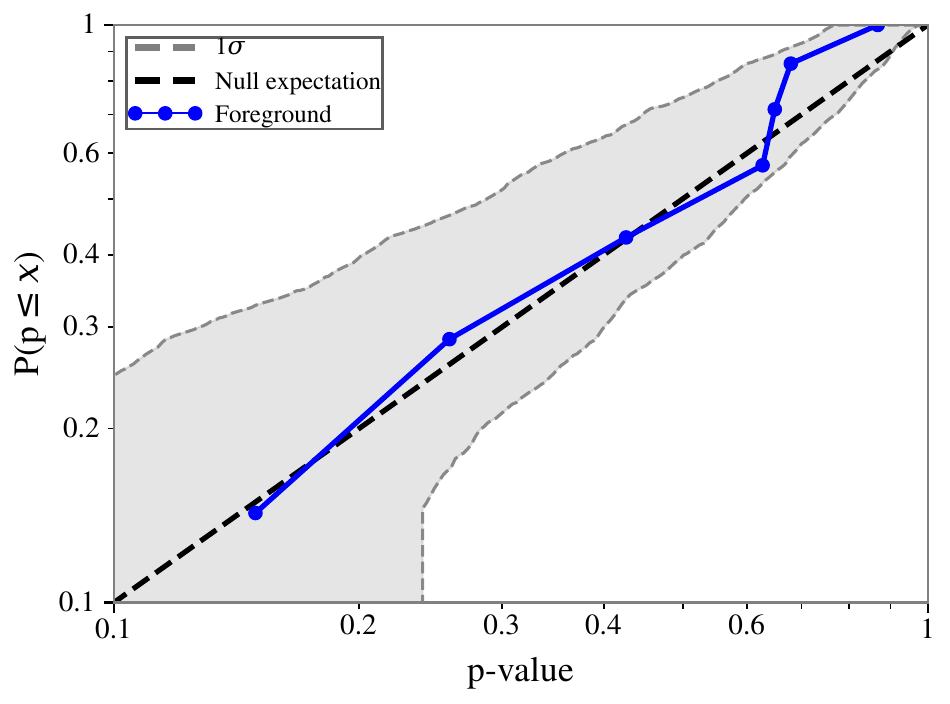}
    \caption{Cumulative p-value distribution of the ranking statistic for the seven EP FXT--GW candidate pairs in Tab.~\ref{rank}. The blue curve shows the observed distribution, the black dashed line shows the expectation for random coincidences, and the shaded gray band indicates the $1\sigma$ interval expected from finite-sample fluctuations. The observed distribution remains consistent with random temporal and spatial coincidences, indicating no statistically significant evidence for a common astrophysical origin in the candidate set.}
    \label{p-val}
\end{figure}

We report the p-value of each FXT-GW candidate in Table~\ref{rank}. A complementary way to test the consistency of the overall population with the null hypothesis is to examine the distribution of foreground $p$-values as shown in Fig.~\ref{p-val}. If EP-FXTs and GW candidates were entirely unrelated, the foreground $p$-values would follow a uniform distribution on the interval $[0,1]$, and their cumulative distribution would be expected to track the diagonal relation $P(p \le x)=x$. Any statistically meaningful excess at small $p$ would then indicate that the foreground sample contains more highly ranked associations than expected from random coincidence alone. To estimate the expected scatter around the null hypothesis, we generate 1000 Monte Carlo realizations with the same number of foreground events, drawing the $p$-value uniformly from $[0,1]$ and computing their cumulative distributions. The gray band in Fig.~\ref{p-val} corresponds to the central 68\% interval of these realizations. The observed distribution is fully contained within the 1$\sigma$ uncertainty band, therefore showing no statistically significant deviation from the null hypothesis.

Having computed the p-value for each GW-FXT association ($p_i$), we can derive a corresponding joint FAR as follows:
\begin{equation}
    \mathrm{FAR}_i = R_{FXT} \left ( 1-e^{-R_{GW}\Delta t_i}\right)p_i,
\end{equation}
where $R_{\rm FXT}$ is the discovery rate of EP fast X-ray transients not associated with GRBs, $R_{\rm GW}$ is the rate of GW triggers, and $\Delta t_i$ is the width of the temporal coincidence window for the $i$th candidate. We estimate $R_{\rm FXT}$ from the 47 EP-FXTs in our sample divided by the O4b observation time, yielding $R_{\rm FXT}=58.5~{\rm yr}^{-1}$. The GW trigger rate is computed from all 1857 GW candidates reported in the GWTC-5 catalog, giving $R_{\rm GW}=2314~{\rm yr}^{-1}$. For each candidate, the coincidence window is defined as $\Delta t_i = 1000~{\rm s} + T_{100,i}$, where $T_{100,i}$ is the duration of the corresponding EP transient. The resulting joint FAR values are reported in Table~\ref{rank}. The candidate with the lowest joint FAR is S241125fj, which is temporally coincident with EP241125a, with a joint FAR of $7.1\times10^{-1}~{\rm yr}^{-1}$, corresponding to one chance coincidence every $\sim1.4$ years.

\begin{table*}
\centering
\caption{Ranking-statistic results for the seven temporally coincident EP FXT--GW candidate pairs identified in Tab.~\ref{ep_gw_candidates}. The columns report the EP event name and the corresponding GW candidate ID, the GW sky-localization overlap probability at the EP position ($I_{\omega}$), the Bayes Factor Coherent vs. Incoherent (BCI), the combined ranking statistic ($\Lambda$), the empirical p-value derived from the background distribution, and the corresponding joint false alarm rate (Joint FAR, in yr$^{-1}$). Larger values of $I_{\omega}$, BCI, and $\Lambda$, together with smaller p-values and Joint FARs, indicate less likely chance alignments. None of the candidate pairs reaches a statistically significant association, and the full p-value distribution is assessed in Fig.~\ref{p-val}}
\begin{tabular}{l l c c c c c}
\toprule
EP event & Superevent\_ID & I$_{\omega}$ & BCI & $\Lambda$ & p-value & Joint FAR \\
 & & & & & & [yr$^{-1}$] \\
\midrule
EP240708a & S240708ex  & 6.0$\times10^{-5}$ & 962.06 & 6.00$\times10^{-5}$ & 0.86 & 7.81 \\
EP240806a & S240806f   & 5.22$\times10^{-5}$ & 12.14 & 4.83$\times10^{-4}$ & 0.67 & 3.17 \\
EP240908a & S240908fo  & 7.63$\times10^{-5}$ & 394.74 & 7.61$\times10^{-4}$ & 0.61 & 4.75 \\
EP240918c & S240918gi  & 6.41$\times10^{-5}$ & 47.71 & 6.28$\times10^{-4}$ & 0.64 & 2.90 \\
EP241125a & S241125fj  & 7.76$\times10^{-2}$ & 3.43 & 6.01$\times10^{-2}$ & 0.15 & 0.71 \\
EP250108a & S250108dc  & 4.02$\times10^{-2}$ & 1.47 & 2.39$\times10^{-2}$ & 0.25 & 3.31 \\
EP250111a & S250111cg  & 6.25$\times10^{-3}$ & 13.05 & 5.81$\times10^{-3}$ & 0.41 & 1.83 \\
\bottomrule
\end{tabular}
\label{rank}
\end{table*}

\section{Targeted Detectability Range for EP-FXTs} \label{Sec:5}

\begin{figure}[h]
    \centering
    \includegraphics[width=1\linewidth]{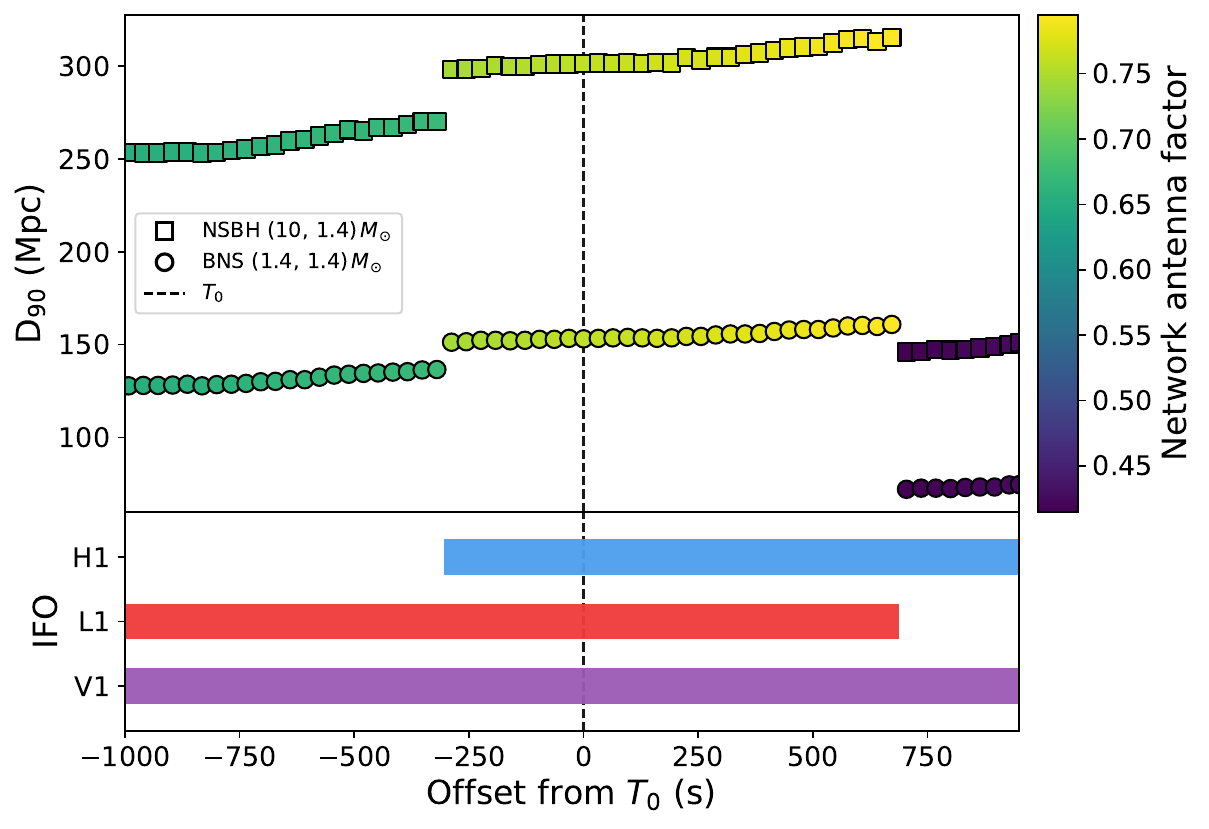}
    \caption{Temporal evolution of the matched-filter 90\% exclusion distance, $D_{90}$, for EP240908a over the search window $[T_0-1000~{\rm s},T_0+T_{100}]$. The upper panel shows the time-dependent $D_{90}$ values for the BNS $(1.4,1.4),M_{\odot}$ and NSBH $(10,1.4),M_{\odot}$ merger scenarios, shown with circles and squares, respectively. Each point corresponds to a scanned TDR time bin, with a nominal bin spacing of 32 s. The marker color gives the network antenna factor at the EP-FXT sky position for that time bin. The vertical dashed line marks the EP trigger time, $T_0$. The lower panel shows the detector availability across the same search window: horizontal rectangular blocks indicate the intervals over which each interferometer contributes to the network configuration used in the TDR calculation. Blank intervals correspond to times when the interferometer is on is observing mode. The standard TDR products for this source at the trigger time are shown in Fig.~\ref{tdr_products}}
    \label{evol_d90}
\end{figure}

\begin{figure*}[t]
    \centering
    \includegraphics[width=0.95\linewidth]{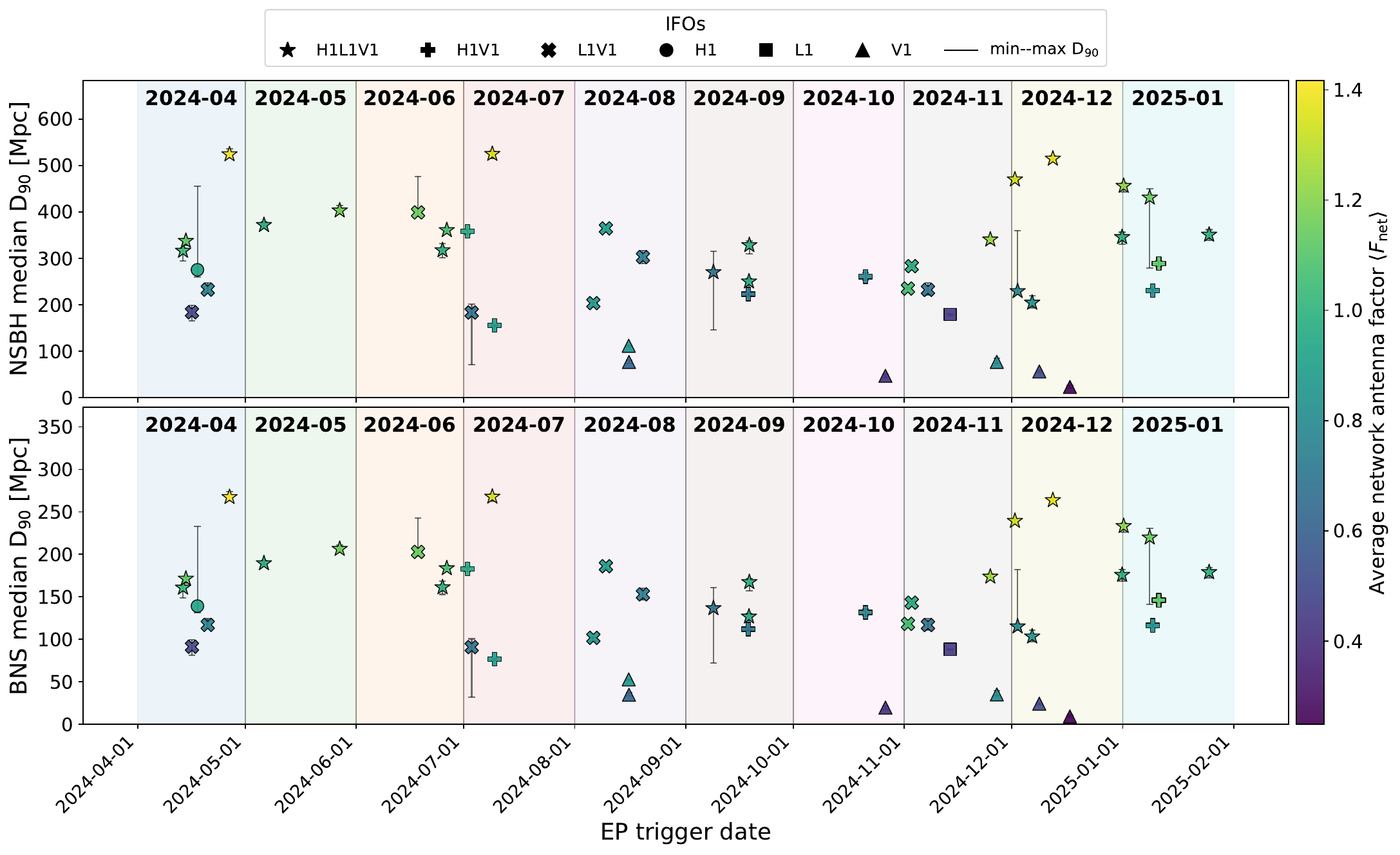}
    \caption{Median 90\% exclusion distance, $D_{90}$, for BNS ($m_1=m_2=1.4\,M_\odot$) and NSBH ($m_1=10\,M_\odot$, $m_2=1.4\,M_\odot$) systems for the EP-FXTs observed during the LVK O4b run evaluated for the Einstein Probe fast X-ray transients in Tab.~\ref{IFO-list}. For each source, the marker indicates the median $D_{90}$ obtained over all analyzed time bins within the search window $[T_{\rm 0}-1000~{\rm s},\,T_{\rm 0}+T_{100}]$, while the vertical error bars denote the minimum and maximum values of $D_{90}$ across the search window. The marker color represents the average network antenna factor at the source position over the search window. The marker type indicates the combination of IFOs that were online for the majority of the search window. The exclusion distances are computed assuming an inclination prior of $0^\circ < \iota < 45^\circ$ and a matched-filter detection threshold of $\rho_{\rm cut}=10$. These minimum--maximum distance range intervals are reported explicitly in Tab.~\ref{app_table}.}
    \label{D90_evol}
\end{figure*}

The lack of a confirmed association between the EP-FXTs and GW candidates motivates a complementary constraint on the possible CBC origin of these transients. For each EP-FXT, we estimate the distance up to which a CBC signal could have been detected by the operating GW detector network at the sky position of the X-ray transient. We use the Targeted Detectability Range\footnote{The tool is open-source and it can be run on a web interface at this link: \url{https://samueleronchini.github.io/gw_tdr/}} (TDR) framework introduced by \citet{2026arXiv260521578R}, which extends the usual LVK range estimate by incorporating information from an external electromagnetic trigger. The TDR evaluates the detectability of a hypothetical CBC signal using the detector sensitivity at the relevant time and the sky position provided by the EM transient.

The TDR is estimated by injecting CBC waveforms into the detector network and computing the corresponding matched-filter signal-to-noise ratio (SNR) using the sensitivity curve of each operating interferometer. At each time considered in the search window, we first identify the interferometers operating in observing mode and estimate their noise power spectral densities from the strain data around that time. In particular, the Power Spectral Density (PSD) is computed from a 256 s data segment centered on the analysis time, using Welch averaging over 16 s chunks with 8 s overlap. Following the prescription of \citet{2026arXiv260521578R}, we consider both BNS and NSBH injections, using TaylorF2 waveforms \citep{PhysRevD.75.124018} for the BNS systems and IMRPhenomNSBH waveforms \citep{PhysRevD.100.044003} for the NSBH systems. For the NSBH configurations, the BH spin and NS tidal deformability are chosen such that 
the baryonic mass remaining outside the merger remnant is nonzero, allowing the production of electromagnetic emission. The component masses are kept fixed for each injection set in order to avoid additional uncertainties from the poorly known compact-object mass distributions. We adopt ($m_1=m_2=1.4~M_{\odot}$) for the BNS injections and ($M_{\rm BH}=10~M_{\odot}; M_{\rm NS}=1.4~M_{\odot}$) for the NSBH injections. Since FXTs are not necessarily produced by an on-axis relativistic jet and may instead originate from off-axis jet emission, shock breakout, or magnetar-powered  processes, we do not adopt the jet-aligned tight inclination prior commonly used for GRBs. We instead extend our analysis to include a moderately off-axis population ($0^\circ<\iota<45^\circ$) when computing the exclusion distances reported in this work. Such an inclination range is compatible with scenarios in which the prompt gamma-ray emission is not observed, while the later X-ray emission remains detectable as an FXT. For a chosen matched-filter SNR threshold, $\rho_{\rm cut}$, we define the quantity $D_{90}$ as the distance at which 90\% of the injected signals are recovered with $\rho_{\rm MF}\geq\rho_{\rm cut}$. In the absence of a GW detection, $D_{90}$ therefore provides an exclusion distance: within this distance, a CBC signal with the assumed source parameters would have been detectable above the chosen SNR threshold with $>$90\% efficiency by the all-sky, all-time searches. As shown by \citep{2026arXiv260521578R}, a value of  $\rho_{\rm cut}=9$ optimizes the match between $D_{90}$ and the 90$\%$ exclusion distance obtained with PyGRB. However,  we adopt a slightly more conservative cut $\rho_{\rm cut}=10$, motivated by our inspection of the GWTC-5 catalog, where above $\mathrm{SNR}\sim10$ more than half of the events have $p_{\rm astro}>0.5$ (see Sec.~\ref{app:snr_threshold}).

The TDR supports searches with sliding time window, namely a time window that is progressively shifted in time. We thus compute the D$_{90}$ for each EP-FXT throughout the full GW search window, from $T_{0}-1000~\mathrm{s}$ to $T_{0}+T_{100}$, using time steps of 32 s. This choice is motivated by two considerations. First, the information about the exact time delay between the putative CBC merger time and the observed X-ray emission is not known a priori. Second, the detector network configuration and sensitivity can vary across the EP-FXT search window. In Fig.~\ref{evol_d90}, we show for EP240908a the time evolution of $D_{90}$ together with the detector availability throughout the interval $[T_0-1000~{\rm s},T_0+T_{100}]$. The lower panel shows that individual interferometers can enter or leave observing mode during the search window, changing the active detector network and directly affecting the corresponding BNS and NSBH $D_{90}$. $D_{90}$ can vary even when the IFO configuration is stable. This can happen because the antenna factor will slightly change as the position of the source changes with respect to the detector reference system over the search window. In addition, changes in the local detector noise, encoded through the PSD estimated around each time bin, can also modify the sensitivity to CBC signals.

\begin{figure}[t]
    \centering
    \includegraphics[width=\linewidth]{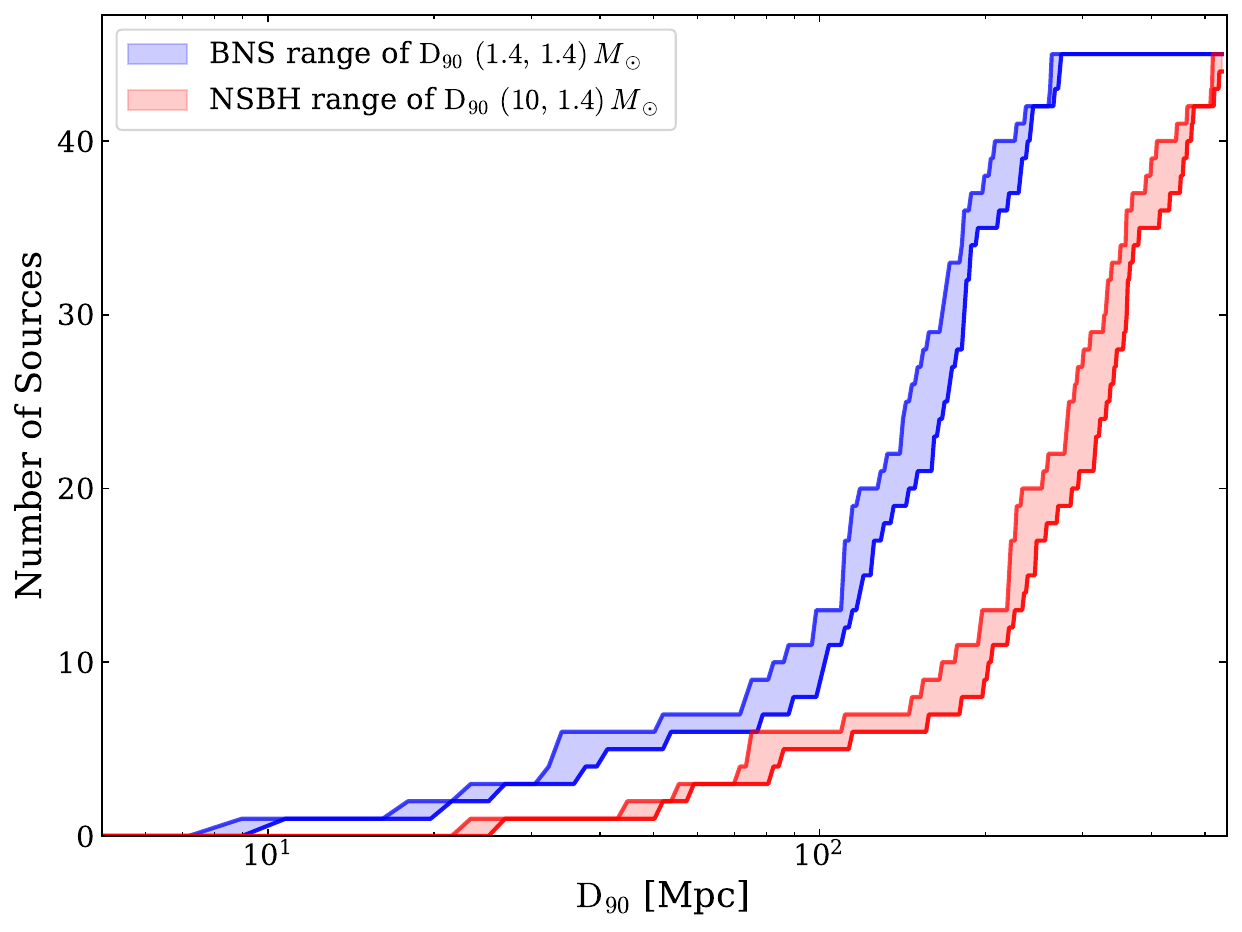}
    \caption{Range of matched-filter 90\% exclusion distances, $D_{90}$, obtained for each Einstein Probe fast X-ray transient over its GW search window. The shaded regions enclose, for the full sample, the interval between the minimum and maximum $D_{90}$ reached during the analyzed time bins for binary neutron star and neutron-star--black-hole assumptions. This figure summarizes the spread shown source by source in Fig.~\ref{D90_evol} and highlights the typical distance scale over which nearby compact-binary origins can be excluded for the sample. The corresponding numerical ranges for each source are listed in Tab.~\ref{app_table}.}
    \label{minmax_bns_nsbh}
\end{figure}

The evolution of the 90\% exclusion distance is shown in Fig.~\ref{D90_evol} for all EP-FXTs. 
The figure shows that, during O4b, sources with the full H1L1V1 network online typically reach median exclusion distances of $\sim178$ Mpc for BNS systems and $\sim349$ Mpc for NSBH systems. Single-detector configurations have a smaller antenna factor, and hence the corresponding distances are lower and depend strongly on the detector. The L1-only cases reach typical median distances of $\sim90$ Mpc for BNS and $\sim182$ Mpc for NSBH systems, while the V1-only cases reach $\sim35$ Mpc and $\sim77$ Mpc, respectively. The H1-only cases reach larger median distances of $\sim127$ Mpc for BNS and $\sim253$ Mpc for NSBH systems. The large spread in some error bars reflects changes in detector availability and sensitivity across the adopted search window, emphasizing the importance of evaluating the exclusion distance as a function of time rather than only at the EP trigger time. The range of $D_{90}$ for each EP source is reported in Tab.~\ref{app_table}, along with the percentage of coverage of each combination of interferometer. The table shows that on average there was a 3 IFO coverage for 34.8 $\%$ of the time, a 2 IFO coverage for 31.1 $\%$ of the time, a 1 IFO coverage for 17.4 $\%$ of the time, and 16.6 $\%$ without any IFO.
The cumulative distribution of the minimum and maximum $D_{90}$ values reached by each EP-FXT over its search window is shown in Fig.~\ref{minmax_bns_nsbh}, for both the BNS and NSBH assumptions.

\begin{figure}[t]
    \centering
    \includegraphics[width=\linewidth]{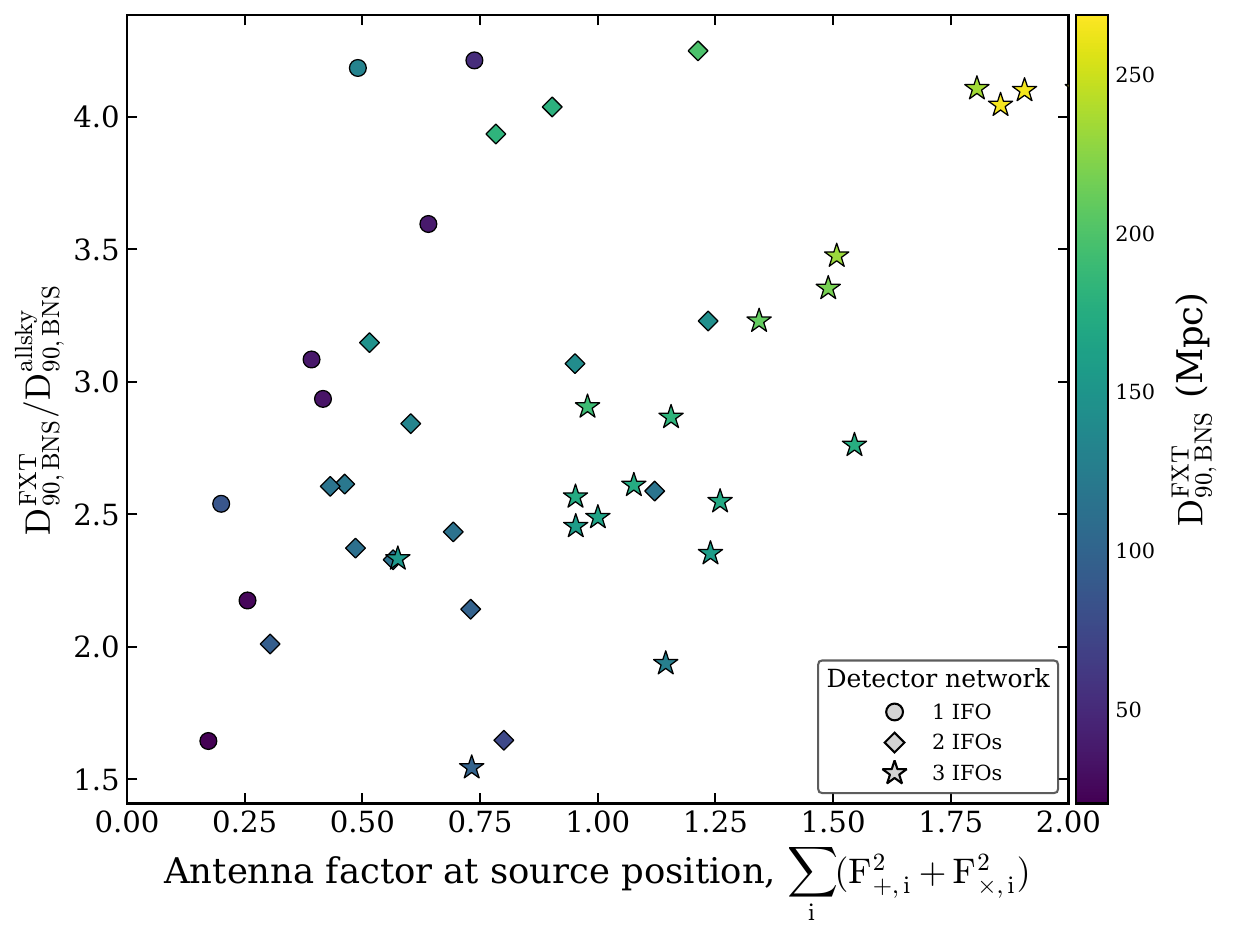}
    \caption{Ratio of the source-position and all-sky matched-filter 90\% exclusion distances, $D_{90}$ computed at T$_{0}$, for BNS systems with $m_1=m_2=1.4,M_\odot$, evaluated for the EP-FXTs at their trigger times. The source-position values are computed using the FXT sky localization and the restricted inclination prior $0^\circ<\iota<45^\circ$, while the all-sky values assume an isotropic inclination distribution, consistent with the publicly reported all-sky BNS range. The horizontal axis shows the network antenna factor at the FXT sky position. Marker shapes indicate the number of operating interferometers, and marker colors show the source-position $D_{90}$ at $T_0$. The ratios range from approximately 1.5 to 4.1 and are generally larger than 1, showing that targeted, externally triggered estimates can provide larger exclusion distances than all-sky averages.}
    \label{all-sky_tdr}
\end{figure}

To further understand how much of the source-dependent exclusion distance is driven by the detector antenna response, Fig.~\ref{all-sky_tdr} compares the BNS $D_{90}^{\rm FXT}$ computed at the FXT sky position with the corresponding all-sky, all-angles BNS range $D_{90,BNS}^{\rm allsky}$ for $m_1=m_2=1.4,M_{\odot}$. The all-sky BNS $D_{90,BNS}^{\rm allsky}$ is defined as the average computed injecting isotropically sources across the sky and with an isotropic inclination angle in the range $0^\circ<\iota<90^\circ$. This definition corresponds to the one used to compute the publicly reported all-sky BNS range\footnote{\url{https://online.ligo.org/grafana/goto/KHuAFEbDz?orgId=1}}. The figure shows the ratio $D_{90}^{\rm FXT}/D_{90,BNS}^{\rm allsky}$  as a function of the network antenna factor at the FXT sky position, with points separated according to the number of online interferometers and color-coded by the value of $D_{90}^{\rm FXT}$. 
This plot illustrates the benefit of using source-specific priors in the detectability estimate, as opposed to relying on the all-sky range alone. A clear positive trend is visible: for a fixed detector configuration, sources located in more favorable sky regions, corresponding to larger antenna factors, reach larger exclusion distances relative to the all-sky average. The ratios are generally larger than unity, reaching a maximum of $\sim$4.1, showing that the targeted, source-position exclusion distances can be substantially larger than the corresponding all-sky values once the transient sky location and external-trigger assumptions are included. The all-sky BNS range therefore provides a useful baseline for detector sensitivity, but it underestimates the detectability distance for favorably located and externally constrained sources.

\section{Discussion} \label{Sec:6}

The absence of a statistically significant GW--FXT association in our sample should be interpreted in the context of both the large number of GW candidates considered and the broad temporal windows required for FXTs. Unlike prompt GRB-triggered GW searches, where the expected delay between the GW signal and the electromagnetic emission is relatively well constrained, the time delay between a CBC merger and a possible soft X-ray transient is uncertain. The wider temporal windows adopted here therefore increase the probability of random temporal overlaps. The seven temporal coincidences found in our sample are compatible with the expected background from random temporal overlap alone. Temporal coincidence by itself is therefore not sufficient evidence for a physical association; the spatial consistency and GW-candidate significance must also be considered.

This motivates the use of a combined ranking statistic, which incorporates the temporal coincidence, the overlap between the GW sky localization and the EP-FXT position, and the significance of the GW candidate. None of the seven temporally coincident pairs shows a statistically significant association under this ranking-statistic framework. The case of EP241125a and S241125fj illustrates the importance of this combined assessment: although the FXT position lies within the $90\%$ credible region of the GW sky map, the GW candidate is very poorly localized, making the apparent spatial overlap inconclusive. The GW non-detections do not rule out a compact-binary origin for the EP-FXT population. Instead, they constrain only those CBC systems that would have been detectable by the current online GW detector network at the relevant sky position and time. \\

In the absence of a significant GW counterpart, we use the Targeted Detectability Range (TDR) framework \citep{2026arXiv260521578R} to quantify the distance up to which a CBC signal from the sky position of each EP-FXT would have been detectable by the online GW detectors. This calculation is complementary to a dedicated coherent targeted GW search. The X-Pipeline is designed to search for GW excess power associated with an external trigger and are more sensitive than all-sky searches for the targeted class of signals, but such an analysis is computationally costly and requires at least two operating interferometers. The PyGRB pipeline is effective and applicable only for targeted GW searches with a narrow temporal on-source window of maximum few seconds, making its use unfeasible for FXT-GW targeted searches where the uncertainty on the on-source window is much larger. The TDR, on the other hand, does not  perform a search for a signal; it provides a detectability range based on the detector sensitivity, the antenna response at the transient sky position, and the assumed CBC source parameters. The range computed by TDR provides an exclusion distance in the absence of any significant GW candidate found by all-sky searches. The TDR tool allows to evaluate a detectability distance even when only one interferometer is online. Its low computational cost makes TDR valuable not only as a post-event diagnostic, but also as a low-latency tool. By providing a rapid estimate of whether the GW network was sensitive enough to constrain a CBC origin for a newly reported FXT, it can help guide follow-up strategies and optimize the allocation of observational resources.
The low computational cost also enables a systematic estimation of the detector sensitivity across the entire FXT on-source time window.
This capability is particular important for astronomical transients such as FXTs, where the time delay between the GW emission and the electromagnetic counterpart is poorly constrained a priori.

For the O4b EP-FXT sample analyzed here, no significant GW counterpart is found. The TDR results exclude a nearby BNS or NSBH interpretation only within the corresponding source-dependent exclusion distances. The median exclusion distances or the binary inspiral scenarios are approximately $180\,\mathrm{Mpc}$ and $350\,\mathrm{Mpc}$ for BNS and NSBH systems, respectively. Eleven sources in our EP-FXT sample have measured redshifts, as reported in Tab.~\ref{IFO-list}. Among these sources, EP240506a is the closest, with $z\simeq0.12$, corresponding to a luminosity distance of $D_L\simeq560\,\mathrm{Mpc}$. At the trigger time of EP240506a, the TDR exclusion distance $D_{90}\simeq597\,\mathrm{Mpc}$ is comparable to the value of $D_L$ only for the most massive NSBH configuration considered in this work, $(m_{\rm BH},m_{\rm NS})=(20,2)\,M_\odot$. For all other BNS and NSBH mass configurations, the exclusion distances are smaller than the luminosity distance of EP240506a, and therefore the GW non-detection does not exclude those CBC scenarios at the measured distance. For all the other ten sources, the distance inferred from the measured redshift is greater than the $D_{90}$ values estimated with TDR, indicating that the source lies beyond the sensitivity range of the GW search and is therefore consistent with the GW non-detection.

For some FXTs, the multi-wavelength follow-up provided direct evidence for a non-CBC origin. EP240414a has been associated with a broad-lined Type Ic supernova  \citep{2025ApJ...982L..47V} and interpreted as the collapse of a massive star, connecting the event to the broader phenomenology of Luminous Fast Blue Optical Transients and long-GRB-like explosions. For EP250108a/SN~2025kg \citep{2025arXiv250417034L,2025ApJ...988L..13R,2025ApJ...988L..14E,2025ApJ...988L..60S}, the optical spectroscopy identifies a broad-lined Type Ic supernova, while the broadband data have been interpreted in terms of a jet-driven or cocoon-producing collapsar \citep{2026arXiv260606213C}, with additional models invoking magnetar energy injection from a rapidly rotating helium-star progenitor in a close binary \citep{2025MNRAS.544L.139Z}. More recently, \citet{2026ApJ...999..239L} reported a supernova candidate associated with EP240506a/AT2024ofs at $z=0.12$. Its X-ray luminosity is similar to that of EP240414a and EP250108a, suggesting that EP240506a may belong to the same class of engine-driven stellar explosions rather than a CBC origin.
For EP240708a, \citet{2026arXiv260627048V} identified two candidate host galaxies at redshifts 0.3734 and 0.3728. Based on the inferred host-galaxy properties, the implied peak X-ray luminosity, and the FXT--host offset, several progenitor scenarios remain viable for EP240708a, including a BNS merger, a relativistic WD--IMBH encounter, and a cocoon emission from a long GRB.
EP241021a \citep{2025A&A...701A.225B,2026MNRAS.545f2064Q} shows complex multi-wavelength evolution, with non-thermal emission at early times and evidence for an additional late-time component that has been interpreted as a possible supernova contribution, supporting a massive-stellar explosion and prolonged central-engine activity. On the other hand, \cite{2025ApJ...990L..29S} support either a merger-triggered magnetar as central engine or a jetted tidal disruption event. EP241217a, at a redshift of 4.59, has been interpreted as a collapsar whose prompt gamma-ray emission was below the sensitivity of available gamma-ray monitors \citep{2026ApJ...998..163Z}. 

One event, EP241107a at redshift 0.456, has been modelled as a GRB afterglow \citep{2026MNRAS.545f2062E}, with the inferred jet geometry favoring an on-axis or mildly off-axis viewing configuration and a relatively low intrinsic energy, which may explain the absence of a detected prompt gamma-ray counterpart. The low isotropic energy and density of interstellar medium inferred from afterglow fit, along with the galactic offset, are also compatible with a CBC origin.

\section{Conclusion} \label{Sec:conclusion}

Fast X-ray transients represent a rapidly growing yet still poorly understood class of high-energy phenomena. Their uncertain origin, coupled with the possibility that a subset may be produced by BNS or NSBH mergers, makes them prime targets for multi-messenger investigations. The first year of Einstein Probe operations has substantially expanded the known population of FXTs, demonstrating the power of the Wide-field X-ray Telescope to discover short-lived high-energy transients. When combined with LVK observations, the Einstein Probe offers a promising opportunity to identify FXTs associated with compact binary mergers.

In this work, we present the first systematic search for GW counterparts to publicly reported Einstein Probe fast X-ray transients during the period overlapping with the second half of the fourth LVK observing run.

The main aspects of our analysis and the resulting conclusions are summarized as follows:
\begin{itemize}
\item We constructed a sample of 47 EP-FXTs by excluding events associated with gamma-ray bursts or likely Galactic stellar flares, thereby focusing on the subset of FXTs whose progenitor remains unconstrained. For each FXT, we defined a source-specific search window extending from 1000 seconds before the EP/WXT trigger time to the end of the reported transient duration, and searched for temporal coincidences with GW candidates reported in GWTC-5. Seven GW candidates were found to be temporally coincident with seven EP-FXTs. Given the cumulative duration of the FXT search windows and the number of GW candidates found by CBC pipelines in O4b, the expected number of random temporal overlaps is \(\lambda=4.017\), corresponding to a Poisson probability \(P(K\ge7)=0.112\). Thus, the observed number of temporal coincidences is compatible with random temporal overlap alone. For the seven temporally coincident FXT--GW candidate pairs, we evaluated the spatial consistency and GW-candidate significance using a combined ranking statistic and estimated the corresponding joint false alarm rate. None of the candidate pairs reaches statistical significance. The p-value distribution is consistent with the expectation from chance alignments, and we find no evidence for a significant physical association between the two messengers.

\item In the absence of a GW detection, we used the Targeted Detectability Range tool to quantify the distance up to which the detector network was sensitive to compact-binary systems at the sky position and time of each transient. This provided the 90\% exclusion distance across the search window for each source, showing that nearby binary neutron star and neutron-star--black-hole origins can be excluded only within the corresponding detector-sensitive distances, with average exclusion distances of about 180 Mpc and 350 Mpc, respectively, for the sample studied here. For the sources with measured redshifts, these exclusion distances allow us to rule out the BNS and NSBH scenarios considered here. The main exception is EP240506a, the closest known-redshift source in the sample, at \(z\simeq0.12\); for this event, only adopting the most massive NSBH configuration for the TDR computation, \((20,2)\,M_\odot\), we reach a comparable exclusion distance at the trigger time, while the other BNS and NSBH configurations have an exclusion distance smaller than the measured distance.

\end{itemize}
The framework developed in this work establishes a methodology for connecting transients discovered by wide-field X-ray satellites with GW observations. A coincident GW detection would provide compelling evidence that at least a subset of FXTs originate from compact binary mergers. Looking ahead, observations by Einstein Probe, together with forthcoming LVK observations during the Interim Run (IR1) and, on longer timescales, with the planned upgrades to Advanced LIGO, Virgo, and KAGRA that will expand the volume accessible to GW observations, will enhance the prospects for joint FXT--GW detections. Such observations will enable increasingly stringent constraints on the physical origin of fast X-ray transients and on the properties of their underlying population.

\begin{acknowledgements}
The authors thank Gor Oganesyan and Matthias Vereecken for fruitful discussions and useful comments. This material is based upon work supported by NSF's LIGO Laboratory which is a major facility fully funded by the National Science Foundation. M.B. and S.R. acknowledge support from the Astrophysics Center for Multi-messenger Studies in Europe (ACME), funded under the European Union’s Horizon Europe Research and Innovation Program, Grant Agreement No. 101131928.      
\end{acknowledgements}

\bibliography{ref}

@ARTICLE{Quirola,
       author = {{Quirola-V{\'a}squez}, J. and {Bauer}, F.~E. and {Jonker}, P.~G. and {Brandt}, W.~N. and {Yang}, G. and {Levan}, A.~J. and {Xue}, Y.~Q. and {Eappachen}, D. and {Zheng}, X.~C. and {Luo}, B.},
        title = "{Extragalactic fast X-ray transient candidates discovered by Chandra (2000-2014)}",
      journal = {\aap},
         year = 2022,
        month = jul,
       volume = {663},
          eid = {A168},
        pages = {A168},
          doi = {10.1051/0004-6361/202243047},
archivePrefix = {arXiv},
       eprint = {2201.07773},
 primaryClass = {astro-ph.HE},
       adsurl = {https://ui.adsabs.harvard.edu/abs/2022A&A...663A.168Q}
}

@ARTICLE{2023A&A...675A..44Q,
       author = {{Quirola-V{\'a}squez}, J. and {Bauer}, F.~E. and {Jonker}, P.~G. and {Brandt}, W.~N. and {Yang}, G. and {Levan}, A.~J. and {Xue}, Y.~Q. and {Eappachen}, D. and {Camacho}, E. and {Ravasio}, M.~E. and {Zheng}, X.~C. and {Luo}, B.},
        title = "{Extragalactic fast X-ray transient candidates discovered by Chandra (2014-2022)}",
      journal = {\aap},
         year = 2023,
        month = jul,
       volume = {675},
          eid = {A44},
        pages = {A44},
          doi = {10.1051/0004-6361/202345912},
archivePrefix = {arXiv},
       eprint = {2304.13795},
 primaryClass = {astro-ph.HE},
       adsurl = {https://ui.adsabs.harvard.edu/abs/2023A&A...675A..44Q}
}

@ARTICLE{2020ApJ...896...39A,
       author = {{Alp}, Dennis and {Larsson}, Josefin},
        title = "{Blasts from the Past: Supernova Shock Breakouts among X-Ray Transients in the XMM-Newton Archive}",
      journal = {\apj},
         year = 2020,
        month = jun,
       volume = {896},
       number = {1},
          eid = {39},
        pages = {39},
          doi = {10.3847/1538-4357/ab91ba},
archivePrefix = {arXiv},
       eprint = {2004.09519},
 primaryClass = {astro-ph.HE},
       adsurl = {https://ui.adsabs.harvard.edu/abs/2020ApJ...896...39A}
}

@ARTICLE{2008Natur.453..469S,
       author = {{Soderberg}, A.~M. and {Berger}, E. and {Page}, K.~L. and {Schady}, P. and {Parrent}, J. and {Pooley}, D. and {Wang}, X.-Y. and {Ofek}, E.~O. and {Cucchiara}, A. and {Rau}, A. and {Waxman}, E. and {Simon}, J.~D. and {Bock}, D.~C.-J. and {Milne}, P.~A. and {Page}, M.~J. and {Barentine}, J.~C. and {Barthelmy}, S.~D. and {Beardmore}, A.~P. and {Bietenholz}, M.~F. and {Brown}, P. and {Burrows}, A. and {Burrows}, D.~N. and {Byrngelson}, G. and {Cenko}, S.~B. and {Chandra}, P. and {Cummings}, J.~R. and {Fox}, D.~B. and {Gal-Yam}, A. and {Gehrels}, N. and {Immler}, S. and {Kasliwal}, M. and {Kong}, A.~K.~H. and {Krimm}, H.~A. and {Kulkarni}, S.~R. and {Maccarone}, T.~J. and {M{\'e}sz{\'a}ros}, P. and {Nakar}, E. and {O'Brien}, P.~T. and {Overzier}, R.~A. and {de Pasquale}, M. and {Racusin}, J. and {Rea}, N. and {York}, D.~G.},
        title = "{An extremely luminous X-ray outburst at the birth of a supernova}",
      journal = {\nat},
         year = 2008,
        month = may,
       volume = {453},
       number = {7194},
        pages = {469-474},
          doi = {10.1038/nature06997},
archivePrefix = {arXiv},
       eprint = {0802.1712},
 primaryClass = {astro-ph},
       adsurl = {https://ui.adsabs.harvard.edu/abs/2008Natur.453..469S}
}

@ARTICLE{2025A&A...693A..62G,
       author = {{Grotova}, I. and {Rau}, A. and {Salvato}, M. and {Buchner}, J. and {Goodwin}, A.~J. and {Liu}, Z. and {Malyali}, A. and {Merloni}, A. and {Tub{\'\i}n-Arenas}, D. and {Homan}, D. and {Krumpe}, M. and {Nandra}, K. and {Shirley}, R. and {Anderson}, G.~E. and {Arcodia}, R. and {Bahic}, S. and {Baldini}, P. and {Buckley}, D.~A.~H. and {Ciroi}, S. and {Kawka}, A. and {Masterson}, M. and {Miller-Jones}, J.~C.~A. and {Di Mille}, F.},
        title = "{eRO-ExTra: eROSITA extragalactic non-AGN X-ray transients and variables in eRASS1 and eRASS2}",
      journal = {\aap},
         year = 2025,
        month = jan,
       volume = {693},
          eid = {A62},
        pages = {A62},
          doi = {10.1051/0004-6361/202451253},
archivePrefix = {arXiv},
       eprint = {2501.04208},
 primaryClass = {astro-ph.HE},
       adsurl = {https://ui.adsabs.harvard.edu/abs/2025A&A...693A..62G}
}

@ARTICLE{2026ApJ...999...75B,
       author = {{Brightman}, Murray and {Jaimes}, Joahan Casta{\~n}eda and {Stern}, Daniel and {Grefenstette}, Brian},
        title = "{Fast X-Ray Transients in NuSTAR Data}",
      journal = {\apj},
         year = 2026,
        month = mar,
       volume = {999},
       number = {1},
          eid = {75},
        pages = {75},
          doi = {10.3847/1538-4357/ae3f31},
archivePrefix = {arXiv},
       eprint = {2601.14375},
 primaryClass = {astro-ph.HE},
       adsurl = {https://ui.adsabs.harvard.edu/abs/2026ApJ...999...75B}
}

@ARTICLE{2019Natur.568..198X,
       author = {{Xue}, Y.~Q. and {Zheng}, X.~C. and {Li}, Y. and {Brandt}, W.~N. and {Zhang}, B. and {Luo}, B. and {Zhang}, B.-B. and {Bauer}, F.~E. and {Sun}, H. and {Lehmer}, B.~D. and {Wu}, X.-F. and {Yang}, G. and {Kong}, X. and {Li}, J.~Y. and {Sun}, M.~Y. and {Wang}, J.-X. and {Vito}, F.},
        title = "{A magnetar-powered X-ray transient as the aftermath of a binary neutron-star merger}",
      journal = {\nat},
         year = 2019,
        month = apr,
       volume = {568},
       number = {7751},
        pages = {198-201},
          doi = {10.1038/s41586-019-1079-5},
archivePrefix = {arXiv},
       eprint = {1904.05368},
 primaryClass = {astro-ph.HE},
       adsurl = {https://ui.adsabs.harvard.edu/abs/2019Natur.568..198X}
}

@ARTICLE{2013ApJ...779...14J,
       author = {{Jonker}, P.~G. and {Glennie}, A. and {Heida}, M. and {Maccarone}, T. and {Hodgkin}, S. and {Nelemans}, G. and {Miller-Jones}, J.~C.~A. and {Torres}, M.~A.~P. and {Fender}, R.},
        title = "{Discovery of a New Kind of Explosive X-Ray Transient near M86}",
      journal = {\apj},
         year = 2013,
        month = dec,
       volume = {779},
       number = {1},
          eid = {14},
        pages = {14},
          doi = {10.1088/0004-637X/779/1/14},
archivePrefix = {arXiv},
       eprint = {1310.7238},
 primaryClass = {astro-ph.HE},
       adsurl = {https://ui.adsabs.harvard.edu/abs/2013ApJ...779...14J}
}

@ARTICLE{2026A&A...705A.233B,
       author = {{Becerra}, R.~L. and {Yang}, Y.-H. and {Troja}, E. and {El Kabir}, M. and {Dichiara}, S. and {Passaleva}, N. and {O'Connor}, B. and {Ricci}, R. and {Fryer}, C. and {Hu}, L. and {Wu}, Q. and {Yadav}, M. and {Watson}, A.~M. and {Tsvetkova}, A. and {Angulo-Valdez}, C. and {Caballero-Garc{\'\i}a}, M.~D. and {Castro-Tirado}, A.~J. and {Cheung}, C.~C. and {Frederiks}, D. and {Gritsevich}, M. and {Grove}, J.~E. and {Kerr}, M. and {Lee}, W.~H. and {Lysenko}, A.~L. and {Pereyra}, M. and {Ridnaia}, A. and {S{\'a}nchez-Ram{\'\i}rez}, R. and {Sun}, H. and {Svinkin}, D. and {Ulanov}, M. and {Woolf}, R. and {Zhang}, B.},
        title = "{Exploring the connection between compact object mergers and fast X-ray transients: The cases of LXT 240402A and EP250207b}",
      journal = {\aap},
         year = 2026,
        month = jan,
       volume = {705},
          eid = {A233},
        pages = {A233},
          doi = {10.1051/0004-6361/202557612},
archivePrefix = {arXiv},
       eprint = {2510.13015},
 primaryClass = {astro-ph.HE},
       adsurl = {https://ui.adsabs.harvard.edu/abs/2026A&A...705A.233B}
}

@ARTICLE{2011PhRvD..83h4002H,
       author = {{Harry}, I.~W. and {Fairhurst}, S.},
        title = "{Targeted coherent search for gravitational waves from compact binary coalescences}",
      journal = {\prd},
         year = 2011,
        month = apr,
       volume = {83},
       number = {8},
          eid = {084002},
        pages = {084002},
          doi = {10.1103/PhysRevD.83.084002},
archivePrefix = {arXiv},
       eprint = {1012.4939},
 primaryClass = {gr-qc},
       adsurl = {https://ui.adsabs.harvard.edu/abs/2011PhRvD..83h4002H}
}

@ARTICLE{2014PhRvD..90l2004W,
       author = {{Williamson}, A.~R. and {Biwer}, C. and {Fairhurst}, S. and {Harry}, I.~W. and {Macdonald}, E. and {Macleod}, D. and {Predoi}, V.},
        title = "{Improved methods for detecting gravitational waves associated with short gamma-ray bursts}",
      journal = {\prd},
         year = 2014,
        month = dec,
       volume = {90},
       number = {12},
          eid = {122004},
        pages = {122004},
          doi = {10.1103/PhysRevD.90.122004},
archivePrefix = {arXiv},
       eprint = {1410.6042},
 primaryClass = {gr-qc},
       adsurl = {https://ui.adsabs.harvard.edu/abs/2014PhRvD..90l2004W}
}

@ARTICLE{2010NJPh...12e3034S,
       author = {{Sutton}, Patrick J. and {Jones}, Gareth and {Chatterji}, Shourov and {Kalmus}, Peter and {Leonor}, Isabel and {Poprocki}, Stephen and {Rollins}, Jameson and {Searle}, Antony and {Stein}, Leo and {Tinto}, Massimo and et al.},
        title = "{X-Pipeline: an analysis package for autonomous gravitational-wave burst searches}",
      journal = {New Journal of Physics},
         year = 2010,
        month = may,
       volume = {12},
       number = {5},
          eid = {053034},
        pages = {053034},
          doi = {10.1088/1367-2630/12/5/053034},
archivePrefix = {arXiv},
       eprint = {0908.3665},
 primaryClass = {gr-qc},
       adsurl = {https://ui.adsabs.harvard.edu/abs/2010NJPh...12e3034S}
}

@ARTICLE{2012PhRvD..86b2003W,
       author = {{W{\k{a}}s}, Micha{\l} and {Sutton}, Patrick J. and {Jones}, Gareth and {Leonor}, Isabel},
        title = "{Performance of an externally triggered gravitational-wave burst search}",
      journal = {\prd},
         year = 2012,
        month = jul,
       volume = {86},
       number = {2},
          eid = {022003},
        pages = {022003},
          doi = {10.1103/PhysRevD.86.022003},
archivePrefix = {arXiv},
       eprint = {1201.5599},
 primaryClass = {gr-qc},
       adsurl = {https://ui.adsabs.harvard.edu/abs/2012PhRvD..86b2003W}
}

@ARTICLE{2025SCPMA..6839501Y,
       author = {{Yuan}, Weimin and {Dai}, Lixin and {Feng}, Hua and {Jin}, Chichuan and {Jonker}, Peter and {Kuulkers}, Erik and {Liu}, Yuan and {Nandra}, Kirpal and {O'Brien}, Paul and {Piro}, Luigi and {Rau}, Arne and {Rea}, Nanda and {Sanders}, Jeremy and {Tao}, Lian and {Wang}, Junfeng and {Wu}, Xuefeng and {Zhang}, Bing and {Zhang}, Shuangnan and {Ai}, Shunke and {Buchner}, Johannes and {Bulbul}, Esra and {Chen}, Hechao and {Chen}, Minghua and {Chen}, Yong and {Chen}, Yu-Peng and {Coleiro}, Alexis and {Coti Zelati}, Francesco and {Dai}, Zigao and {Fan}, Xilong and {Fan}, Zhou and {Friedrich}, Susanne and {Gao}, He and {Ge}, Chong and {Ge}, Mingyu and {Geng}, Jinjun and {Ghirlanda}, Giancarlo and {Gianfagna}, Giulia and {Gou}, Lijun and {Guillot}, S{\'e}bastien and {Hou}, Xian and {Hu}, Jingwei and {Huang}, Yongfeng and {Ji}, Long and {Jia}, Shumei and {Komossa}, S. and {Kong}, Albert K.~H. and {Lan}, Lin and {Li}, An and {Li}, Ang and {Li}, Chengkui and {Li}, Dongyue and {Li}, Jian and {Li}, Zhaosheng and {Ling}, Zhixing and {Liu}, Ang and {Liu}, Jinzhong and {Liu}, Liangduan and {Liu}, Zhu and {Luo}, Jiawei and {Ma}, Ruican and {Maggi}, Pierre and {Maitra}, Chandreyee and {Marino}, Alessio and {Ng}, Stephen Chi-Yung and {Pan}, Haiwu and {Rukdee}, Surangkhana and {Soria}, Roberto and {Sun}, Hui and {Tam}, Pak-Hin Thomas and {Thakur}, Aishwarya Linesh and {Tian}, Hui and {Troja}, Eleonora and {Wang}, Wei and {Wang}, Xiangyu and {Wang}, Yanan and {Wei}, Junjie and {Wen}, Sixiang and {Wu}, Jianfeng and {Wu}, Ting and {Xiao}, Di and {Xu}, Dong and {Xu}, Renxin and {Xu}, Yanjun and {Xu}, Yu and {Yang}, Haonan and {You}, Bei and {Yu}, Heng and {Yu}, Yunwei and {Zhang}, Binbin and {Zhang}, Chen and {Zhang}, Guobao and {Zhang}, Liang and {Zhang}, Wenda and {Zhang}, Yu and {Zhou}, Ping and {Zou}, Zecheng},
        title = "{Science objectives of the Einstein Probe mission}",
      journal = {Science China Physics, Mechanics, and Astronomy},
         year = 2025,
        month = mar,
       volume = {68},
       number = {3},
          eid = {239501},
        pages = {239501},
          doi = {10.1007/s11433-024-2600-3},
archivePrefix = {arXiv},
       eprint = {2501.07362},
 primaryClass = {astro-ph.HE},
       adsurl = {https://ui.adsabs.harvard.edu/abs/2025SCPMA..6839501Y}
}

@ARTICLE{2017ApJ...841...89A,
       author = {{Abbott}, B.~P. and {Abbott}, R. and {Abbott}, T.~D. and {Abernathy}, M.~R. and {Acernese}, F. and {Ackley}, K. and {Adams}, C. and {Adams}, T. and {Addesso}, P. and {Adhikari}, R.~X. and et al.},
        title = "{Search for Gravitational Waves Associated with Gamma-Ray Bursts during the First Advanced LIGO Observing Run and Implications for the Origin of GRB 150906B}",
      journal = {\apj},
         year = 2017,
        month = jun,
       volume = {841},
       number = {2},
          eid = {89},
        pages = {89},
          doi = {10.3847/1538-4357/aa6c47},
archivePrefix = {arXiv},
       eprint = {1611.07947},
 primaryClass = {astro-ph.HE},
       adsurl = {https://ui.adsabs.harvard.edu/abs/2017ApJ...841...89A}
}

@ARTICLE{2019ApJ...886...75A,
       author = {{Abbott}, B.~P. and {Abbott}, R. and {Abbott}, T.~D. and {Abraham}, S. and {Acernese}, F. and {Ackley}, K. and {Adams}, C. and {Adhikari}, R.~X. and {Adya}, V.~B. and {Affeldt}, C. and et al.},
        title = "{Search for Gravitational-wave Signals Associated with Gamma-Ray Bursts during the Second Observing Run of Advanced LIGO and Advanced Virgo}",
      journal = {\apj},
         year = 2019,
        month = nov,
       volume = {886},
       number = {1},
          eid = {75},
        pages = {75},
          doi = {10.3847/1538-4357/ab4b48},
archivePrefix = {arXiv},
       eprint = {1907.01443},
 primaryClass = {astro-ph.HE},
       adsurl = {https://ui.adsabs.harvard.edu/abs/2019ApJ...886...75A}
}

@ARTICLE{2021ApJ...915...86A,
       author = {{Abbott}, R. and {Abbott}, T.~D. and {Abraham}, S. and {Acernese}, F. and {Ackley}, K. and {Adams}, C. and {Adhikari}, R.~X. and {Adya}, V.~B. and {Affeldt}, C. and {Agathos}, M. and et al.},
        title = "{Search for Gravitational Waves Associated with Gamma-Ray Bursts Detected by Fermi and Swift during the LIGO-Virgo Run O3a}",
      journal = {\apj},
         year = 2021,
        month = jul,
       volume = {915},
       number = {2},
          eid = {86},
        pages = {86},
          doi = {10.3847/1538-4357/abee15},
archivePrefix = {arXiv},
       eprint = {2010.14550},
 primaryClass = {astro-ph.HE},
       adsurl = {https://ui.adsabs.harvard.edu/abs/2021ApJ...915...86A}
}

@ARTICLE{2022ApJ...928..186A,
       author = {{Abbott}, R. and {Abbott}, T.~D. and {Acernese}, F. and {Ackley}, K. and {Adams}, C. and {Adhikari}, N. and {Adhikari}, R.~X. and {Adya}, V.~B. and {Affeldt}, C. and {Agarwal}, D. and et al.},
        title = "{Search for Gravitational Waves Associated with Gamma-Ray Bursts Detected by Fermi and Swift during the LIGO-Virgo Run O3b}",
      journal = {\apj},
         year = 2022,
        month = apr,
       volume = {928},
       number = {2},
          eid = {186},
        pages = {186},
          doi = {10.3847/1538-4357/ac532b},
archivePrefix = {arXiv},
       eprint = {2111.03608},
 primaryClass = {astro-ph.HE},
       adsurl = {https://ui.adsabs.harvard.edu/abs/2022ApJ...928..186A}
}

@ARTICLE{2026MNRAS.545f2021J,
       author = {{Jonker}, P.~G. and {Levan}, A.~J. and {Liu}, Xing and {Xu}, Dong and {Liu}, Yuan and {Xu}, Xinpeng and {Li}, An and {Sarin}, N. and {Tanvir}, N.~R. and {Lamb}, G.~P. and {Ravasio}, M.~E. and {S{\'a}nchez-Sierras}, J. and {Quirola-V{\'a}squez}, J.~A. and {Rayson}, B.~C. and {van Dalen}, J.~N.~D. and {Malesani}, D.~B. and {van Hoof}, A.~P.~C. and {Bauer}, F.~E. and {Chac{\'o}n}, J. and {Smartt}, S.~J. and {Martin-Carrillo}, A. and {Corcoran}, G. and {Cotter}, L. and {Rossi}, A. and {Onori}, F. and {Fraser}, M. and {O'Brien}, P.~T. and {Eyles-Ferris}, R.~A.~J. and {Hjorth}, J. and {Chen}, T.-W. and {Leloudas}, G. and {Tomasella}, L. and {Schulze}, S. and {De Pasquale}, M. and {Carotenuto}, F. and {Bright}, J. and {Wang}, Chenwei and {Xiong}, Shaolin and {Zhang}, Jinpeng and {Xue}, Wangchen and {Liu}, Jiacong and {Li}, Chengkui and {Mata S{\'a}nchez}, D. and {Torres}, M.~A.~P.},
        title = "{EP250207b is not a collapsar fast X-ray transient. Is it due to a binary compact object merger?}",
      journal = {\mnras},
         year = 2026,
        month = jan,
       volume = {545},
       number = {2},
          eid = {staf2021},
        pages = {staf2021},
          doi = {10.1093/mnras/staf2021},
archivePrefix = {arXiv},
       eprint = {2508.13039},
 primaryClass = {astro-ph.HE},
       adsurl = {https://ui.adsabs.harvard.edu/abs/2026MNRAS.545f2021J}
}

@ARTICLE{2025NatAs...9..564L,
       author = {{Liu}, Y. and {Sun}, H. and {Xu}, D. and {Svinkin}, D.~S. and {Delaunay}, J. and {Tanvir}, N.~R. and {Gao}, H. and {Zhang}, C. and {Chen}, Y. and {Wu}, X.-F. and {Zhang}, B. and {Yuan}, W. and {An}, J. and {Bruni}, G. and {Frederiks}, D.~D. and {Ghirlanda}, G. and {Hu}, J.-W. and {Li}, A. and {Li}, C.-K. and {Li}, J.-D. and {Malesani}, D.~B. and {Piro}, L. and {Raman}, G. and {Ricci}, R. and {Troja}, E. and {Vergani}, S.~D. and {Wu}, Q.-Y. and {Yang}, J. and {Zhang}, B.-B. and {Zhu}, Z.-P. and {de Ugarte Postigo}, A. and {Demin}, A.~G. and {Dobie}, D. and {Fan}, Z. and {Fu}, S.-Y. and {Fynbo}, J.~P.~U. and {Geng}, J.-J. and {Gianfagna}, G. and {Hu}, Y.-D. and {Huang}, Y.-F. and {Jiang}, S.-Q. and {Jonker}, P.~G. and {Julakanti}, Y. and {Kennea}, J.~A. and {Kokomov}, A.~A. and {Kuulkers}, E. and {Lei}, W.-H. and {Leung}, J.~K. and {Levan}, A.~J. and {Li}, D.-Y. and {Li}, Y. and {Littlefair}, S.~P. and {Liu}, X. and {Lysenko}, A.~L. and {Ma}, Y.-N. and {Martin-Carrillo}, A. and {O'Brien}, P. and {Parsotan}, T. and {Quirola-V{\'a}squez}, J. and {Ridnaia}, A.~V. and {Ronchini}, S. and {Rossi}, A. and {Mata-S{\'a}nchez}, D. and {Schneider}, B. and {Shen}, R.-F. and {Thakur}, A.~L. and {Tohuvavohu}, A. and {Torres}, M.~A.~P. and {Tsvetkova}, A.~E. and {Ulanov}, M.~V. and {Wei}, J.-J. and {Xiao}, D. and {Yin}, Y.-H.~I. and {Bai}, M. and {Burwitz}, V. and {Cai}, Z.-M. and {Chen}, F.-S. and {Chen}, H.-L. and {Chen}, T.-X. and {Chen}, W. and {Chen}, Y.-F. and {Chen}, Y.-H. and {Cheng}, H.-Q. and {Cordier}, B. and {Cui}, C.-Z. and {Cui}, W.-W. and {Dai}, Y.-F. and {Dai}, Z.-G. and {Eder}, J. and {Eyles-Ferris}, R.~A.~J. and {Fan}, D.-W. and {Feldman}, C. and {Feng}, H. and {Feng}, Z. and {Friedrich}, P. and {Gao}, X. and {Gonzalez}, J.-F. and {Guan}, J. and {Han}, D.-W. and {Han}, J. and {Hou}, D.-J. and {Hu}, H.-B. and {Hu}, T. and {Huang}, M.-H. and {Huo}, J. and {Hutchinson}, I. and {Ji}, Z. and {Jia}, S.-M. and {Jia}, Z.-Q. and {Jiang}, B.-W. and {Jin}, C.-C. and {Jin}, G. and {Jin}, J.-J. and {Keereman}, A. and {Lerman}, H. and {Li}, J.-F. and {Li}, L.-H. and {Li}, M.-S. and {Li}, W. and {Li}, Z.-D. and {Lian}, T.-Y. and {Liang}, E.-W. and {Ling}, Z.-X. and {Liu}, C.-Z. and {Liu}, H.-Y. and {Liu}, H.-Q. and {Liu}, M.-J. and {Liu}, Y.-R. and {Lu}, F.-J. and {L{\"u}}, H.-J. and {Luo}, L.-D. and {Ma}, F.~L. and {Ma}, J. and {Mao}, J.-R. and {Mao}, X. and {McHugh}, M. and {Meidinger}, N. and {Nandra}, K. and {Osborne}, J.~P. and {Pan}, H.-W. and {Pan}, X. and {Ravasio}, M.~E. and {Rau}, A. and {Rea}, N. and {Rehman}, U. and {Sanders}, J. and {Santovincenzo}, A. and {Song}, L.-M. and {Su}, J. and {Sun}, L.-J. and {Sun}, S.-L. and {Sun}, X.-J. and {Tan}, Y.-Y. and {Tang}, Q.-J. and {Tao}, Y.-H. and {Tong}, J.-Z. and {Wang}, C.-Y. and {Wang}, H. and {Wang}, J. and {Wang}, L. and {Wang}, W.-X. and {Wang}, X.-F. and {Wang}, X.-Y. and {Wang}, Y.-L. and {Wang}, Y.-S. and {Wei}, D.-M. and {Willingale}, R. and {Xiong}, S.-L. and {Xu}, H.-T. and {Xu}, J.-J. and {Xu}, X.-P. and {Xu}, Y.-F. and {Xu}, Z. and {Xue}, C.-B. and {Xue}, Y.-L. and {Yan}, A.-L. and {Yang}, F. and {Yang}, H.-N. and {Yang}, X.-T. and {Yang}, Y.-J. and {Yu}, Y.-W. and {Zhang}, J. and {Zhang}, M. and {Zhang}, S.-N. and {Zhang}, W.-D. and {Zhang}, W.-J. and {Zhang}, Y.-H. and {Zhang}, Z. and {Zhang}, Z. and {Zhang}, Z.-L. and {Zhao}, D.-H. and {Zhao}, H.-S. and {Zhao}, X.-F. and {Zhao}, Z.-J. and {Zhou}, L.-X. and {Zhou}, Y.-L. and {Zhu}, Y.-X. and {Zhu}, Z.-C. and {Zuo}, X.-X.},
        title = "{Soft X-ray prompt emission from the high-redshift gamma-ray burst EP240315a}",
      journal = {Nature Astronomy},
         year = 2025,
        month = apr,
       volume = {9},
        pages = {564-576},
          doi = {10.1038/s41550-024-02449-8},
archivePrefix = {arXiv},
       eprint = {2404.16425},
 primaryClass = {astro-ph.HE},
       adsurl = {https://ui.adsabs.harvard.edu/abs/2025NatAs...9..564L}
}

@ARTICLE{2025ApJ...988L..34J,
       author = {{Jiang}, Shuai-Qing and {Xu}, Dong and {van Hoof}, Agnes P.~C. and {Lei}, Wei-Hua and {Liu}, Yuan and {Zhou}, Hao and {Chen}, Yong and {Fu}, Shao-Yu and {Yang}, Jun and {Liu}, Xing and {Zhu}, Zi-Pei and {Filippenko}, Alexei V. and {Jonker}, Peter G. and {Pozanenko}, A.~S. and {Gao}, He and {Wu}, Xue-Feng and {Zhang}, Bing and {Lamb}, Gavin P. and {De Pasquale}, Massimiliano and {Kobayashi}, Shiho and {Bauer}, Franz Erik and {Sun}, Hui and {Pugliese}, Giovanna and {An}, Jie and {D'Elia}, Valerio and {Fynbo}, Johan P.~U. and {Zheng}, WeiKang and {Castro-Tirado}, Alberto J. and {Yin}, Yi-Han Iris and {Zou}, Yuan-Chuan and {Deller}, Adam T. and {Pankov}, N.~S. and {Volnova}, A.~A. and {Moskvitin}, A.~S. and {Spiridonova}, O.~I. and {Oparin}, D.~V. and {Rumyantsev}, V. and {Burkhonov}, O.~A. and {Egamberdiyev}, Sh. A. and {Kim}, V. and {Krugov}, M. and {Tatarnikov}, A.~M. and {Inasaridze}, R. and {Levan}, Andrew J. and {Malesani}, Daniele Bj{\o}rn and {Ravasio}, Maria E. and {Quirola-V{\'a}squez}, Jonathan and {van Dalen}, Joyce N.~D. and {S{\'a}nchez-Sierras}, Javi and {Mata S{\'a}nchez}, Daniel and {Littlefair}, Stuart P. and {Chac{\'o}n}, Jennifer A. and {Torres}, Manuel A.~P. and {Chrimes}, Ashley A. and {Sarin}, Nikhil and {Martin-Carrillo}, Antonio and {Dhillon}, Vik and {Yang}, Yi and {Brink}, Thomas G. and {Davies}, Rebecca L. and {Yang}, Sheng and {Aryan}, Amar and {Chen}, Ting-Wan and {Kong}, Albert K.~H. and {Li}, Wen-Xiong and {Li}, Rui-Zhi and {Mao}, Jirong and {P{\'e}rez-Garc{\'\i}a}, Ignacio and {Fern{\'a}ndez-Garc{\'\i}a}, Emilio J. and {Andrews}, Moira and {Farah}, Joseph and {Fan}, Zhou and {Padilla Gonzalez}, Estefania and {Howell}, D. Andrew and {Hartmann}, Dieter and {Hu}, Jing-Wei and {Jakobsson}, P{\'a}ll and {Li}, Cheng-Kui and {Ling}, Zhi-Xing and {McCully}, Curtis and {Newsome}, Megan and {Schneider}, Benjamin and {Tinyanont}, Kaew Samaporn and {Sun}, Ning-Chen and {Terreran}, Giacomo and {Tang}, Qing-Wen and {Wang}, Wen-Xin and {Xu}, Jing-Jing and {Yuan}, Wei-Min and {Zhang}, Bin-Bin and {Zhao}, Hai-Sheng and {Zhang}, Juan},
        title = "{EP240801a/XRF 240801B: An X-Ray Flash Detected by the Einstein Probe and the Implications of Its Multiband Afterglow}",
      journal = {\apjl},
         year = 2025,
        month = jul,
       volume = {988},
       number = {1},
          eid = {L34},
        pages = {L34},
          doi = {10.3847/2041-8213/addebf},
archivePrefix = {arXiv},
       eprint = {2503.04306},
 primaryClass = {astro-ph.HE},
       adsurl = {https://ui.adsabs.harvard.edu/abs/2025ApJ...988L..34J}
}

@ARTICLE{2024A&A...683A.243Q,
       author = {{Quirola-V{\'a}squez}, J. and {Bauer}, F.~E. and {Jonker}, P.~G. and {Brandt}, W.~N. and {Eappachen}, D. and {Levan}, A.~J. and {L{\'o}pez}, E. and {Luo}, B. and {Ravasio}, M.~E. and {Sun}, H. and {Xue}, Y.~Q. and {Yang}, G. and {Zheng}, X.~C.},
        title = "{Probing a magnetar origin for the population of extragalactic fast X-ray transients detected by Chandra}",
      journal = {\aap},
         year = 2024,
        month = mar,
       volume = {683},
          eid = {A243},
        pages = {A243},
          doi = {10.1051/0004-6361/202347629},
archivePrefix = {arXiv},
       eprint = {2401.01415},
 primaryClass = {astro-ph.HE},
       adsurl = {https://ui.adsabs.harvard.edu/abs/2024A&A...683A.243Q}
}

@ARTICLE{2022ApJ...927..211L,
       author = {{Lin}, Dacheng and {Irwin}, Jimmy A. and {Berger}, Edo and {Nguyen}, Ronny},
        title = "{Discovery of Three Candidate Magnetar-powered Fast X-Ray Transients from Chandra Archival Data}",
      journal = {\apj},
         year = 2022,
        month = mar,
       volume = {927},
       number = {2},
          eid = {211},
        pages = {211},
          doi = {10.3847/1538-4357/ac4fc6},
archivePrefix = {arXiv},
       eprint = {2201.06754},
 primaryClass = {astro-ph.HE},
       adsurl = {https://ui.adsabs.harvard.edu/abs/2022ApJ...927..211L}
}

@ARTICLE{2026arXiv260610002M,
       author = {{Martin-Carrillo}, Antonio and {Th{\"o}ne}, Christina C. and {Leung}, James K. and {Corcoran}, Gregory and {de Ugarte Postigo}, Antonio and {Jonker}, Peter G. and {Izzo}, Luca and {Levan}, Andrew J. and {Gompertz}, Benjamin P. and {Basa}, St{\'e}phane and {Sarin}, Nikhil and {Quirola-V{\'a}squez}, Jonathan and {Eyles-Ferris}, Rob A.~J. and {Brivio}, Riccardo and {Watson}, Alan M. and {Cotter}, Laura and {Chac{\'o}n}, Jennifer Alexandra and {Rossi}, Andrea and {Melandri}, Andrea and {Kumnurdmanee}, Piramon and {Tanvir}, Nial R. and {Gupta}, Anshika and {Bauer}, Franz E. and {Ducoin}, Jean-Gr{\'e}goire and {Reguitti}, Andrea and {Misra}, Kuntal and {Xu}, Dong and {Vergani}, Susanna D. and {Fong}, Wen-fai and {Ackley}, Kendall and {Aguilar-Ruiz}, Edilberto and {Akl}, Dalya and {Aloy}, Miguel {\'A}ngel and {An}, Jie and {Angulo-Valdez}, Camila and {Antier}, Sarah and {Atteia}, Jean-Luc and {Becerra}, Rosa L. and {Breton}, Rene P. and {Butler}, Nathaniel R. and {Campana}, Sergio and {Carotenuto}, Francesco and {Casares Vel{\'a}zquez}, Jorge and {Chrimes}, Ashley A. and {D'Elia}, Valerio and {van Dalen}, Joyce N.~D. and {De Colle}, Fabio and {De Pasquale}, Massimiliano and {Dhillon}, Vik S. and {Dornic}, Damien and {Dyer}, Martin J. and {Ferro}, Matteo and {Fraser}, Morgan and {Fruchter}, Andrew S. and {Fortin}, Francis and {Galloway}, Duncan K. and {Garc{\'\i}a-Garc{\'\i}a}, Leonardo and {Geier}, Stefan and {Gill}, Ramandeep and {Globus}, No{\'e}mie and {Gualandi}, Roberto and {Guelfand}, Marion and {Guidolin}, Francesco and {Hartmann}, Dieter H. and {van Hoof}, Agnes P.~C. and {Jakobsson}, Pall and {Janghel}, Divyanshu and {Killestein}, Tom L. and {Klose}, Sylvio and {Kobayashi}, Shiho and {Kotak}, Rubina and {Kumar}, Amit and {Kuwata}, Asuka and {Laskar}, Tanmoy and {Lee}, William H. and {Lincetto}, Massimiliano and {Lombardi}, Gianluca and {L{\'o}pez-C{\'a}mara}, Diego and {Lyman}, Joseph D. and {Maiorano}, Elisabetta and {Maeda}, Keiichi and {Mandarakas}, Nikos and {Magnani}, Francesco and {Mao}, Jirong and {Moreno M{\'e}ndez}, Enrique and {Mar{\'\i}a Nicuesa Guelbenzu}, Ana and {Noysena}, Kanthanakorn and {Nuttall}, Laura K. and {O'Brien}, Paul T. and {O'Neill}, David and {Ochner}, Paolo and {Pereyra}, Margarita and {Pugliese}, Giovanna and {Ramsay}, Gavin and {Rhodes}, Lauren and {Saccardi}, Andrea and {Salvaterra}, Ruben and {S{\'a}nchez {\'A}lvarez}, Fredd and {Schneider}, Benjamin and {Schulze}, Steve and {Starling}, Rhaana L.~C. and {Steeghs}, Danny and {Ulaczyk}, Kzrysztof and {Ventura}, Chiara and {Zafar}, Tayyaba and {Zhu}, Zi-Pei},
        title = "{Failed jet breakout in the metal-poor broad-lined type Ic supernova 2026gzf}",
      journal = {arXiv e-prints},
         year = 2026,
        month = jun,
          eid = {arXiv:2606.10002},
        pages = {arXiv:2606.10002},
          doi = {10.48550/arXiv.2606.10002},
archivePrefix = {arXiv},
       eprint = {2606.10002},
 primaryClass = {astro-ph.HE},
       adsurl = {https://ui.adsabs.harvard.edu/abs/2026arXiv260610002M}
}

@ARTICLE{2026arXiv260527223T,
       author = {{The LIGO Scientific Collaboration} and {the Virgo Collaboration} and {the KAGRA Collaboration} and {Abac}, A.~G. and {Abe}, A. and {Abouelfettouh}, I. and {Acernese}, F. and {Ackley}, K. and {Adam}, A. and {Adhicary}, S. and {Adhikari}, D. and {Adhikari}, R.~X. and {Adkins}, V.~K. and {Afroz}, S. and {Agapito}, A. and {Agarwal}, D. and {Agathos}, M. and {Aggarwal}, N. and {Aggarwal}, S. and {Aguiar}, O.~D. and {Ahrend}, I.-L. and {Aiello}, L. and {Ain}, A. and {Ajith}, P. and {Akutsu}, T. and {Albers}, L. and {Ali}, W. and {Al-Kershi}, S. and {Allene}, C. and {Allocca}, A. and {Al-Shammari}, S. and {Alvarez}, J.~A. and {Alvarez-Lopez}, S. and {Amar}, W. and {Amarasinghe}, O. and {Amato}, A. and {Amicucci}, F. and {Amra}, C. and {Anand}, A.~B. and {Anand}, C. and {Ananyeva}, A. and {Anderson}, S.~B. and {Anderson}, W.~G. and {Andia}, M. and {Ando}, M. and {Andrade-Oliveira}, F. and {Andr{\'e}s-Carcasona}, M. and {Andrey}, J.~L. and {Andri{\'c}}, T. and {Anglin}, J. and {Anna}, J. and {Antelis}, J.~M. and {Antier}, S. and {Aoki}, T. and {Aoumi}, M. and {Appavuravther}, E.~Z. and {Appelt}, E.~A. and {Appert}, S. and {Apple}, S.~K. and {Arai}, K. and {Araya}, A. and {Araya}, M.~C. and {Arca Sedda}, M. and {Arciprete}, F. and {Areeda}, J.~S. and {Aritomi}, N. and {Armato}, F. and {Armstrong}, S. and {Arnaud}, N. and {Arogeti}, M. and {Aronson}, S.~M. and {Ashton}, G. and {Aso}, Y. and {Asprea}, L. and {Assiduo}, M. and {Assis de Souza Melo}, S. and {Aston}, S.~M. and {Astone}, P. and {Aswathi}, P.~S. and {Attadio}, F. and {Aubin}, F. and {AultONeal}, K. and {Avallone}, G. and {Avdeev}, N. and {Avila}, E.~A. and {Babak}, S. and {Badger}, C. and {Bae}, S. and {Bagnasco}, S. and {Baimukhametova}, S. and {Baiotti}, L. and {Baka}, T. and {Baker}, K.~A. and {Baker}, T. and {Balbi}, G. and {Baldi}, G. and {Baldicchi}, N. and {Ball}, M. and {Ballardin}, G. and {Ballelli}, M. and {Ballmer}, S.~W. and {Banagiri}, S. and {Banerjee}, B. and {Bankar}, D. and {Baptiste}, T.~M. and {Baral}, P. and {Baratti}, M. and {Barayoga}, J.~C. and {Baric}, K. and {Barish}, B.~C. and {Barker}, D. and {Barman}, N. and {Barone}, F. and {Barr}, B. and {Barrios}, M. and {Barsotti}, L. and {Barsuglia}, M. and {Barta}, D. and {Barton}, M.~A. and {Bartos}, I. and {Basalaev}, A. and {Bassiri}, R. and {Basti}, A. and {Bawaj}, M. and {Bayley}, J.~C. and {Baylor}, A.~C. and {Baynard}, II, P.~A. and {Bazzan}, M. and {Bedakihale}, V.~M. and {Beirnaert}, F. and {Bejger}, M. and {Bell}, A.~S. and {Bellani}, C. and {Bellie}, D.~S. and {Beltran-Martinez}, D. and {Benedetti}, E. and {Benoit}, W. and {Bentara}, I. and {Ben Yaala}, M. and {Bera}, S. and {Bergamin}, F. and {Berger}, B.~K. and {Beroiz}, M. and {Berry}, C.~P.~L. and {Berry}, I. and {Bersanetti}, D. and {Bertheas}, T. and {Bertolini}, A. and {Betzwieser}, J. and {Beveridge}, D. and {Bevins}, N. and {Bezerra-Sobrinho}, J. and {Bhandare}, R. and {Bhatt}, R. and {Bhattacharjee}, A. and {Bhattacharjee}, D. and {Bhattacharyya}, S. and {Bhaumik}, S. and {Biancalana}, V. and {Bianchi}, F. and {Bilenko}, I.~A. and {Bilicki}, M. and {Billingsley}, G. and {Binetti}, A. and {Bini}, S. and {Biot}, S. and {Birnholtz}, O. and {Biscoveanu}, S. and {Bisht}, A. and {Bitossi}, M. and {Bizouard}, M.-A. and {Blaber}, S. and {Blackburn}, J.~K. and {Blagg}, L.~A. and {Blair}, C.~D. and {Blair}, D.~G. and {Bloch}, M. and {Bode}, N. and {Boettner}, N. and {Bogdan}, P. and {Boileau}, G. and {Boldrini}, M. and {Bolingbroke}, G.~N. and {Bonavena}, L.~D. and {Bonhomme}, V.~A. and {Bonilla}, E. and {Bonilla}, M.~S. and {Bonino}, A. and {Bonnand}, R. and {Borchers}, A. and {Borghi}, N. and {Boschi}, V. and {Bose}, S. and {Bossilkov}, V. and {Bothra}, Y. and {Boudon}, A. and {Boybeyi}, T.~D. and {Boyle}, M. and {Bozzi}, A. and {Bradaschia}, C.},
        title = "{GWTC-5.0: An Introduction to Version 5.0 of the Gravitational-Wave Transient Catalog}",
      journal = {arXiv e-prints},
         year = 2026,
        month = may,
          eid = {arXiv:2605.27223},
        pages = {arXiv:2605.27223},
          doi = {10.48550/arXiv.2605.27223},
archivePrefix = {arXiv},
       eprint = {2605.27223},
 primaryClass = {gr-qc},
       adsurl = {https://ui.adsabs.harvard.edu/abs/2026arXiv260527223T}
}

@ARTICLE{2026arXiv260606213C,
       author = {{Cotter}, L. and {Martin-Carrillo}, A. and {Eyles-Ferris}, R.~A.~J. and {Izzo}, L. and {Malesani}, D.~B. and {Julakanti}, Y. and {Corcoran}, G. and {Saccardi}, A. and {Jonker}, P.~G. and {Levan}, A.~J. and {Carotenuto}, F. and {O'Brien}, P.~T. and {Gillanders}, J.~H. and {van Dalen}, J.~N.~D. and {Ravasio}, M.~E. and {Schulze}, S. and {Sarin}, N. and {Bauer}, F.~E. and {Fraser}, M. and {Quirola-Vasquez}, J. and {van Hoof}, A.~P.~C. and {Smartt}, S.~J. and {Gall}, C. and {Rest}, A. and {Murphey}, C.~T. and {Tanvir}, N. and {Chen}, T.-W. and {Campana}, S. and {Ashall}, C. and {Anderson}, J.~P. and {Chacon}, J.~A. and {Cowie}, F.~J. and {D'Elia}, V. and {Galbany}, L. and {Gutierrez}, C.~P. and {Hartmann}, D.~H. and {Jakobsson}, P. and {Kobayashi}, S. and {Kong}, A.~H. and {Mazalli}, P. and {Muller-Bravo}, T.~E. and {De Pasquale}, M. and {Rhodes}, L. and {Rossi}, A. and {Sanchez-Sierras}, J. and {Sollerman}, J. and {Andersson}, A. and {Aryan}, A. and {de Boer}, T. and {Bright}, J.~S. and {Chambers}, K.~C. and {Gromadzki}, M. and {Huber}, M.~E. and {Inserra}, C. and {Lowe}, T. and {Minguez}, P. and {Narayan}, G.~S. and {Nicholl}, M. and {Paek}, G.~S.~H. and {Sedgewick}, A. and {Smith}, K.~W. and {Tweddle}, J.~W. and {Yang}, S.},
        title = "{Probing a new subclass of llGRB-SN transients: Insights from EP250304a and its associated supernova}",
      journal = {arXiv e-prints},
         year = 2026,
        month = jun,
          eid = {arXiv:2606.06213},
        pages = {arXiv:2606.06213},
          doi = {10.48550/arXiv.2606.06213},
archivePrefix = {arXiv},
       eprint = {2606.06213},
 primaryClass = {astro-ph.HE},
       adsurl = {https://ui.adsabs.harvard.edu/abs/2026arXiv260606213C}
}

@ARTICLE{2025ApJ...982L..47V,
       author = {{van Dalen}, Joyce N.~D. and {Levan}, Andrew J. and {Jonker}, Peter G. and {Malesani}, Daniele Bj{\o}rn and {Izzo}, Luca and {Sarin}, Nikhil and {Quirola-V{\'a}squez}, Jonathan and {Mata S{\'a}nchez}, Daniel and {de Ugarte Postigo}, Antonio and {van Hoof}, Agnes P.~C. and {Torres}, Manuel A.~P. and {Schulze}, Steve and {Littlefair}, Stuart P. and {Chrimes}, Ashley and {Ravasio}, Maria E. and {Bauer}, Franz E. and {Martin-Carrillo}, Antonio and {Fraser}, Morgan and {van der Horst}, Alexander J. and {Jakobsson}, Pall and {O'Brien}, Paul and {De Pasquale}, Massimiliano and {Pugliese}, Giovanna and {Sollerman}, Jesper and {Tanvir}, Nial R. and {Zafar}, Tayyaba and {Anderson}, Joseph P. and {Galbany}, Llu{\'\i}s and {Gal-Yam}, Avishay and {Gromadzki}, Mariusz and {M{\"u}ller-Bravo}, Tom{\'a}s E. and {Ragosta}, Fabio and {Terwel}, Jacco H.},
        title = "{The Einstein Probe Transient EP240414a: Linking Fast X-Ray Transients, Gamma-Ray Bursts, and Luminous Fast Blue Optical Transients}",
      journal = {\apjl},
         year = 2025,
        month = apr,
       volume = {982},
       number = {2},
          eid = {L47},
        pages = {L47},
          doi = {10.3847/2041-8213/adbc7e},
archivePrefix = {arXiv},
       eprint = {2409.19056},
 primaryClass = {astro-ph.HE},
       adsurl = {https://ui.adsabs.harvard.edu/abs/2025ApJ...982L..47V}
}

@ARTICLE{2026MNRAS.545f2064Q,
       author = {{Quirola-V{\'a}squez}, J. and {Jonker}, P.~G. and {Levan}, A.~J. and {Malesani}, D.~B. and {Bauer}, F.~E. and {Sarin}, N. and {Lamb}, G.~P. and {Martin-Carrillo}, A. and {S{\'a}nchez-Sierras}, J. and {Fraser}, M. and {Izzo}, L. and {Ravasio}, M.~E. and {Mata S{\'a}nchez}, D. and {Torres}, M.~A.~P. and {van Dalen}, J.~N.~D. and {van Hoof}, A.~P.~C. and {Chac{\'o}n}, J.~A. and {Littlefair}, S. and {Dhillon}, V.~S. and {Cotter}, L. and {Corcoran}, G. and {Eyles-Ferris}, R.~A.~J. and {O'Brien}, P.~T. and {Stern}, D. and {Harrison}, F. and {D'Elia}, V. and {Hartmann}, D.~H.},
        title = "{Unveiling the nature of the Einstein Probe transient EP241021a}",
      journal = {\mnras},
         year = 2026,
        month = feb,
       volume = {545},
       number = {4},
          eid = {staf2064},
        pages = {staf2064},
          doi = {10.1093/mnras/staf2064},
archivePrefix = {arXiv},
       eprint = {2511.13314},
 primaryClass = {astro-ph.HE},
       adsurl = {https://ui.adsabs.harvard.edu/abs/2026MNRAS.545f2064Q}
}

@ARTICLE{2024GCN.36405....1L,
       author = {{Li}, D.~Y. and {Yang}, J. and {Li}, A. and {Yuan}, W. and {Ling}, Z.~X. and {Liu}, Y. and {Zhang}, C. and {Cheng}, H.~Q. and {Chen}, W. and {Cui}, C.~Z. and {Fan}, D.~W. and {Hu}, H.~B. and {Hu}, J.~W. and {Huang}, M.~H. and {Liu}, H.~Y. and {Liu}, M.~J. and {Lv}, Z.~Z. and {Lian}, T.~Y. and {Mao}, X. and {Pan}, H.~W. and {Sun}, H. and {Wang}, W.~X. and {Wang}, Y.~L. and {Wu}, Q.~Y. and {Xu}, X.~P. and {Xu}, Y.~F. and {Yang}, H.~N. and {Zhang}, M. and {Zhang}, W.~D. and {Zhang}, W.~J. and {Zhang}, Z. and {Chen}, Y. and {Jia}, S.~M. and {Zhang}, S.~N. and {Kuulkers}, E. and {Santovincenzo}, A. and {O'Brien}, P. and {Nandra}, K. and {Rau}, A. and {Cordier}, B. and {Einstein Probe Team}},
        title = "{EP240506a: EP-WXT detection of a new fast X-ray transient}",
      journal = {GRB Coordinates Network},
         year = 2024,
        month = may,
       volume = {36405},
        pages = {1},
       adsurl = {https://ui.adsabs.harvard.edu/abs/2024GCN.36405....1L}
}

@ARTICLE{2024Natur.626..737L,
       author = {{Levan}, Andrew J. and {Gompertz}, Benjamin P. and {Salafia}, Om Sharan and {Bulla}, Mattia and {Burns}, Eric and {Hotokezaka}, Kenta and {Izzo}, Luca and {Lamb}, Gavin P. and {Malesani}, Daniele B. and {Oates}, Samantha R. and {Ravasio}, Maria Edvige and {Rouco Escorial}, Alicia and {Schneider}, Benjamin and {Sarin}, Nikhil and {Schulze}, Steve and {Tanvir}, Nial R. and {Ackley}, Kendall and {Anderson}, Gemma and {Brammer}, Gabriel B. and {Christensen}, Lise and {Dhillon}, Vikram S. and {Evans}, Phil A. and {Fausnaugh}, Michael and {Fong}, Wen-fai and {Fruchter}, Andrew S. and {Fryer}, Chris and {Fynbo}, Johan P.~U. and {Gaspari}, Nicola and {Heintz}, Kasper E. and {Hjorth}, Jens and {Kennea}, Jamie A. and {Kennedy}, Mark R. and {Laskar}, Tanmoy and {Leloudas}, Giorgos and {Mandel}, Ilya and {Martin-Carrillo}, Antonio and {Metzger}, Brian D. and {Nicholl}, Matt and {Nugent}, Anya and {Palmerio}, Jesse T. and {Pugliese}, Giovanna and {Rastinejad}, Jillian and {Rhodes}, Lauren and {Rossi}, Andrea and {Saccardi}, Andrea and {Smartt}, Stephen J. and {Stevance}, Heloise F. and {Tohuvavohu}, Aaron and {van der Horst}, Alexander and {Vergani}, Susanna D. and {Watson}, Darach and {Barclay}, Thomas and {Bhirombhakdi}, Kornpob and {Breedt}, Elm{\'e} and {Breeveld}, Alice A. and {Brown}, Alexander J. and {Campana}, Sergio and {Chrimes}, Ashley A. and {D'Avanzo}, Paolo and {D'Elia}, Valerio and {De Pasquale}, Massimiliano and {Dyer}, Martin J. and {Galloway}, Duncan K. and {Garbutt}, James A. and {Green}, Matthew J. and {Hartmann}, Dieter H. and {Jakobsson}, P{\'a}ll and {Kerry}, Paul and {Kouveliotou}, Chryssa and {Langeroodi}, Danial and {Le Floc'h}, Emeric and {Leung}, James K. and {Littlefair}, Stuart P. and {Munday}, James and {O'Brien}, Paul and {Parsons}, Steven G. and {Pelisoli}, Ingrid and {Sahman}, David I. and {Salvaterra}, Ruben and {Sbarufatti}, Boris and {Steeghs}, Danny and {Tagliaferri}, Gianpiero and {Th{\"o}ne}, Christina C. and {de Ugarte Postigo}, Antonio and {Kann}, David Alexander},
        title = "{Heavy-element production in a compact object merger observed by JWST}",
      journal = {\nat},
         year = 2024,
        month = feb,
       volume = {626},
       number = {8000},
        pages = {737-741},
          doi = {10.1038/s41586-023-06759-1},
archivePrefix = {arXiv},
       eprint = {2307.02098},
 primaryClass = {astro-ph.HE},
       adsurl = {https://ui.adsabs.harvard.edu/abs/2024Natur.626..737L}
}

@ARTICLE{2025ApJ...988L..60S,
       author = {{Srinivasaragavan}, Gokul P. and {Hamidani}, Hamid and {Schroeder}, Genevieve and {Sarin}, Nikhil and {Ho}, Anna Y.~Q. and {Piro}, Anthony L. and {Cenko}, S. Bradley and {Anand}, Shreya and {Sollerman}, Jesper and {Perley}, Daniel A. and {Maeda}, Keiichi and {O'Connor}, Brendan and {Kuncarayakti}, Hanindyo and {Miller}, M. Coleman and {Ahumada}, Tom{\'a}s and {Vail}, Jada L. and {Duffell}, Paul and {Dastidar}, Ranadeep and {Andreoni}, Igor and {Bochenek}, Aleksandra and {Brennan}, Se{\'a}n. J. and {Carney}, Jonathan and {Chen}, Ping and {Freeburn}, James and {Gal-Yam}, Avishay and {Jacobson-Gal{\'a}n}, Wynn and {Kasliwal}, Mansi M. and {Li}, Jiaxuan and {Li}, Maggie L. and {Sravan}, Niharika and {Warshofsky}, Daniel E.},
        title = "{EP250108a/SN 2025kg: A Jet-driven Stellar Explosion Interacting with Circumstellar Material}",
      journal = {\apjl},
         year = 2025,
        month = aug,
       volume = {988},
       number = {2},
          eid = {L60},
        pages = {L60},
          doi = {10.3847/2041-8213/ade870},
archivePrefix = {arXiv},
       eprint = {2504.17516},
 primaryClass = {astro-ph.HE},
       adsurl = {https://ui.adsabs.harvard.edu/abs/2025ApJ...988L..60S}
}

@ARTICLE{2025ApJ...990L..28L,
       author = {{Levan}, Andrew J. and {Martin-Carrillo}, Antonio and {Laskar}, Tanmoy and {Eyles-Ferris}, Rob A.~J. and {Sneppen}, Albert and {Ravasio}, Maria Edvige and {Rastinejad}, Jillian C. and {Bright}, Joe S. and {Carotenuto}, Francesco and {Chrimes}, Ashley A. and {Corcoran}, Gregory and {Gompertz}, Benjamin P. and {Jonker}, Peter G. and {Lamb}, Gavin P. and {Malesani}, Daniele B. and {Saccardi}, Andrea and {S{\'a}nchez-Sierras}, Javier and {Schneider}, Benjamin and {Schulze}, Steve and {Tanvir}, Nial R. and {Vergani}, Susanna D. and {Watson}, Darach and {An}, Jie and {Bauer}, Franz E. and {Campana}, Sergio and {Cotter}, Laura and {van Dalen}, Joyce N.~D. and {D'Elia}, Valerio and {De Pasquale}, Massimiliano and {de Ugarte Postigo}, Antonio and {Dimple} and {Hartmann}, Dieter H. and {Hjorth}, Jens and {Izzo}, Luca and {Jakobsson}, P{\'a}ll and {Kumar}, Amit and {Melandri}, Andrea and {O'Brien}, Paul and {Piranomonte}, Silvia and {Pugliese}, Giovanna and {Quirola-V{\'a}squez}, Jonathan and {Starling}, Rhaana and {Tagliaferri}, Gianpiero and {Xu}, Dong and {Wortley}, Makenzie E.},
        title = "{The Day-long, Repeating GRB 250702B: A Unique Extragalactic Transient}",
      journal = {\apjl},
         year = 2025,
        month = sep,
       volume = {990},
       number = {1},
          eid = {L28},
        pages = {L28},
          doi = {10.3847/2041-8213/adf8e1},
archivePrefix = {arXiv},
       eprint = {2507.14286},
 primaryClass = {astro-ph.HE},
       adsurl = {https://ui.adsabs.harvard.edu/abs/2025ApJ...990L..28L}
}

@ARTICLE{2026SciBu..71..538L,
       author = {{Li}, Dongyue and {Zhang}, Wenda and {Yang}, Jun and {Chen}, Jin-Hong and {Yuan}, Weimin and {Cheng}, Huaqing and {Xu}, Fan and {Shu}, Xinwen and {Shen}, Rong-Feng and {Jiang}, Ning and {Zhu}, Jiazheng and {Zhou}, Chang and {Lei}, Weihua and {Sun}, Hui and {Jin}, Chichuan and {Dai}, Lixin and {Zhang}, Bing and {Yang}, Yu-Han and {Zhang}, Wenjie and {Feng}, Hua and {Liu}, Bifang and {Zhou}, Hongyan and {Pan}, Haiwu and {Liu}, Mingjun and {Corbel}, St{\'e}phane and {Jagan}, Sitha K. and {Baglio}, Maria Cristina and {Burns}, Christopher R. and {Cangemi}, Floriane and {Chen}, Chun and {Cheng}, Yehao and {Coleiro}, Alexis and {Coti Zelati}, Francesco and {Das}, Sourya R. and {Dong}, Zhongnan and {Galbany}, Luis and {Grollimund}, Noa and {Kelson}, Daniel and {Lai}, Dong and {Li}, Xia and {Liu}, Yuan and {Marino}, Alessio and {Mockler}, Brenna and {O'Brien}, Paul and {Qiao}, Erlin and {Rea}, Nanda and {Resmi}, L. and {Rodriguez}, J{\'e}rome and {Saxton}, Richard and {Sun}, Luming and {Tao}, Lian and {Wang}, Tinggui and {Wang}, Yilong and {Wu}, Xuefeng and {Xu}, Dong and {Zhang}, Yijia and {Zhao}, Guoying and {Bao}, Congying and {Cai}, Zhiming and {Chen}, Yehai and {Chen}, Yong and {Cordier}, Bertrand and {Cui}, Chenzhou and {Cui}, Weiwei and {Fan}, Zhou and {Gao}, He and {Ghirlanda}, Giancarlo and {Guan}, Ju and {Han}, Dawei and {Hao}, Jinxin and {Hu}, Jingwei and {Huang}, Maohai and {Huang}, Yong-Feng and {Jia}, Shumei and {Jin}, Ge and {Komossa}, Stefanie and {Li}, Chengkui and {Ling}, Zhixing and {Liu}, Congzhan and {Liu}, Heyang and {Liu}, Huaqiu and {Lu}, Fangjun and {Nandra}, Kirpal and {Ness}, Jan-Uwe and {Rau}, Arne and {Sanders}, Jeremy and {Song}, Liming and {Soria}, Roberto and {Sun}, Shengli and {Sun}, Xiaojin and {Tan}, Yuyin and {Troja}, Eleonora and {Wen}, Sixiang and {Xu}, Haitao and {Xue}, Changbin and {Xue}, Yongquan and {Yin}, Yi-Han Iris and {Zhang}, Chen and {Zhang}, Shuang-Nan and {Zhang}, Yonghe},
        title = "{A fast powerful X-ray transient from possible tidal disruption of a white dwarf}",
      journal = {Science Bulletin},
         year = 2026,
        month = feb,
       volume = {71},
       number = {3},
        pages = {538-546},
          doi = {10.1016/j.scib.2025.12.050},
archivePrefix = {arXiv},
       eprint = {2509.25877},
 primaryClass = {astro-ph.HE},
       adsurl = {https://ui.adsabs.harvard.edu/abs/2026SciBu..71..538L}
}

@ARTICLE{2026arXiv260521578R,
       author = {{Ronchini}, S. and {Chopra}, A. and {Dal Canton}, T. and {Banerjee}, B. and {De Santis}, A.~L. and {Branchesi}, M.},
        title = "{Gravitational wave detectability range informed by external messengers}",
      journal = {arXiv e-prints},
         year = 2026,
        month = may,
          eid = {arXiv:2605.21578},
        pages = {arXiv:2605.21578},
archivePrefix = {arXiv},
       eprint = {2605.21578},
 primaryClass = {astro-ph.HE},
       adsurl = {https://ui.adsabs.harvard.edu/abs/2026arXiv260521578R}
}

@ARTICLE{2021PhRvD.104l2004A,
       author = {{Abbott}, R. and {Abbott}, T.~D. and {Acernese}, F. and {Ackley}, K. and {Adams}, C. and {Adhikari}, N. and {Adhikari}, R.~X. and {Adya}, V.~B. and {Affeldt}, C. and {Agarwal}, D. and {Agathos}, M. and {Agatsuma}, K. and {Aggarwal}, N. and {Aguiar}, O.~D. and {Aiello}, L. and {Ain}, A. and {Ajith}, P. and {Akutsu}, T. and {Albanesi}, S. and {Allocca}, A. and {Altin}, P.~A. and {Amato}, A. and {Anand}, C. and {Anand}, S. and {Ananyeva}, A. and {Anderson}, S.~B. and {Anderson}, W.~G. and {Ando}, M. and {Andrade}, T. and {Andres}, N. and {Andri{\'c}}, T. and {Angelova}, S.~V. and {Ansoldi}, S. and {Antelis}, J.~M. and {Antier}, S. and {Appert}, S. and {Arai}, Koji and {Arai}, Koya and {Arai}, Y. and {Araki}, S. and {Araya}, A. and {Araya}, M.~C. and {Areeda}, J.~S. and {Ar{\`e}ne}, M. and {Aritomi}, N. and {Arnaud}, N. and {Aronson}, S.~M. and {Arun}, K.~G. and {Asada}, H. and {Asali}, Y. and {Ashton}, G. and {Aso}, Y. and {Assiduo}, M. and {Aston}, S.~M. and {Astone}, P. and {Aubin}, F. and {Austin}, C. and {Babak}, S. and {Badaracco}, F. and {Bader}, M.~K.~M. and {Badger}, C. and {Bae}, S. and {Bae}, Y. and {Baer}, A.~M. and {Bagnasco}, S. and {Bai}, Y. and {Baiotti}, L. and {Baird}, J. and {Bajpai}, R. and {Ball}, M. and {Ballardin}, G. and {Ballmer}, S.~W. and {Balsamo}, A. and {Baltus}, G. and {Banagiri}, S. and {Bankar}, D. and {Barayoga}, J.~C. and {Barbieri}, C. and {Barish}, B.~C. and {Barker}, D. and {Barneo}, P. and {Barone}, F. and {Barr}, B. and {Barsotti}, L. and {Barsuglia}, M. and {Barta}, D. and {Bartlett}, J. and {Barton}, M.~A. and {Bartos}, I. and {Bassiri}, R. and {Basti}, A. and {Bawaj}, M. and {Bayley}, J.~C. and {Baylor}, A.~C. and {Bazzan}, M. and {B{\'e}csy}, B. and {Bedakihale}, V.~M. and {Bejger}, M. and {Belahcene}, I. and {Benedetto}, V. and {Beniwal}, D. and {Bennett}, T.~F. and {Bentley}, J.~D. and {Benyaala}, M. and {Bergamin}, F. and {Berger}, B.~K. and {Bernuzzi}, S. and {Berry}, C.~P.~L. and {Bersanetti}, D. and {Bertolini}, A. and {Betzwieser}, J. and {Beveridge}, D. and {Bhandare}, R. and {Bhardwaj}, U. and {Bhattacharjee}, D. and {Bhaumik}, S. and {Bilenko}, I.~A. and {Billingsley}, G. and {Bini}, S. and {Birney}, R. and {Birnholtz}, O. and {Biscans}, S. and {Bischi}, M. and {Biscoveanu}, S. and {Bisht}, A. and {Biswas}, B. and {Bitossi}, M. and {Bizouard}, M.-A. and {Blackburn}, J.~K. and {Blair}, C.~D. and {Blair}, D.~G. and {Blair}, R.~M. and {Bobba}, F. and {Bode}, N. and {Boer}, M. and {Bogaert}, G. and {Boldrini}, M. and {Bonavena}, L.~D. and {Bondu}, F. and {Bonilla}, E. and {Bonnand}, R. and {Booker}, P. and {Boom}, B.~A. and {Bork}, R. and {Boschi}, V. and {Bose}, N. and {Bose}, S. and {Bossilkov}, V. and {Boudart}, V. and {Bouffanais}, Y. and {Bozzi}, A. and {Bradaschia}, C. and {Brady}, P.~R. and {Bramley}, A. and {Branch}, A. and {Branchesi}, M. and {Brau}, J.~E. and {Breschi}, M. and {Briant}, T. and {Briggs}, J.~H. and {Brillet}, A. and {Brinkmann}, M. and {Brockill}, P. and {Brooks}, A.~F. and {Brooks}, J. and {Brown}, D.~D. and {Brunett}, S. and {Bruno}, G. and {Bruntz}, R. and {Bryant}, J. and {Bulik}, T. and {Bulten}, H.~J. and {Buonanno}, A. and {Buscicchio}, R. and {Buskulic}, D. and {Buy}, C. and {Byer}, R.~L. and {Cadonati}, L. and {Cagnoli}, G. and {Cahillane}, C. and {Bustillo}, J. Calder{\'o}n and {Callaghan}, J.~D. and {Callister}, T.~A. and {Calloni}, E. and {Cameron}, J. and {Camp}, J.~B. and {Canepa}, M. and {Canevarolo}, S. and {Cannavacciuolo}, M. and {Cannon}, K.~C. and {Cao}, H. and {Cao}, Z. and {Capocasa}, E. and {Capote}, E. and {Carapella}, G. and {Carbognani}, F. and {Carlin}, J.~B. and {Carney}, M.~F. and {Carpinelli}, M. and {Carrillo}, G.},
        title = "{All-sky search for short gravitational-wave bursts in the third Advanced LIGO and Advanced Virgo run}",
      journal = {\prd},
         year = 2021,
        month = dec,
       volume = {104},
       number = {12},
          eid = {122004},
        pages = {122004},
          doi = {10.1103/PhysRevD.104.122004},
archivePrefix = {arXiv},
       eprint = {2107.03701},
 primaryClass = {gr-qc},
       adsurl = {https://ui.adsabs.harvard.edu/abs/2021PhRvD.104l2004A}
}

@ARTICLE{2025PhRvD.112j2005A,
       author = {{Abac}, A.~G. and {Abouelfettouh}, I. and {Acernese}, F. and {Ackley}, K. and {Adamcewicz}, C. and {Adhicary}, S. and {Adhikari}, D. and {Adhikari}, N. and {Adhikari}, R.~X. and {Adkins}, V.~K. and {Afroz}, S. and {Agapito}, A. and {Agarwal}, D. and {Agathos}, M. and {Aggarwal}, N. and {Aggarwal}, S. and {Aguiar}, O.~D. and {Ahrend}, I.-L. and {Aiello}, L. and {Ain}, A. and {Ajith}, P. and {Akutsu}, T. and {Albanesi}, S. and {Ali}, W. and {Al-Kershi}, S. and {All{\'e}n{\'e}}, C. and {Allocca}, A. and {Al-Shammari}, S. and {Altin}, P.~A. and {Alvarez-Lopez}, S. and {Amar}, W. and {Amarasinghe}, O. and {Amato}, A. and {Amicucci}, F. and {Amra}, C. and {Ananyeva}, A. and {Anderson}, S.~B. and {Anderson}, W.~G. and {Andia}, M. and {Ando}, M. and {Andr{\'e}s-Carcasona}, M. and {Andri{\'c}}, T. and {Anglin}, J. and {Ansoldi}, S. and {Antelis}, J.~M. and {Antier}, S. and {Aoumi}, M. and {Appavuravther}, E.~Z. and {Appert}, S. and {Apple}, S.~K. and {Arai}, K. and {Araya}, A. and {Araya}, M.~C. and {Arca Sedda}, M. and {Areeda}, J.~S. and {Aritomi}, N. and {Armato}, F. and {Armstrong}, S. and {Arnaud}, N. and {Arogeti}, M. and {Aronson}, S.~M. and {Ashton}, G. and {Aso}, Y. and {Asprea}, L. and {Assiduo}, M. and {Melo}, S. Assis De Souza and {Aston}, S.~M. and {Astone}, P. and {Attadio}, F. and {Aubin}, F. and {Aultoneal}, K. and {Avallone}, G. and {Avila}, E.~A. and {Babak}, S. and {Badger}, C. and {Bae}, S. and {Bagnasco}, S. and {Baiotti}, L. and {Bajpai}, R. and {Baka}, T. and {Baker}, A.~M. and {Baker}, K.~A. and {Baldi}, G. and {Baldicchi}, N. and {Ball}, M. and {Ballardin}, G. and {Ballmer}, S.~W. and {Banagiri}, S. and {Banerjee}, B. and {Bankar}, D. and {Baptiste}, T.~M. and {Baral}, P. and {Baratti}, M. and {Barayoga}, J.~C. and {Barish}, B.~C. and {Barker}, D. and {Barman}, N. and {Barneo}, P. and {Barone}, F. and {Barr}, B. and {Barsotti}, L. and {Barsuglia}, M. and {Barta}, D. and {Bartoletti}, A.~M. and {Barton}, M.~A. and {Bartos}, I. and {Basalaev}, A. and {Bassiri}, R. and {Basti}, A. and {Bawaj}, M. and {Baxi}, P. and {Bayley}, J.~C. and {Baylor}, A.~C. and {Baynard}, II, P.~A. and {Bazzan}, M. and {Bedakihale}, V.~M. and {Beirnaert}, F. and {Bejger}, M. and {Belardinelli}, D. and {Bell}, A.~S. and {Bellie}, D.~S. and {Bellizzi}, L. and {Benoit}, W. and {Bentara}, I. and {Bentley}, J.~D. and {Yaala}, M. Ben and {Bera}, S. and {Bergamin}, F. and {Berger}, B.~K. and {Bernuzzi}, S. and {Beroiz}, M. and {Bersanetti}, D. and {Bertheas}, T. and {Bertolini}, A. and {Betzwieser}, J. and {Beveridge}, D. and {Bevilacqua}, G. and {Bevins}, N. and {Bhandare}, R. and {Bhatt}, R. and {Bhattacharjee}, D. and {Bhattacharyya}, S. and {Bhaumik}, S. and {Biancalana}, V. and {Bianchi}, A. and {Bilenko}, I.~A. and {Billingsley}, G. and {Binetti}, A. and {Bini}, S. and {Binu}, C. and {Biot}, S. and {Birnholtz}, O. and {Biscoveanu}, S. and {Bisht}, A. and {Bitossi}, M. and {Bizouard}, M.-A. and {Blaber}, S. and {Blackburn}, J.~K. and {Blagg}, L.~A. and {Blair}, C.~D. and {Blair}, D.~G. and {Bode}, N. and {Boettner}, N. and {Boileau}, G. and {Boldrini}, M. and {Bolingbroke}, G.~N. and {Bolliand}, A. and {Bonavena}, L.~D. and {Bondarescu}, R. and {Bondu}, F. and {Bonilla}, E. and {Bonilla}, M.~S. and {Bonino}, A. and {Bonnand}, R. and {Borchers}, A. and {Borhanian}, S. and {Boschi}, V. and {Bose}, S. and {Bossilkov}, V. and {Bothra}, Y. and {Boudon}, A. and {Bourg}, L. and {Boyle}, M. and {Bozzi}, A. and {Bradaschia}, C. and {Brady}, P.~R. and {Branch}, A. and {Branchesi}, M. and {Braun}, I. and {Briant}, T. and {Brillet}, A. and {Brinkmann}, M. and {Brockill}, P. and {Brockmueller}, E. and {Brooks}, A.~F. and {Brown}, B.~C. and {Brown}, D.~D. and {Brozzetti}, M.~L. and {Brunett}, S. and {Bruno}, G.},
        title = "{All-sky search for short gravitational-wave bursts in the first part of the fourth LIGO-Virgo-KAGRA observing run}",
      journal = {\prd},
         year = 2025,
        month = nov,
       volume = {112},
       number = {10},
          eid = {102005},
        pages = {102005},
          doi = {10.1103/wjdz-jdby},
archivePrefix = {arXiv},
       eprint = {2507.12374},
 primaryClass = {astro-ph.HE},
       adsurl = {https://ui.adsabs.harvard.edu/abs/2025PhRvD.112j2005A}
}

@ARTICLE{2017PhRvD..95d2003A,
       author = {{Abbott}, B.~P. and {Abbott}, R. and {Abbott}, T.~D. and {Abernathy}, M.~R. and {Acernese}, F. and {Ackley}, K. and {Adams}, C. and {Adams}, T. and {Addesso}, P. and {Adhikari}, R.~X. and {Adya}, V.~B. and {Affeldt}, C. and {Agathos}, M. and {Agatsuma}, K. and {Aggarwal}, N. and {Aguiar}, O.~D. and {Aiello}, L. and {Ain}, A. and {Allen}, B. and {Allocca}, A. and {Altin}, P.~A. and {Ananyeva}, A. and {Anderson}, S.~B. and {Anderson}, W.~G. and {Appert}, S. and {Arai}, K. and {Araya}, M.~C. and {Areeda}, J.~S. and {Arnaud}, N. and {Arun}, K.~G. and {Ascenzi}, S. and {Ashton}, G. and {Ast}, M. and {Aston}, S.~M. and {Astone}, P. and {Aufmuth}, P. and {Aulbert}, C. and {Avila-Alvarez}, A. and {Babak}, S. and {Bacon}, P. and {Bader}, M.~K.~M. and {Baker}, P.~T. and {Baldaccini}, F. and {Ballardin}, G. and {Ballmer}, S.~W. and {Barayoga}, J.~C. and {Barclay}, S.~E. and {Barish}, B.~C. and {Barker}, D. and {Barone}, F. and {Barr}, B. and {Barsotti}, L. and {Barsuglia}, M. and {Barta}, D. and {Bartlett}, J. and {Bartos}, I. and {Bassiri}, R. and {Basti}, A. and {Batch}, J.~C. and {Baune}, C. and {Bavigadda}, V. and {Bazzan}, M. and {Beer}, C. and {Bejger}, M. and {Belahcene}, I. and {Belgin}, M. and {Bell}, A.~S. and {Berger}, B.~K. and {Bergmann}, G. and {Berry}, C.~P.~L. and {Bersanetti}, D. and {Bertolini}, A. and {Betzwieser}, J. and {Bhagwat}, S. and {Bhandare}, R. and {Bilenko}, I.~A. and {Billingsley}, G. and {Billman}, C.~R. and {Birch}, J. and {Birney}, I.~A. and {Birnholtz}, O. and {Biscans}, S. and {Bisht}, A. and {Bitossi}, M. and {Biwer}, C. and {Bizouard}, M.~A. and {Blackburn}, J.~K. and {Blackman}, J. and {Blair}, C.~D. and {Blair}, D.~G. and {Blair}, R.~M. and {Bloemen}, S. and {Bock}, O. and {Boer}, M. and {Bogaert}, G. and {Bohe}, A. and {Bondu}, F. and {Bonnand}, R. and {Boom}, B.~A. and {Bork}, R. and {Boschi}, V. and {Bose}, S. and {Bouffanais}, Y. and {Bozzi}, A. and {Bradaschia}, C. and {Brady}, P.~R. and {Braginsky}, V.~B. and {Branchesi}, M. and {Brau}, J.~E. and {Briant}, T. and {Brillet}, A. and {Brinkmann}, M. and {Brisson}, V. and {Brockill}, P. and {Broida}, J.~E. and {Brooks}, A.~F. and {Brown}, D.~A. and {Brown}, D.~D. and {Brown}, N.~M. and {Brunett}, S. and {Buchanan}, C.~C. and {Buikema}, A. and {Bulik}, T. and {Bulten}, H.~J. and {Buonanno}, A. and {Buskulic}, D. and {Buy}, C. and {Byer}, R.~L. and {Cabero}, M. and {Cadonati}, L. and {Cagnoli}, G. and {Cahillane}, C. and {Calder{\'o}n Bustillo}, J. and {Callister}, T.~A. and {Calloni}, E. and {Camp}, J.~B. and {Canepa}, M. and {Cannon}, K.~C. and {Cao}, H. and {Cao}, J. and {Capano}, C.~D. and {Capocasa}, E. and {Carbognani}, F. and {Caride}, S. and {Casanueva Diaz}, J. and {Casentini}, C. and {Caudill}, S. and {Cavagli{\`a}}, M. and {Cavalier}, F. and {Cavalieri}, R. and {Cella}, G. and {Cepeda}, C.~B. and {Cerboni Baiardi}, L. and {Cerretani}, G. and {Cesarini}, E. and {Chamberlin}, S.~J. and {Chan}, M. and {Chao}, S. and {Charlton}, P. and {Chassande-Mottin}, E. and {Cheeseboro}, B.~D. and {Chen}, H.~Y. and {Chen}, Y. and {Cheng}, H.-P. and {Chincarini}, A. and {Chiummo}, A. and {Chmiel}, T. and {Cho}, H.~S. and {Cho}, M. and {Chow}, J.~H. and {Christensen}, N. and {Chu}, Q. and {Chua}, A.~J.~K. and {Chua}, S. and {Chung}, S. and {Ciani}, G. and {Clara}, F. and {Clark}, J.~A. and {Cleva}, F. and {Cocchieri}, C. and {Coccia}, E. and {Cohadon}, P.-F. and {Colla}, A. and {Collette}, C.~G. and {Cominsky}, L. and {Constancio}, M. and {Conti}, L. and {Cooper}, S.~J. and {Corbitt}, T.~R. and {Cornish}, N. and {Corsi}, A. and {Cortese}, S. and {Costa}, C.~A. and {Coughlin}, M.~W. and {Coughlin}, S.~B. and {Coulon}, J.-P. and {Countryman}, S.~T. and {Couvares}, P. and {Covas}, P.~B. and {Cowan}, E.~E.},
        title = "{All-sky search for short gravitational-wave bursts in the first Advanced LIGO run}",
      journal = {\prd},
         year = 2017,
        month = feb,
       volume = {95},
       number = {4},
          eid = {042003},
        pages = {042003},
          doi = {10.1103/PhysRevD.95.042003},
archivePrefix = {arXiv},
       eprint = {1611.02972},
 primaryClass = {gr-qc},
       adsurl = {https://ui.adsabs.harvard.edu/abs/2017PhRvD..95d2003A}
}

@ARTICLE{2019PhRvD.100b4017A,
       author = {{Abbott}, B.~P. and {Abbott}, R. and {Abbott}, T.~D. and {Abraham}, S. and {Acernese}, F. and {Ackley}, K. and {Adams}, C. and {Adhikari}, R.~X. and {Adya}, V.~B. and {Affeldt}, C. and {Agathos}, M. and {Agatsuma}, K. and {Aggarwal}, N. and {Aguiar}, O.~D. and {Aiello}, L. and {Ain}, A. and {Ajith}, P. and {Allen}, G. and {Allocca}, A. and {Aloy}, M.~A. and {Altin}, P.~A. and {Amato}, A. and {Anand}, S. and {Ananyeva}, A. and {Anderson}, S.~B. and {Anderson}, W.~G. and {Angelova}, S.~V. and {Antier}, S. and {Appert}, S. and {Arai}, K. and {Araya}, M.~C. and {Areeda}, J.~S. and {Ar{\`e}ne}, M. and {Arnaud}, N. and {Aronson}, S.~M. and {Ascenzi}, S. and {Ashton}, G. and {Aston}, S.~M. and {Astone}, P. and {Aubin}, F. and {Aufmuth}, P. and {AultONeal}, K. and {Austin}, C. and {Avendano}, V. and {Avila-Alvarez}, A. and {Babak}, S. and {Bacon}, P. and {Badaracco}, F. and {Bader}, M.~K.~M. and {Bae}, S. and {Baird}, J. and {Baker}, P.~T. and {Baldaccini}, F. and {Ballardin}, G. and {Ballmer}, S.~W. and {Bals}, A. and {Banagiri}, S. and {Barayoga}, J.~C. and {Barbieri}, C. and {Barclay}, S.~E. and {Barish}, B.~C. and {Barker}, D. and {Barkett}, K. and {Barnum}, S. and {Barone}, F. and {Barr}, B. and {Barsotti}, L. and {Barsuglia}, M. and {Barta}, D. and {Bartlett}, J. and {Bartos}, I. and {Bassiri}, R. and {Basti}, A. and {Bawaj}, M. and {Bayley}, J.~C. and {Bazzan}, M. and {B{\'e}csy}, B. and {Bejger}, M. and {Belahcene}, I. and {Bell}, A.~S. and {Beniwal}, D. and {Benjamin}, M.~G. and {Berger}, B.~K. and {Bergmann}, G. and {Bernuzzi}, S. and {Berry}, C.~P.~L. and {Bersanetti}, D. and {Bertolini}, A. and {Betzwieser}, J. and {Bhandare}, R. and {Bidler}, J. and {Biggs}, E. and {Bilenko}, I.~A. and {Bilgili}, S.~A. and {Billingsley}, G. and {Birney}, R. and {Birnholtz}, O. and {Biscans}, S. and {Bischi}, M. and {Biscoveanu}, S. and {Bisht}, A. and {Bitossi}, M. and {Bizouard}, M.~A. and {Blackburn}, J.~K. and {Blackman}, J. and {Blair}, C.~D. and {Blair}, D.~G. and {Blair}, R.~M. and {Bloemen}, S. and {Bobba}, F. and {Bode}, N. and {Boer}, M. and {Boetzel}, Y. and {Bogaert}, G. and {Bondu}, F. and {Bonnand}, R. and {Booker}, P. and {Boom}, B.~A. and {Bork}, R. and {Boschi}, V. and {Bose}, S. and {Bossilkov}, V. and {Bosveld}, J. and {Bouffanais}, Y. and {Bozzi}, A. and {Bradaschia}, C. and {Brady}, P.~R. and {Bramley}, A. and {Branchesi}, M. and {Brau}, J.~E. and {Breschi}, M. and {Briant}, T. and {Briggs}, J.~H. and {Brighenti}, F. and {Brillet}, A. and {Brinkmann}, M. and {Brockill}, P. and {Brooks}, A.~F. and {Brooks}, J. and {Brown}, D.~D. and {Brunett}, S. and {Buikema}, A. and {Bulik}, T. and {Bulten}, H.~J. and {Buonanno}, A. and {Buskulic}, D. and {Buy}, C. and {Byer}, R.~L. and {Cabero}, M. and {Cadonati}, L. and {Cagnoli}, G. and {Cahillane}, C. and {Calder{\'o}n Bustillo}, J. and {Callister}, T.~A. and {Calloni}, E. and {Camp}, J.~B. and {Campbell}, W.~A. and {Canepa}, M. and {Cannon}, K.~C. and {Cao}, H. and {Cao}, J. and {Carapella}, G. and {Carbognani}, F. and {Caride}, S. and {Carney}, M.~F. and {Carullo}, G. and {Casanueva Diaz}, J. and {Casentini}, C. and {Caudill}, S. and {Cavagli{\`a}}, M. and {Cavalier}, F. and {Cavalieri}, R. and {Cella}, G. and {Cerd{\'a}-Dur{\'a}n}, P. and {Cesarini}, E. and {Chaibi}, O. and {Chakravarti}, K. and {Chamberlin}, S.~J. and {Chan}, M. and {Chao}, S. and {Charlton}, P. and {Chase}, E.~A. and {Chassande-Mottin}, E. and {Chatterjee}, D. and {Chaturvedi}, M. and {Chatziioannou}, K. and {Cheeseboro}, B.~D. and {Chen}, H.~Y. and {Chen}, X. and {Chen}, Y. and {Cheng}, H.-P. and {Cheong}, C.~K. and {Chia}, H.~Y. and {Chiadini}, F. and {Chincarini}, A. and {Chiummo}, A. and {Cho}, G. and {Cho}, H.~S. and {Cho}, M. and {Christensen}, N.},
        title = "{All-sky search for short gravitational-wave bursts in the second Advanced LIGO and Advanced Virgo run}",
      journal = {\prd},
         year = 2019,
        month = jul,
       volume = {100},
       number = {2},
          eid = {024017},
        pages = {024017},
          doi = {10.1103/PhysRevD.100.024017},
archivePrefix = {arXiv},
       eprint = {1904.08976},
 primaryClass = {astro-ph.CO},
       adsurl = {https://ui.adsabs.harvard.edu/abs/2019PhRvD.100b4017A}
}

@ARTICLE{2013PhRvD..87f4033D,
       author = {{Dietz}, Alexander and {Fotopoulos}, Nickolas and {Singer}, Leo and {Cutler}, Curt},
        title = "{Outlook for detection of GW inspirals by GRB-triggered searches in the advanced detector era}",
      journal = {\prd},
         year = 2013,
        month = mar,
       volume = {87},
       number = {6},
          eid = {064033},
        pages = {064033},
          doi = {10.1103/PhysRevD.87.064033},
archivePrefix = {arXiv},
       eprint = {1210.3095},
 primaryClass = {gr-qc},
       adsurl = {https://ui.adsabs.harvard.edu/abs/2013PhRvD..87f4033D}
}

@ARTICLE{2022CQGra..39h5010P,
       author = {{Piotrzkowski}, Brandon and {Baylor}, Amanda and {Hernandez}, Ignacio Maga{\~n}a},
        title = "{A joint ranking statistic for multi-messenger astronomical searches with gravitational waves}",
      journal = {Classical and Quantum Gravity},
         year = 2022,
        month = apr,
       volume = {39},
       number = {8},
          eid = {085010},
        pages = {085010},
          doi = {10.1088/1361-6382/ac5c00},
archivePrefix = {arXiv},
       eprint = {2111.12814},
 primaryClass = {astro-ph.IM},
       adsurl = {https://ui.adsabs.harvard.edu/abs/2022CQGra..39h5010P}
}

@ARTICLE{2026A&A...708A.190I,
       author = {{Ierardi}, Annarita and {Oganesyan}, Gor and {Ascenzi}, Stefano and {Branchesi}, Marica and {Banerjee}, Biswajit and {Ronchini}, Samuele},
        title = "{Early X-ray emission of short gamma-ray bursts: Insights into physics and multi-messenger prospects}",
      journal = {\aap},
         year = 2026,
        month = apr,
       volume = {708},
          eid = {A190},
        pages = {A190},
          doi = {10.1051/0004-6361/202557744},
archivePrefix = {arXiv},
       eprint = {2510.16108},
 primaryClass = {astro-ph.HE},
       adsurl = {https://ui.adsabs.harvard.edu/abs/2026A&A...708A.190I}
}

@ARTICLE{2025ApJ...988L..13R,
       author = {{Rastinejad}, Jillian C. and {Levan}, Andrew J. and {Jonker}, Peter G. and {Kilpatrick}, Charles D. and {Fryer}, Christopher L. and {Sarin}, Nikhil and {Gompertz}, Benjamin P. and {Liu}, Chang and {Eyles-Ferris}, Rob A.~J. and {Fong}, Wen-fai and {Burns}, Eric and {Gillanders}, James H. and {Mandel}, Ilya and {Malesani}, Daniele Bj{\o}rn and {O'Brien}, Paul T. and {Tanvir}, Nial R. and {Ackley}, Kendall and {Aryan}, Amar and {Bauer}, Franz E. and {Bloemen}, Steven and {de Boer}, Thomas and {Bom}, Cl{\'e}cio R. and {Chac{\'o}n}, Jennifer A. and {Chambers}, Ken and {Chen}, Ting-Wan and {Chrimes}, Ashley A. and {van Dalen}, Joyce N.~D. and {D'Elia}, Valerio and {De Pasquale}, Massimiliano and {Fulton}, Michael D. and {Groot}, Paul J. and {Gupta}, Rahul and {Hartmann}, Dieter H. and {van Hoof}, Agnes P.~C. and {Huber}, Mark E. and {Izzo}, Luca and {Jacobson-Galan}, Wynn and {Jakobsson}, P{\'a}ll and {Kong}, Albert and {Laskar}, Tanmoy and {Lowe}, Thomas B. and {Magnier}, Eugene A. and {Maiorano}, Elisabetta and {Martin-Carrillo}, Antonio and {Mas-Ribas}, Lluis and {Mata S{\'a}nchez}, Daniel and {Nicholl}, Matt and {Nixon}, Christopher J. and {Oates}, Samantha R. and {Paek}, Gregory and {Palmerio}, Jesse and {Paris}, Diego and {Pieterse}, Dani{\"e}lle L.~A. and {Pugliese}, Giovanna and {Quirola Vasquez}, Jonathan A. and {van Roestel}, Jan and {Rossi}, Andrea and {Rouco Escorial}, Alicia and {Salvaterra}, Ruben and {Schneider}, Benjamin and {Smartt}, Stephen J. and {Smith}, Ken and {Smith}, Ian A. and {Srivastav}, Shubham and {Torres}, Manuel A.~P. and {Ventura}, Chiara and {Vreeswijk}, Paul and {Wainscoat}, Richard and {Yang}, Yi-Jung and {Yang}, Sheng},
        title = "{EP 250108a/SN 2025kg: Observations of the Most Nearby Broad-line Type Ic Supernova Following an Einstein Probe Fast X-Ray Transient}",
      journal = {\apjl},
         year = 2025,
        month = jul,
       volume = {988},
       number = {1},
          eid = {L13},
        pages = {L13},
          doi = {10.3847/2041-8213/ade7f9},
archivePrefix = {arXiv},
       eprint = {2504.08889},
 primaryClass = {astro-ph.HE},
       adsurl = {https://ui.adsabs.harvard.edu/abs/2025ApJ...988L..13R}
}

@ARTICLE{2025ApJ...988L..14E,
       author = {{Eyles-Ferris}, Rob A.~J. and {Jonker}, Peter G. and {Levan}, Andrew J. and {Malesani}, Daniele Bj{\o}rn and {Sarin}, Nikhil and {Fryer}, Christopher L. and {Rastinejad}, Jillian C. and {Burns}, Eric and {Tanvir}, Nial R. and {O'Brien}, Paul T. and {Fong}, Wen-fai and {Mandel}, Ilya and {Gompertz}, Benjamin P. and {Kilpatrick}, Charles D. and {Bloemen}, Steven and {Bright}, Joe S. and {Carotenuto}, Francesco and {Corcoran}, Gregory and {Cotter}, Laura and {Groot}, Paul J. and {Izzo}, Luca and {Laskar}, Tanmoy and {Martin-Carrillo}, Antonio and {Palmerio}, Jesse and {Ravasio}, Maria E. and {van Roestel}, Jan and {Saccardi}, Andrea and {Starling}, Rhaana L.~C. and {Thakur}, Aishwarya Linesh and {Vergani}, Susanna D. and {Vreeswijk}, Paul M. and {Bauer}, Franz E. and {Campana}, Sergio and {Chac{\'o}n}, Jennifer A. and {Chrimes}, Ashley A. and {Covino}, Stefano and {van Dalen}, Joyce N.~D. and {D'Elia}, Valerio and {De Pasquale}, Massimiliano and {Habeeb}, Nusrin and {Hartmann}, Dieter H. and {van Hoof}, Agnes P.~C. and {Jakobsson}, P{\'a}ll and {Julakanti}, Yashaswi and {Leloudas}, Giorgos and {Mata S{\'a}nchez}, Daniel and {Nixon}, Christopher J. and {Pieterse}, Dani{\"e}lle L.~A. and {Pugliese}, Giovanna and {Quirola-V{\'a}squez}, Jonathan and {Rayson}, Ben C. and {Salvaterra}, Ruben and {Schneider}, Ben and {Torres}, Manuel A.~P. and {Zafar}, Tayyaba},
        title = "{The Kangaroo's First Hop: The Early Fast Cooling Phase of EP250108a/SN 2025kg}",
      journal = {\apjl},
         year = 2025,
        month = jul,
       volume = {988},
       number = {1},
          eid = {L14},
        pages = {L14},
          doi = {10.3847/2041-8213/ade1d9},
archivePrefix = {arXiv},
       eprint = {2504.08886},
 primaryClass = {astro-ph.HE},
       adsurl = {https://ui.adsabs.harvard.edu/abs/2025ApJ...988L..14E}
}

@ARTICLE{2025MNRAS.544L.139Z,
       author = {{Zhu}, Jin-Ping and {Zheng}, Jian-He and {Zhang}, Bing},
        title = "{EP 250108a/SN 2025kg: a magnetar-powered gamma-ray burst supernova originating from a close helium-star binary via isolated binary evolution}",
      journal = {\mnras},
         year = 2025,
        month = nov,
       volume = {544},
       number = {1},
        pages = {L139-L149},
          doi = {10.1093/mnrasl/slaf114},
archivePrefix = {arXiv},
       eprint = {2507.18544},
 primaryClass = {astro-ph.HE},
       adsurl = {https://ui.adsabs.harvard.edu/abs/2025MNRAS.544L.139Z}
}

@ARTICLE{2025A&A...701A.225B,
       author = {{Busmann}, Malte and {O'Connor}, Brendan and {Sommer}, Julian and {Gruen}, Daniel and {Beniamini}, Paz and {Gill}, Ramandeep and {Moss}, Michael J. and {Palmese}, Antonella and {Riffeser}, Arno and {Yang}, Yu-Han and {Troja}, Eleonora and {Dichiara}, Simone and {Ricci}, Roberto and {Klingler}, Noel and {G{\"o}ssl}, Claus and {Hu}, Lei and {Rau}, Arne and {Ries}, Christoph and {Ryan}, Geoffrey and {Schmidt}, Michael and {Yadav}, Muskan and {Zeimann}, Gregory R.},
        title = "{The curious case of EP241021a: Unraveling the mystery of its exceptional rebrightening}",
      journal = {\aap},
         year = 2025,
        month = sep,
       volume = {701},
          eid = {A225},
        pages = {A225},
          doi = {10.1051/0004-6361/202554626},
archivePrefix = {arXiv},
       eprint = {2503.14588},
 primaryClass = {astro-ph.HE},
       adsurl = {https://ui.adsabs.harvard.edu/abs/2025A&A...701A.225B}
}

@ARTICLE{2026MNRAS.545f2062E,
       author = {{Eappachen}, D. and {Balasubramanian}, A. and {Swain}, Vishwajeet and {Anupama}, G.~C. and {Sahu}, D.~K. and {Bhalerao}, V. and {Ahumada}, T. and {Andreoni}, I. and {Barway}, Sudhanshu and {Carney}, J. and {Freeburn}, J. and {Kasliwal}, M.~M. and {Mohan}, Tanishk and {Rodriguez}, A.~C. and {Waratkar}, G.},
        title = "{Characterizing EP241107a: multiwavelength observations of an Einstein Probe-detected fast X-ray transient}",
      journal = {\mnras},
         year = 2026,
        month = jan,
       volume = {545},
       number = {1},
          eid = {staf2062},
        pages = {staf2062},
          doi = {10.1093/mnras/staf2062},
archivePrefix = {arXiv},
       eprint = {2511.02562},
 primaryClass = {astro-ph.HE},
       adsurl = {https://ui.adsabs.harvard.edu/abs/2026MNRAS.545f2062E}
}

@ARTICLE{2026ApJ...998..163Z,
       author = {{Zhou}, Hao and {Ren}, Jia and {Wang}, Chen-Wei and {Liu}, Xing and {Liu}, Bin-Yang and {Levan}, Andrew J. and {Rastinejad}, Jillian C. and {Geng}, Jin-Jun and {Wang}, Hao and {Blanchard}, Peter K. and {Fong}, Wen-fai and {Gompertz}, Benjamin P. and {Malesani}, Daniele B. and {Kilpatrick}, Charles D. and {Lamb}, Gavin P. and {Metzger}, Brian D. and {Nicholl}, Matt and {Tanvir}, Nial R. and {Wang}, Yun and {Rong}, Yu and {Liang}, Run-Duo and {Ling}, Zhi-Xing and {Xu}, Dong and {Jin}, Zhi-Ping and {Wei}, Da-Ming},
        title = "{EP241217a: A Likely Type II GRB with an Achromatic Bump at z = 4.59}",
      journal = {\apj},
         year = 2026,
        month = feb,
       volume = {998},
       number = {1},
          eid = {163},
        pages = {163},
          doi = {10.3847/1538-4357/ae29b5},
archivePrefix = {arXiv},
       eprint = {2512.07233},
 primaryClass = {astro-ph.HE},
       adsurl = {https://ui.adsabs.harvard.edu/abs/2026ApJ...998..163Z}
}

@ARTICLE{2026arXiv260115732F,
       author = {{Fraija}, Nissim and {Galv{\'a}}, Antonio and {Betancourt Kamenetskaia}, Boris and {Dainotti}, Maria G},
        title = "{GRB\raisebox{-0.5ex}\textasciitilde250704B/EP250704a a Short Gamma-Ray Burst Powered by a Magnetar}",
      journal = {arXiv e-prints},
         year = 2026,
        month = jan,
          eid = {arXiv:2601.15732},
        pages = {arXiv:2601.15732},
          doi = {10.48550/arXiv.2601.15732},
archivePrefix = {arXiv},
       eprint = {2601.15732},
 primaryClass = {astro-ph.HE},
       adsurl = {https://ui.adsabs.harvard.edu/abs/2026arXiv260115732F}
}

@ARTICLE{2026GCN.44722....1S,
       author = {{Svinkin}, D. and {Frederiks}, D. and {Lysenko}, A. and {Ridnaia}, A. and {Tsvetkova}, A. and {Ulanov}, M. and {Cline}, T. and {Konus-Wind Team}},
        title = "{Konus-Wind detection of GRB 260527A (short/hard) \raisebox{-0.5ex}\textasciitilde4 min before EP260527a}",
      journal = {GRB Coordinates Network},
         year = 2026,
        month = may,
       volume = {44722},
        pages = {1},
       adsurl = {https://ui.adsabs.harvard.edu/abs/2026GCN.44722....1S}
}

@ARTICLE{2026arXiv260627048V,
       author = {{van Hoof}, Agnes P.~C. and {Jonker}, Peter G. and {Tommel}, Lieke and {Quirola-V{\'a}squez}, Jonathan A. and {Mata S{\'a}nchez}, Daniel and {van Dalen}, Joyce N.~D. and {Levan}, Andrew J. and {Fraser}, Morgan and {Zabludoff}, Ann and {Torres}, Manuel A.~P. and {S{\'a}nchez-Sierras}, Javi and {Martin-Carrillo}, Antonio and {Wevers}, Thomas and {Berton}, Marco and {Yue}, Minghao and {Dhillon}, Vik S. and {Bauer}, Franz E. and {Littlefair}, Stuart P.},
        title = "{Optical observations of candidate host galaxies of eight fast X-ray transients}",
      journal = {arXiv e-prints},
         year = 2026,
        month = jun,
          eid = {arXiv:2606.27048},
        pages = {arXiv:2606.27048},
          doi = {10.48550/arXiv.2606.27048},
archivePrefix = {arXiv},
       eprint = {2606.27048},
 primaryClass = {astro-ph.HE},
       adsurl = {https://ui.adsabs.harvard.edu/abs/2026arXiv260627048V}
}

@ARTICLE{2026arXiv260617230E,
       author = {{Eyles-Ferris}, Rob A.~J.},
        title = "{The properties of tidal disruption event infrared counterparts produced by dust rings and inference of the observing angle}",
      journal = {arXiv e-prints},
         year = 2026,
        month = jun,
          eid = {arXiv:2606.17230},
        pages = {arXiv:2606.17230},
          doi = {10.48550/arXiv.2606.17230},
archivePrefix = {arXiv},
       eprint = {2606.17230},
 primaryClass = {astro-ph.HE},
       adsurl = {https://ui.adsabs.harvard.edu/abs/2026arXiv260617230E}
}

@ARTICLE{2025A&A...703L...2O,
       author = {{Oganesyan}, Gor and {Kammoun}, Elias and {Ierardi}, Annarita and {De Santis}, Alessio Ludovico and {Banerjee}, Biswajit and {Sobacchi}, Emanuele and {Aharonian}, Felix and {Macera}, Samanta and {Tiwari}, Pawan and {Mei}, Alessio and {Mohnani}, Shraddha and {Ascenzi}, Stefano and {Ronchini}, Samuele and {Branchesi}, Marica},
        title = "{Ultra-long MeV transient from a relativistic jet: A tidal disruption event candidate}",
      journal = {\aap},
         year = 2025,
        month = oct,
       volume = {703},
          eid = {L2},
        pages = {L2},
          doi = {10.1051/0004-6361/202556591},
archivePrefix = {arXiv},
       eprint = {2507.18694},
 primaryClass = {astro-ph.HE},
       adsurl = {https://ui.adsabs.harvard.edu/abs/2025A&A...703L...2O}
}

@INPROCEEDINGS{2022ASSL..465..245D,
       author = {{Dall'Osso}, Simone and {Stella}, Luigi},
        title = "{Millisecond Magnetars}",
    booktitle = {Astrophysics and Space Science Library},
         year = 2022,
       editor = {{Bhattacharyya}, Sudip and {Papitto}, Alessandro and {Bhattacharya}, Dipankar},
       series = {Astrophysics and Space Science Library},
       volume = {465},
        month = jan,
        pages = {245-280},
          doi = {10.1007/978-3-030-85198-9_8},
archivePrefix = {arXiv},
       eprint = {2103.10878},
 primaryClass = {astro-ph.HE},
       adsurl = {https://ui.adsabs.harvard.edu/abs/2022ASSL..465..245D}
}

@ARTICLE{2025arXiv250417034L,
       author = {{Li}, W.-X. and {Zhu}, Z.-P. and {Zou}, X.-Z. and {Geng}, J.-J. and {Liu}, L.-D. and {Wang}, Y.-H. and {Li}, R.-Z. and {Xu}, D. and {Sun}, H. and {Wang}, X.-F. and {Yu}, Y.-W. and {Zhang}, B. and {Wu}, X.-F. and {Yang}, Y. and {Filippenko}, A.~V. and {Liu}, X.-W. and {Yuan}, W.-M. and {Aguado}, D. and {An}, J. and {An}, T. and {Buckley}, D.~A.~H. and {Castro-Tirado}, A.~J. and {Fu}, S.-Y. and {Fynbo}, J.~P.~U. and {Howell}, D.~A. and {Hu}, J.-W. and {Jiang}, S.-Q. and {Kumar}, A. and {Mao}, J.-R. and {Maund}, J.~R. and {Liu}, X. and {Mockler}, B. and {Moskvitin}, A. and {Andrews}, M. and {Bom}, C.~R. and {Brink}, T.~G. and {Chatterjee}, K. and {Chen}, Y. and {Cheng}, H.-Q. and {Cooke}, J. and {Dai}, J.~L. and {Du}, G.-W. and {Erasmus}, N. and {Fang}, Y. and {Farah}, J. and {Goranskij}, V. and {Gritsevich}, M. and {Gu}, M. and {Guo}, Z. and {Hsiao}, E. and {Hu}, Y.-D. and {Hua}, Y.-L. and {Jacobson-Gal{\'a}n}, W. and {Jia}, S.-M. and {Jin}, C.-C. and {Kasliwal}, M.~M. and {Kilpatrick}, C.~D. and {Kumar}, B. and {Lei}, W.-H. and {Li}, C.-K. and {Li}, D.-Y. and {Li}, L.-P. and {Ling}, Z.-X. and {Liu}, Q.-C. and {Liu}, Y. and {Liu}, Y.-Q. and {L{\'o}pez-Oramas}, A. and {Maslennikova}, O. and {McCully}, C. and {Monageng}, I. and {Newsone}, M. and {Padilla Gonzalez}, E. and {Pan}, H.-W. and {Peng}, H.-W. and {Pignata}, G. and {Poidevin}, F. and {Potter}, S.~B. and {P{\'e}rez-Fournon}, I. and {Santana-Silva}, L. and {Santos}, A. and {Song}, C.-Y. and {Song}, F.-F. and {Spiridonova}, O. and {Sun}, N.-C. and {Sun}, X.-J. and {Terreran}, G. and {Wang}, L.-Z. and {Wang}, L.-F. and {Wang}, B.-T. and {Wang}, Z.-Y. and {Wu}, G.-L. and {Xiang}, D.-F. and {Xiao}, H.-F. and {Xu}, Y.-F. and {Xue}, S.-J. and {Yan}, S.-Y. and {Yang}, Y.-P. and {Yu}, L.-X. and {Zhang}, Y.-H. and {Zhang}, Y.-H. and {Zhang}, C. and {Zhang}, J.-H. and {Zhang}, J.-J. and {Zheng}, W. and {Zou}, H.},
        title = "{An extremely soft and weak fast X-ray transient associated with a luminous supernova}",
      journal = {arXiv e-prints},
         year = 2025,
        month = apr,
          eid = {arXiv:2504.17034},
        pages = {arXiv:2504.17034},
          doi = {10.48550/arXiv.2504.17034},
archivePrefix = {arXiv},
       eprint = {2504.17034},
 primaryClass = {astro-ph.HE},
       adsurl = {https://ui.adsabs.harvard.edu/abs/2025arXiv250417034L}
}

@article{PhysRevD.75.124018,
  title = {Inspiral, merger, and ring-down of equal-mass black-hole binaries},
  author = {Buonanno, Alessandra and Cook, Gregory B. and Pretorius, Frans},
  journal = {Phys. Rev. D},
  volume = {75},
  issue = {12},
  pages = {124018},
  numpages = {42},
  year = {2007},
  month = {Jun},
  publisher = {American Physical Society},
  doi = {10.1103/PhysRevD.75.124018},
  url = {https://link.aps.org/doi/10.1103/PhysRevD.75.124018}
}

@article{PhysRevD.100.044003,
  title = {Improving the NRTidal model for binary neutron star systems},
  author = {Dietrich, Tim and Samajdar, Anuradha and Khan, Sebastian and Johnson-McDaniel, Nathan K. and Dudi, Reetika and Tichy, Wolfgang},
  journal = {Phys. Rev. D},
  volume = {100},
  issue = {4},
  pages = {044003},
  numpages = {19},
  year = {2019},
  month = {Aug},
  publisher = {American Physical Society},
  doi = {10.1103/PhysRevD.100.044003},
  url = {https://link.aps.org/doi/10.1103/PhysRevD.100.044003}
}

@ARTICLE{2020CQGra..37q5001S,
       author = {{Stachie}, C. and {Canton}, T. Dal and {Burns}, E. and {Christensen}, N. and {Hamburg}, R. and {Briggs}, M. and {Broida}, J. and {Goldstein}, A. and {Hayes}, F. and {Littenberg}, T. and {Shawhan}, P. and {Veitch}, J. and {Veres}, P. and {Wilson-Hodge}, C.~A.},
        title = "{Search for advanced LIGO single interferometer compact binary coalescence signals in coincidence with Gamma-ray events in Fermi-GBM}",
      journal = {Classical and Quantum Gravity},
         year = 2020,
        month = sep,
       volume = {37},
       number = {17},
          eid = {175001},
        pages = {175001},
          doi = {10.1088/1361-6382/aba28a},
archivePrefix = {arXiv},
       eprint = {2001.01462},
 primaryClass = {gr-qc},
       adsurl = {https://ui.adsabs.harvard.edu/abs/2020CQGra..37q5001S}
}

@ARTICLE{2024PhRvD.109d2008E,
       author = {{Ewing}, Becca and {Huxford}, Rachael and {Singh}, Divya and {Tsukada}, Leo and {Hanna}, Chad and {Huang}, Yun-Jing and {Joshi}, Prathamesh and {Li}, Alvin K.~Y. and {Magee}, Ryan and {Messick}, Cody and {Pace}, Alex and {Ray}, Anarya and {Sachdev}, Surabhi and {Sakon}, Shio and {Tapia}, Ron and {Adhicary}, Shomik and {Baral}, Pratyusava and {Baylor}, Amanda and {Cannon}, Kipp and {Caudill}, Sarah and {Chaudhary}, Sushant Sharma and {Coughlin}, Michael W. and {Cousins}, Bryce and {Creighton}, Jolien D.~E. and {Essick}, Reed and {Fong}, Heather and {George}, Richard N. and {Godwin}, Patrick and {Harada}, Reiko and {Kennington}, James and {Kuwahara}, Soichiro and {Meacher}, Duncan and {Morisaki}, Soichiro and {Mukherjee}, Debnandini and {Niu}, Wanting and {Posnansky}, Cort and {Toivonen}, Andrew and {Tsutsui}, Takuya and {Ueno}, Koh and {Viets}, Aaron and {Wade}, Leslie and {Wade}, Madeline and {Waratkar}, Gaurav},
        title = "{Performance of the low-latency GstLAL inspiral search towards LIGO, Virgo, and KAGRA's fourth observing run}",
      journal = {\prd},
         year = 2024,
        month = feb,
       volume = {109},
       number = {4},
          eid = {042008},
        pages = {042008},
          doi = {10.1103/PhysRevD.109.042008},
archivePrefix = {arXiv},
       eprint = {2305.05625},
 primaryClass = {gr-qc},
       adsurl = {https://ui.adsabs.harvard.edu/abs/2024PhRvD.109d2008E}
}

@ARTICLE{2021ApJ...923..254D,
       author = {{Dal Canton}, Tito and {Nitz}, Alexander H. and {Gadre}, Bhooshan and {Cabourn Davies}, Gareth S. and {Villa-Ortega}, Ver{\'o}nica and {Dent}, Thomas and {Harry}, Ian and {Xiao}, Liting},
        title = "{Real-time Search for Compact Binary Mergers in Advanced LIGO and Virgo's Third Observing Run Using PyCBC Live}",
      journal = {\apj},
         year = 2021,
        month = dec,
       volume = {923},
       number = {2},
          eid = {254},
        pages = {254},
          doi = {10.3847/1538-4357/ac2f9a},
archivePrefix = {arXiv},
       eprint = {2008.07494},
 primaryClass = {astro-ph.HE},
       adsurl = {https://ui.adsabs.harvard.edu/abs/2021ApJ...923..254D}
}

\begin{appendix}

\section{Supplementary Material}

\subsection{cWB candidate temporally coincident with EP240918c}
\label{app}

\begin{figure}[h]
    \centering
    \includegraphics[width=1\linewidth]{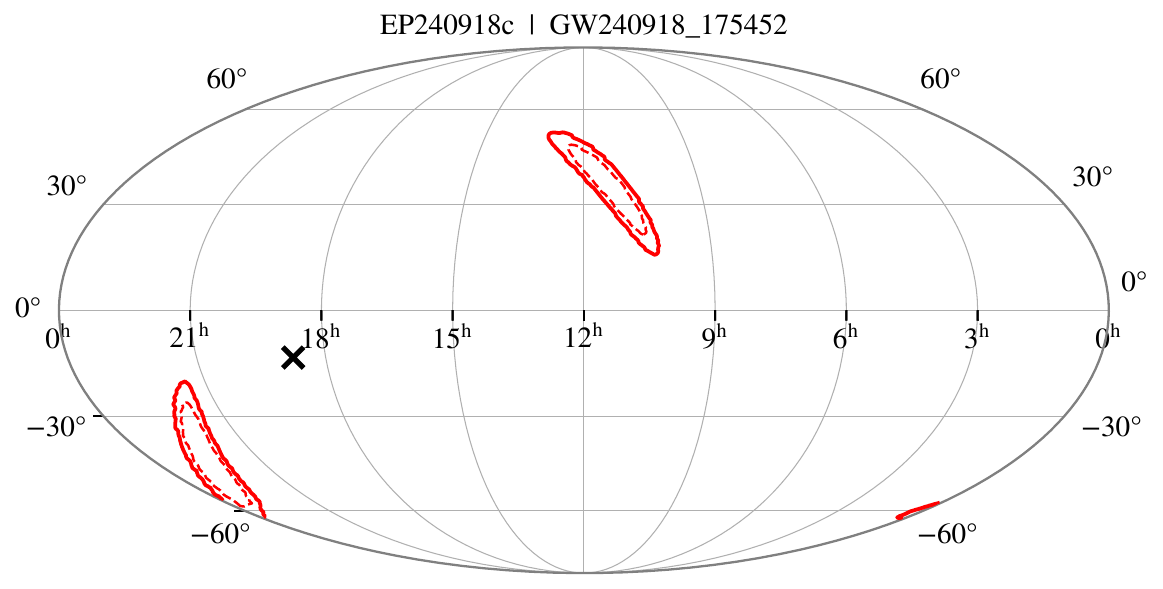}
    \caption{Sky localization of the cWB candidate GW240918\_175452, which is temporally coincident with EP240918c within the search window adopted in this work. The EP-FXT position is shown with the black cross, together with the dashed and solid contours showing the 50\% and 90\% credible regions of the GW localization. Although the candidate occurs within the temporal window, the EP position does not overlap with the high-probability region of the cWB skymap, and the event is therefore not considered spatially coincident.}
    \label{cWB}
\end{figure}

One additional GW candidate, GW240918\_175452 (superevent\_id S240918gl), falls within the search window of EP240918c, with a temporal offset of $-714.47~{\rm s}$ relative to the EP trigger time. This candidate was reported by the coherent WaveBurst (cWB) pipeline, with network SNR $=9.70$, FAR $=1.38\times10^{-5}$Hz (or $\sim$1.2/day), with H1--L1 online detector network. The source-classification probabilities reported for this candidate are $p_{\rm BBH}=0.0095$, $p_{\rm BNS}=0$, $p_{\rm NSBH}=0$, and $p_{\rm terr}=0.9905$, indicating that the event is most likely terrestrial in origin. cWB is an unmodelled burst search; this candidate is not identified by the CBC pipelines, and the skymap provided for the cWB triggers does not have the Bayes' factor for coherent and incoherent signal (BCI) used in our ranking-statistic analysis in Sec.~\ref{ranking statistic}. Therefore, we do not include it in the main FXT--GW association sample. We nevertheless inspect its sky localization, shown in Fig.~\ref{cWB}. Although the event is temporally coincident with EP240918c under our adopted search window, the EP-FXT position does not overlap with the high-probability region of the cWB localization. We therefore find no evidence for a spatial association between EP240918c and this cWB candidate.

\subsection{Choice of SNR threshold for the Targeted Detectability Range} \label{app:snr_threshold}

\begin{figure}[h]
    \centering
    \includegraphics[width=1\linewidth]{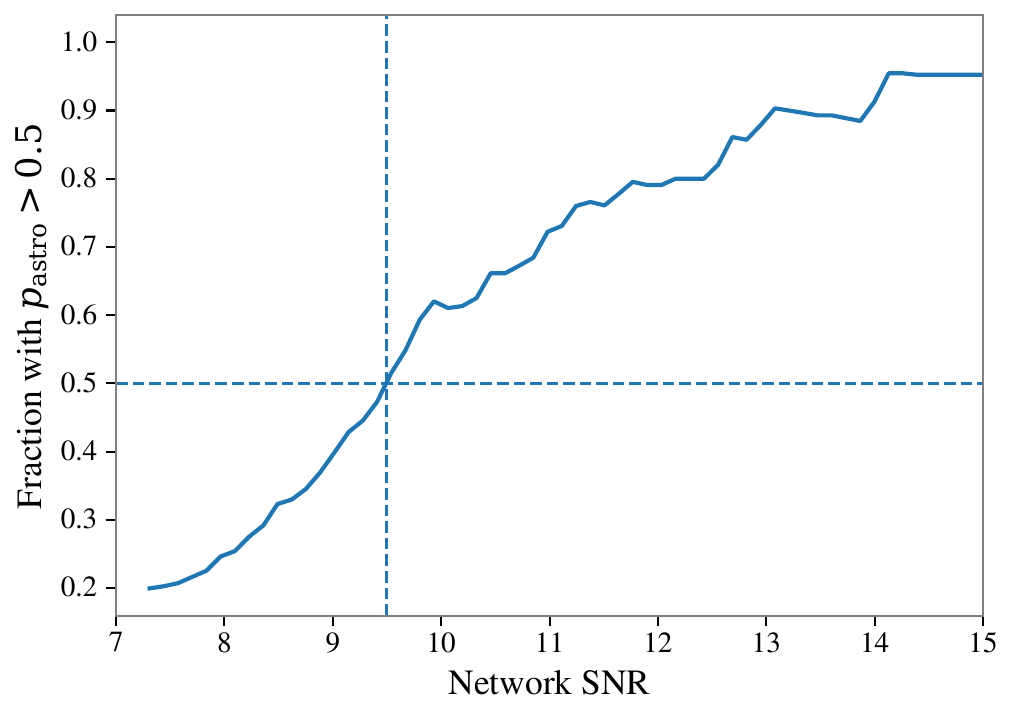}
    \caption{Fraction of GWTC-5 candidates with $p_\mathrm{astro}>0.5$ as a function of network SNR. The horizontal dashed line marks the 50\% level, while the vertical dashed line indicates the corresponding SNR threshold, $\rho \simeq 10.03$, motivating the choice of $\rho_\mathrm{cut}=10$ used for the TDR calculations.}
    \label{p-astro}
\end{figure}

The TDR exclusion distances in this work are computed at $\rho_\mathrm{cut} = 10$. To motivate this choice, we examine the astrophysical probability of GW candidates in the GWTC-5 catalog as a function of network SNR. $p_\mathrm{astro}$ represents the probability that a GW candidate is of astrophysical origin rather than being produced by terrestrial noise. It is inferred from the statistical significance of the event and the expected distributions of astrophysical and background triggers. Therefore, candidates with $p_\mathrm{astro}>0.5$ are more likely to have an astrophysical origin. Figure~\ref{p-astro} shows the fraction of sources with $p_\mathrm{astro} > 0.5$ as a function of network SNR, with the 50\% mark reached at SNR\,$\approx 9.5$. Below this value, fewer than half the detected candidates are more likely astrophysical than not, making it a natural lower bound for a meaningful detectability estimate. This is further supported by \citet{2026arXiv260521578R}, who compared TDR and PyGRB exclusion distances for short and ambiguous GRBs across O1, O2, and O3, finding the two methods agree most closely at $\rho_\mathrm{cut} \approx 9$. Our adopted threshold of $\rho_\mathrm{cut} = 10$ is consistent with this finding, and the small offset reflects the slightly more conservative choice driven by the $p_\mathrm{astro}$ analysis.

\subsection{TDR results on Zenodo} \label{A3}

The full set of TDR products generated for the EP-FXT sample is made available on Zenodo at \href{https://doi.org/10.5281/zenodo.20753673}{https://doi.org/10.5281/zenodo.20753673}. For each source, we provide the evolution of $D_{90}$ over the search window adopted in this work, $[T_0-1000~{\rm s},T_0+T_{100}]$. We also provide the BNS range map for the canonical configuration $m_1=m_2=1.4,M_\odot$, computed assuming the inclination prior $0^\circ<\iota<45^\circ$, together with the distance-efficiency curves for the BNS and NSBH configurations considered in this analysis. The efficiency curves are shown both for the restricted inclination prior $0^\circ<\iota<45^\circ$ and for an isotropic inclination distribution over $0^\circ<\iota<90^\circ$. The range maps and distance-efficiency curves are computed at the EP trigger time reported in Tab.~\ref{IFO-list}. These data products for each EP-FXT in our sample have been uploaded to Zenodo.

\begin{figure}
\centering
\includegraphics[width=0.5\textwidth]{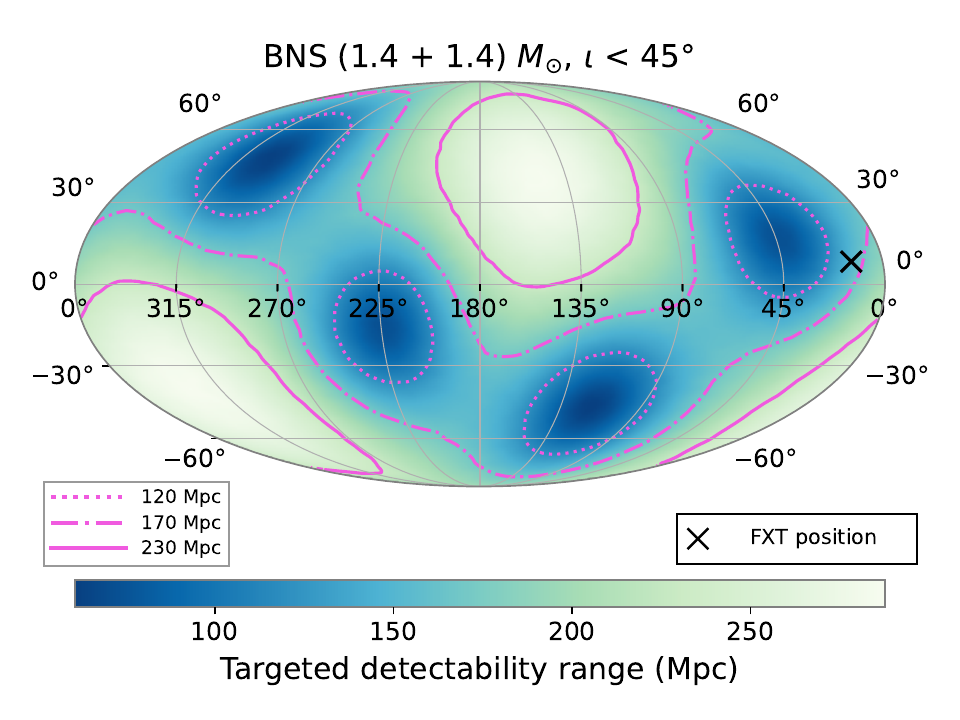}

\includegraphics[width=0.5\textwidth]{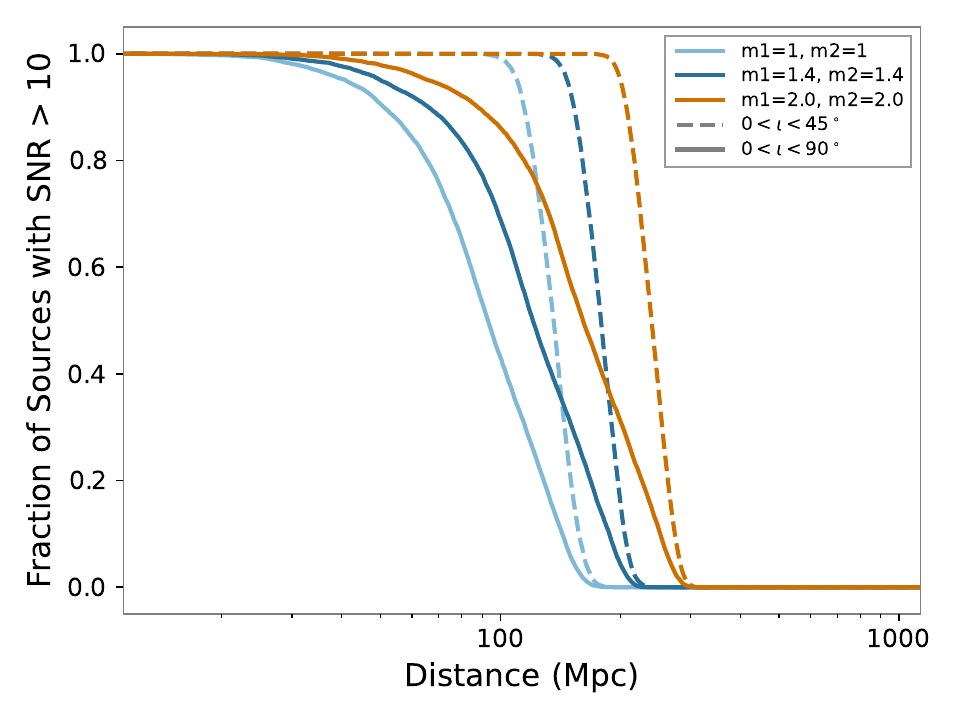}

\includegraphics[width=0.5\textwidth]{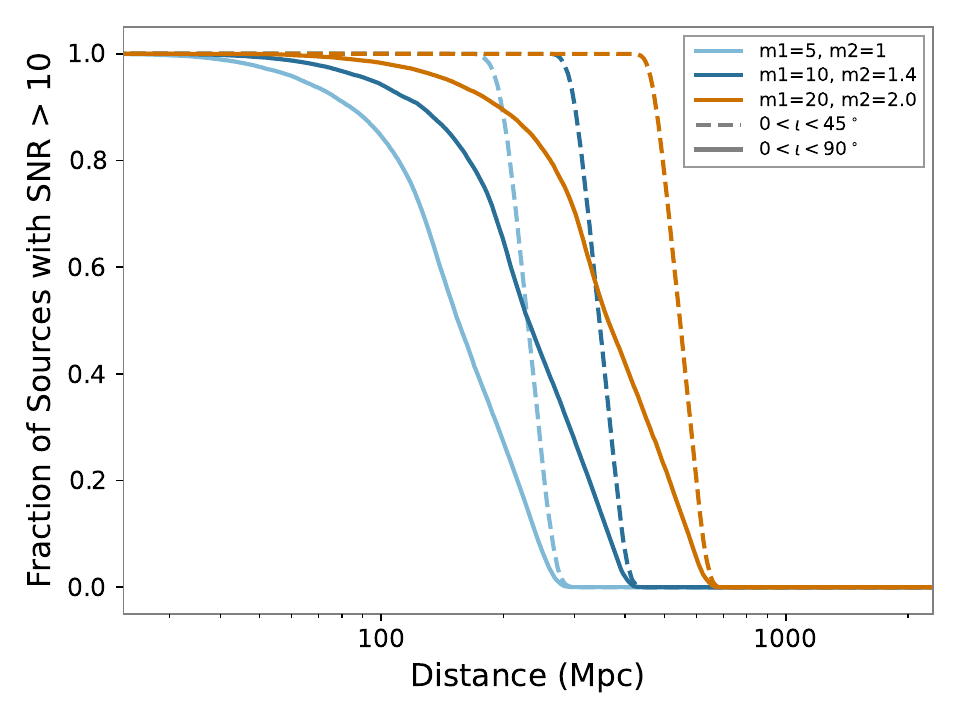}

\caption{Example of standard TDR products for EP240908a at the time of trigger. The upper panel shows the BNS range map for the canonical $m_1=m_2=1.4,M_\odot$ configuration at the FXT trigger time, assuming an inclination angle isotropically distributed in $0^\circ<\iota<45^\circ$. 
The lower panels show the BNS and NSBH distance-efficiency curves at the EP trigger time, for the restricted inclination prior $0^\circ<\iota<45^\circ$ and for an isotropic inclination distribution over $0^\circ<\iota<90^\circ$. In addition, a plot like Fig.~\ref{evol_d90} is also included in the Zenodo data release.}
\label{tdr_products}

\end{figure}

Figure~\ref{tdr_products} shows representative examples of the products included in the Zenodo release for EP240908a, the EP-FXT with the highest value of $D_{90}$ in our sample at the trigger time. The first panel shows the time evolution of $D_{90}$ across the full search window for the BNS and NSBH merger scenarios. Each point corresponds to a scanned TDR time bin, with a nominal bin spacing of 32 s, and the marker color gives the network antenna factor at the EP-FXT sky position. The detector-availability panel below shows which interferometers contribute to the usable detector network throughout the same search window. The remaining three panels are evaluated at the EP trigger time: the BNS and NSBH distance-efficiency curves, and the BNS range map for the canonical $m_1=m_2=1.4,M_\odot$ configuration.

\newpage

\begin{table*}[t]
\centering
\small
\setlength{\tabcolsep}{8pt}
\renewcommand{\arraystretch}{0.95}

\caption{Range of matched-filter 90\% exclusion distances, $D_{90}$, obtained over the GW search window for each Einstein Probe fast X-ray transient. The binary neutron star column assumes $m_1=m_2=1.4\,M_\odot$, while the neutron-star--black-hole column assumes $m_1=10\,M_\odot$ and $m_2=1.4\,M_\odot$. For each source, the quoted interval gives the minimum and maximum $D_{90}$ values across the analyzed time bins in $[T_0-1000~{\rm s},\,T_0+T_{100}]$. The detector columns give the percentage of the same search window for which each interferometer combination was available; None indicates the fraction of the window with no usable detector combination. These source-wise ranges correspond to the vertical extent shown in Fig.~\ref{D90_evol} and their distributions are summarized in Fig.~\ref{minmax_bns_nsbh}.}
\label{app_table}

\begin{tabular}{lccrrrrrrrr}
\toprule
Source & BNS & NSBH & H1  & L1  & V1  & H1L1  & H1V1  & L1V1  & H1L1V1  & None  \\
EP- & [Mpc] & [Mpc] & [\%] & [\%] & [\%] & [\%] & [\%] & [\%] & [\%] & [\%]\\
\midrule
EP240413a & 149--163 & 294--322 & 0.0 & 0.0 & 0.0 & 21.3 & 0.0 & 0.0 & 78.7 & 0.0 \\
EP240414a & 166--173 & 328--342 & 0.0 & 0.0 & 0.0 & 0.0 & 0.0 & 0.0 & 100.0 & 0.0 \\
EP240416a & 82--100 & 165--198 & 0.0 & 44.3 & 0.0 & 0.0 & 0.0 & 55.7 & 0.0 & 0.0 \\
EP240417a & 131--233 & 259--456 & 43.5 & 0.0 & 0.0 & 25.6 & 0.0 & 0.0 & 8.6 & 22.2 \\
EP240420a & 113--124 & 226--247 & 0.0 & 0.0 & 0.0 & 0.0 & 0.0 & 100.0 & 0.0 & 0.0 \\
EP240426b & 263--274 & 516--536 & 0.0 & 0.0 & 0.0 & 0.0 & 0.0 & 0.0 & 100.0 & 0.0 \\
EP240506a & 188--193 & 369--379 & 0.0 & 0.0 & 0.0 & 0.0 & 0.0 & 0.0 & 100.0 & 0.0 \\
EP240527a & 204--211 & 399--414 & 0.0 & 0.0 & 0.0 & 0.0 & 0.0 & 0.0 & 100.0 & 0.0 \\
EP240618a & 198--243 & 390--476 & 0.0 & 0.0 & 0.0 & 0.0 & 0.0 & 74.5 & 25.5 & 0.0 \\
EP240625a & 153--168 & 301--331 & 0.0 & 0.0 & 0.0 & 0.0 & 0.0 & 0.0 & 100.0 & 0.0 \\
EP240626a & 182--187 & 359--365 & 0.0 & 0.0 & 0.0 & 0.0 & 0.0 & 0.0 & 100.0 & 0.0 \\
EP240702a & 179--185 & 351--362 & 0.0 & 0.0 & 0.0 & 0.0 & 100.0 & 0.0 & 0.0 & 0.0 \\
EP240703b & 32--101 & 71--201 & 0.0 & 18.0 & 38.0 & 0.0 & 0.0 & 44.0 & 0.0 & 0.0 \\
EP240703c & - & - & 0.0 & 0.0 & 0.0 & 0.0 & 0.0 & 0.0 & 0.0 & 100.0 \\
EP240708a & 263--271 & 516--531 & 0.0 & 0.0 & 0.0 & 0.0 & 0.0 & 0.0 & 100.0 & 0.0 \\
EP240709a & 75--77 & 153--156 & 0.0 & 0.0 & 0.0 & 0.0 & 100.0 & 0.0 & 0.0 & 0.0 \\
EP240806a & 98--103 & 196--206 & 0.0 & 0.0 & 0.0 & 0.0 & 0.0 & 100.0 & 0.0 & 0.0 \\
EP240809a & 183--188 & 359--370 & 0.0 & 0.0 & 0.0 & 0.0 & 0.0 & 100.0 & 0.0 & 0.0 \\
EP240816a & 33--37 & 74--81 & 0.0 & 0.0 & 100.0 & 0.0 & 0.0 & 0.0 & 0.0 & 0.0 \\
EP240816b & 52--54 & 110--113 & 0.0 & 0.0 & 100.0 & 0.0 & 0.0 & 0.0 & 0.0 & 0.0 \\
EP240820a & 146--160 & 289--316 & 0.0 & 0.0 & 0.0 & 0.0 & 0.0 & 100.0 & 0.0 & 0.0 \\
EP240908a & 72--161 & 146--315 & 0.0 & 0.0 & 0.0 & 0.0 & 13.4 & 35.7 & 50.9 & 0.0 \\
EP240918a & 110--113 & 220--225 & 0.0 & 0.0 & 0.0 & 0.0 & 100.0 & 0.0 & 0.0 & 0.0 \\
EP240918b & 110--130 & 219--257 & 0.0 & 0.0 & 0.0 & 21.3 & 0.0 & 0.0 & 78.7 & 0.0 \\
EP240918c & 157--172 & 309--337 & 0.0 & 0.0 & 0.0 & 7.6 & 0.0 & 0.0 & 92.4 & 0.0 \\
EP241021a & 127--136 & 253--269 & 0.0 & 0.0 & 0.0 & 0.0 & 100.0 & 0.0 & 0.0 & 0.0 \\
EP241026b & 18--21 & 43--50 & 0.0 & 0.0 & 100.0 & 0.0 & 0.0 & 0.0 & 0.0 & 0.0 \\
EP241101a & 117--119 & 233--238 & 0.0 & 0.0 & 0.0 & 0.0 & 0.0 & 100.0 & 0.0 & 0.0 \\
EP241103a & 141--145 & 280--286 & 0.0 & 0.0 & 0.0 & 0.0 & 0.0 & 100.0 & 0.0 & 0.0 \\
EP241107a & 111--124 & 221--247 & 0.0 & 0.0 & 0.0 & 0.0 & 0.0 & 100.0 & 0.0 & 0.0 \\
EP241113a & 87--89 & 177--180 & 0.0 & 100.0 & 0.0 & 0.0 & 0.0 & 0.0 & 0.0 & 0.0 \\
EP241119a & - & - & 0.0 & 0.0 & 0.0 & 0.0 & 0.0 & 0.0 & 0.0 & 100.0 \\
EP241125a & 170--176 & 332--346 & 0.0 & 0.0 & 0.0 & 0.0 & 0.0 & 0.0 & 100.0 & 0.0 \\
EP241126a & 34--39 & 74--86 & 0.0 & 0.0 & 100.0 & 0.0 & 0.0 & 0.0 & 0.0 & 0.0 \\
EP241201a & 236--241 & 464--473 & 0.0 & 0.0 & 0.0 & 0.0 & 0.0 & 0.0 & 100.0 & 0.0 \\
EP241202b & 111--182 & 222--359 & 38.9 & 0.0 & 0.0 & 0.0 & 16.8 & 0.0 & 44.2 & 0.0 \\
EP241206a & 97--111 & 194--219 & 0.0 & 0.0 & 0.0 & 0.0 & 0.0 & 0.0 & 100.0 & 0.0 \\
EP241208a & 23--25 & 54--58 & 0.0 & 0.0 & 100.0 & 0.0 & 0.0 & 0.0 & 0.0 & 0.0 \\
EP241212a & 261--266 & 511--520 & 0.0 & 0.0 & 0.0 & 0.0 & 0.0 & 0.0 & 100.0 & 0.0 \\
EP241217a & 9--11 & 22--26 & 0.0 & 0.0 & 100.0 & 0.0 & 0.0 & 0.0 & 0.0 & 0.0 \\
EP241223a & - & - & 0.0 & 0.0 & 0.0 & 0.0 & 0.0 & 0.0 & 0.0 & 100.0 \\
EP241231b & 168--182 & 331--356 & 0.0 & 0.0 & 0.0 & 24.2 & 0.0 & 0.0 & 75.8 & 0.0 \\
EP250101a & 226--237 & 443--463 & 0.0 & 0.0 & 0.0 & 0.0 & 0.0 & 0.0 & 100.0 & 0.0 \\
EP250108a & 141--230 & 279--450 & 0.0 & 0.0 & 0.0 & 15.5 & 17.9 & 0.0 & 66.5 & 0.0 \\
EP250109b & 114--117 & 227--233 & 0.0 & 0.0 & 0.0 & 0.0 & 100.0 & 0.0 & 0.0 & 0.0 \\
EP250111a & 142--149 & 283--295 & 0.0 & 0.0 & 0.0 & 0.0 & 100.0 & 0.0 & 0.0 & 0.0 \\
EP250125a & 172--184 & 339--362 & 0.0 & 0.0 & 0.0 & 0.0 & 0.0 & 0.0 & 100.0 & 0.0 \\[2pt]
\hline 
\addlinespace
average & 133 - 150 & 264 - 297 & 1.75	&3.45	&13.57	&2.46	&13.79	&19.36	&38.75 & 6.86 \\
\bottomrule
\end{tabular}
\end{table*}

\end{appendix}
\end{document}